# Evidence for diffuse molecular gas and dust in the hearts of gamma-ray burst host galaxies

## Unveiling the nature of high-redshift damped Lyman-$\alpha$ systems

J. Bolmer[1,2], C. Ledoux[1], P. Wiseman[3], A. De Cia[4], J. Selsing[5], P. Schady[2,6], J. Greiner[2], S. Savaglio[7], J. M. Burgess[2],
V. D'Elia[8], J. P. U. Fynbo[5], P. Goldoni[9], D. Hartmann[10], K. E. Heintz[11], P. Jakobsson[11], J. Japelj[12], L. Kaper[12],
N. R. Tanvir[13], P. M. Vreeswijk[14], and T. Zafar[15]

*(Affiliations can be found after the references)*

Received; accepted

**ABSTRACT**

*Context.* Damped Lyman-$\alpha$ absorption-line systems at the redshifts of gamma-ray burst afterglows offer a unique way to probe the physical conditions within star-forming galaxies in the early Universe.
*Aims.* Here we built up a large sample of 22 GRBs at redshifts $z > 2$ observed with VLT/X-shooter in order to determine the abundances of hydrogen, metals, dust, and molecular species. This allows us to study the metallicity and dust depletion effects in the neutral interstellar medium at high redshift and to answer the question whether (and why) there might be a lack of $H_2$ in GRB-DLAs.
*Methods.* We developed new, state-of-the-art methods based on the Bayesian inference package, PyMC, to fit absorption lines and measure the column densities of different metal species as well as atomic and molecular hydrogen. The derived relative abundances are used to fit dust depletion sequences and determine the dust-to-metals ratio and the host-galaxy intrinsic visual extinction. Additionally, we searched for the absorption signatures of vibrationally-excited $H_2$ and carbon monoxide.
*Results.* There is no lack of $H_2$-bearing GRB-DLAs. We detect absorption lines from molecular hydrogen in 6 out of 22 GRB afterglow spectra, with molecular fractions ranging between $f \simeq 5 \cdot 10^{-5}$ and $f \simeq 0.04$, and claim tentative detections in three other cases. For the remaining sample, we measure, depending on S/N, spectral coverage and instrumental resolution, more or less stringent upper limits. The GRB-DLAs in the present sample have on average low metallicities ($\overline{[X/H]} \approx -1.3$), comparable to the rare population of extremely strong QSO-DLAs (log $N(H_I) > 21.5$). Furthermore, $H_2$-bearing GRB-DLAs are found to be associated with significant dust extinction, $A_V > 0.1$ mag, and have dust-to-metals ratios $\mathcal{DTM} > 0.4$. All of these systems exhibit neutral hydrogen column densities of log $N(H_I) > 21.7$. The overall fraction of $H_2$ detections is $\geq 27\%$ (41% including tentative detections), which is three to four times larger than in the general population of QSO-DLAs. For $2 < z < 4$, and considering column densities log $N(H_I) > 21.7$, the $H_2$ detection fraction in GRB-DLAs as well as in extremely strong QSO-DLAs is 60–80% in both cases. This is likely a consequence of the fact that both GRB- and QSO-DLAs with high neutral hydrogen column densities probe sight-lines with small impact parameters that indicate that the absorbing gas is associated with the inner regions of the absorbing galaxy, where the gas pressure is higher and the conversion of $H_I$ to $H_2$ takes place. In the case of GRB hosts, this diffuse molecular gas is located at distances $\gtrsim 500$ pc and hence is unrelated to the star-forming region where the GRB explosion occurred.

## 1. Introduction

Luminous background sources like quasars (QSOs) or gamma-ray bursts (GRBs) offer a unique way to probe star-forming regions in high redshift galaxies. Usually, strong Lyman-$\alpha$ absorption features, the so-called damped Lyman-$\alpha$ absorbers (DLAs) with log $N(H_I) > 20.3$ (Wolfe et al. 2005), are evident in the optical spectra of gamma-ray burst afterglows (Fynbo et al. 2009; Tanvir et al. 2018). DLAs contain most of the neutral gas in the Universe and represent a reservoir of gas available for star formation (McKee & Ostriker 2007; Altay et al. 2011). Spectra of GRB afterglows can thus be used to study the different phases of the gas detected in absorption, including circumgalactic and the interstellar medium (ISM), as well as dust and molecular phases (for a recent review see Schady 2017). Molecular hydrogen is of particular interest because its presence is tightly correlated with the formation of stars (Bigiel et al. 2008, 2011; Leroy et al. 2013; Krumholz et al. 2012). However, due to the lacking dipole moment and the low mass of the molecule, it is hard to study $H_2$ in emission at high redshift ($z \gtrsim 2$), because the rotational transitions require high temperatures to be excited (Kennicutt & Evans 2012). Therefore carbon monoxide is commonly used as a tracer of molecular gas, but with the drawback that the conversion factor to $H_2$ is still uncertain and likely also varies on different scales and for different gas properties (Tacconi et al. 2008; Bolatto et al. 2013; Gong et al. 2018). Fortunately, for DLAs at redshifts larger than $z > 2$ the Lyman and Werner bands of molecular hydrogen are shifted into the observed UV band and $H_2$ absorption lines can thus be detected in spectra of GRBs and QSOs obtained with ground-based instruments like VLT/X-shooter (Vernet et al. 2011).

The $H_2$ content of a star-forming region is the result of the balance between its formation on dust grains and its dissociation by UV photons from the interstellar radiation field. For GRB-DLAs also the UV flux from the GRB itself could potentially photodissociate the molecular gas or pump it to its vibrationally-excited states (Draine 2000; Draine & Hao 2002). Additionally, large dust masses or the molecular hydrogen itself can provide (self-)shielding against the destructive UV radiation (Krumholz et al. 2009). Dust, however, also efficiently attenuates the background UV flux from the GRB, which makes it difficult to obtain high-resolution spectra of very dusty sight lines ($A_V \gtrsim 1$ mag, Greiner et al. 2011), and adds to the difficulty of identifying the $H_2$ absorption lines and distinguish them from the Lyman-$\alpha$ forest. So, while it remains difficult to study dark molecular clouds ($A_V > 0.5$ mag), GRBs offer an ideal probe of the diffuse or translucent gas in high redshift galaxies.

Absorption from $H_2$ is generally detected in 10% or less of QSO-DLA systems (Ledoux et al. 2003; Noterdaeme et al. 2008; Balashev et al. 2014; Jorgenson et al. 2014), but with a strong dependence on the neutral hydrogen column density, with the fraction increasing when selecting QSO-DLAs with large neutral hydrogen column densities (Noterdaeme et al. 2015a). Recently Balashev et al. (2017) and Ranjan et al. (2018) also reported the detection of very large $H_2$ column densities in two other extremely strong DLAs (ESDLAs, defined as having log $N(H_I) \gtrsim 21.5$). Since GRBs are linked to star formation and originate from the inner regions of their host galaxies (Fruchter et al. 2006; Lyman et al. 2017), where they are usually behind very high column densities of gas (Jakobsson et al. 2006; Pontzen et al. 2008), one would expect to find a higher fraction of $H_2$ bearing systems in GRB-DLAs compared to QSO-DLAs (Zwaan & Prochaska 2006); also because the absorbing gas is located at distances from the explosion site where the influence of the ionizing UV radiation of the GRB itself should be negligible (50 to more than several hundred parsec; Prochaska et al. 2006; Vreeswijk et al. 2007, 2011; Ledoux et al. 2009; D'Elia et al. 2009; Hartoog et al. 2013). Yet, first searches for absorption from molecular hydrogen in small samples of GRB-DLAs were surprisingly unsuccessful. For example, from the lack of molecular hydrogen in a sample of 5 GRB afterglows, Tumlinson et al. (2007) argued that a deficiency of molecular gas in GRB-DLAs compared to QSO-DLAs, could be the result of a combination of lower metallicity and a stronger UV radiation field in GRB host galaxies. Later, from the analysis of the physical conditions in a sample of 7 GRB-DLAs, Ledoux et al. (2009) concluded that the lack of $H_2$ can be explained by the low metallicities and depletion factors, but also by the moderate particle densities in the systems: large ($\gtrsim 100$ pc), metal-poor atomic clouds with high temperatures ($T > 1000$ K). Since then, absorption from molecular hydrogen has only been identified in a handful of GRB-DLAs: GRB080607 (Prochaska et al. 2009), 120327A (D'Elia et al. 2014), 120815A (Krühler et al. 2013), and 121024A (Friis et al. 2015) (possibly also in GRB 060206 Fynbo et al. 2006).[1]

Recently, Selsing et al. (2018) published the X-shooter GRB optical afterglow legacy sample, which is the result of an extensive follow-up of 103 GRB afterglows with VLT/X-shooter between 2009 and 2017. Based on their work, we here aim to use a sub-sample of 22 GRBs at redshifts $z > 2$ to perform a systematic search for absorption lines from molecular hydrogen, in order to answer the question whether, and if so why, there is a lack of $H_2$ in GRB-DLAs. A full exploration of the column densities of the entire X-shooter sample will be presented in Thöne (2018, in prep.). For the purpose of the present paper, we only use such measurements to analyze the metal and dust depletion characteristics of the GRB-DLAs where $H_2$ can be searched for.

---

[1] Recently, also the molecule $CH^+$ was detected in the spectrum of GRB 140506A at $z = 0.889$ (Fynbo et al. 2014).

**Table 1.** Our sample of 22 GRBs afterglows at $z > 2$ observed with X-shooter. From the initially 33 selected bursts, 11 were excluded because of poor S/N, which are additionally listed at the bottom of the table

| GRB yymmdd# | Redshift (z) | Instrumental Resolution (FWHM) | | | S/N |
|---|---|---|---|---|---|
| | | UVB (km/s) | VIS (km/s) | NIR (km/s) | |
| 090809A | 2.7373 | 50.8 | 29.4 | 44.8 | 3 |
| 090926A | 2.1069 | 49.1 | 28.4 | 42.2 | 14 |
| 100219A | 4.6676 | 48.7 | 28.7 | 46.9 | 3 |
| 111008A | 4.9910 | - | 22.7 | 38.5 | 5 |
| 111107A | 2.8930 | 49.3 | 20.6 | 40.5 | 4 |
| 120327A | 2.8143 | 47.9 | 29.0 | 41.0 | 15 |
| 120712A | 4.1719 | 53.2 | 34.5 | 55.6 | 2 |
| 120716A | 2.4874 | 56.1 | 33.0 | 48.9 | 4 |
| 120815A | 2.3582 | 48.6 | 26.8 | 48.4 | 6 |
| 120909A | 3.9290 | 60.1 | 32.0 | 55.6 | 11 |
| 121024A | 2.3005 | 50.2 | 24.6 | 38.0 | 5 |
| 130408A | 3.7579 | 50.6 | 22.9 | 38.5 | 12 |
| 130606A | 5.9127 | - | 24.2 | 46.2 | 15 |
| 140311A | 4.9550 | - | 28.5 | 40.4 | 3 |
| 141028A | 2.3333 | 51.2 | 30.6 | 44.7 | 4 |
| 141109A | 2.9940 | 51.7 | 29.9 | 44.9 | 9 |
| 150403A | 2.0571 | 51.9 | 29.9 | 44.9 | 6 |
| 151021A | 2.3297 | 52.7 | 28.3 | 46.9 | 3 |
| 151027B | 4.0650 | 54.3 | 31.2 | 47.3 | 8 |
| 160203A | 3.5187 | 51.5 | 23.1 | 38.5 | 14 |
| 161023A | 2.7100 | 50.7 | 29.3 | 44.0 | 16 |
| 170202A | 3.6456 | 46.9 | 27.3 | 40.0 | 9 |

| Excluded | | Comment | |
|---|---|---|---|
| 100728B | 2.106 | obs. $T_0$ + 22.0 hrs | < 2 |
| 110128A | 2.339 | *dark* burst? high $A_V$? | < 2 |
| 120404A | 2.876 | obs. $T_0$ + 15.7 hrs, $A_V \sim 0.07$ mag | < 2 |
| 121201A | 3.385 | obs. $T_0$ + 12.0 hrs, $A_V \sim 0.17$ mag | < 2 |
| 121229A | 2.707 | poor seeing | < 2 |
| 130427B | 2.780 | obs. $T_0$ + 20.6 hrs, twilight | < 2 |
| 130612A | 2.007 | *dark* burst? high $A_V$? | < 2 |
| 131117A | 4.042 | $A_V$ < 0.20 mag, $z > 4$ | < 2 |
| 140515A | 6.327 | obs. $T_0$ + 15.5 hrs, $z > 6$ | < 2 |
| 140614A | 4.233 | $A_V$ < 0.50 mag, $z > 4$ | < 2 |
| 161014A | 2.823 | *dark* burst? anomalous extinction? | < 2 |

**Notes.** Columns from left to right: Name of the GBR; redshift; FWHM of the instrumental resolution in km/s inferred in the UVB, VIS, and NIR arm of the X-shooter spectra (as taken from Selsing et al. 2018), and the S/N ratio in the range of the expected $H_2$ absorption lines. For the 11 bursts at the end of the table that were excluded from the sample, we write a small comment indicating the reason for the low S/N.

The paper is organized as follows. In Sect. 2, we describe the $H_2$ GRB afterglow X-shooter sub-sample, and in Sect. 3 the absorption line analysis methods we have developed. Our main results are summarized in Sect. 4 and discussed in Sect. 5. We conclude in Sect. 6. Throughout the paper, when referring to GRBs, we refer to the class of long-duration GRBs, and column densities are given in $cm^{-2}$.

## 2. Sample selection

The GRB sample presented in this paper was selected based on the complete sample of GRB afterglows observed with X-shooter as presented in Selsing et al. (2018). This complete sample is selected to be unbiased with respect to the parent *Swift*

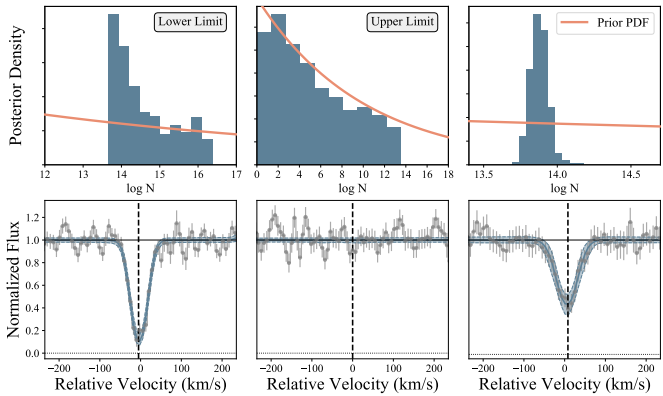

**Fig. 1.** Typical posterior distributions for a saturated line, a non-detection and a line which can be used to measure a column density. In case of a non-detection, the posterior distribution is completely determined by the prior PDF, which is expected in case there is no information about the assumed model, in this case a Voigt profile like absorption, in the data.

GRB sample, while at the same time optimising observability. This was done on the basis of a set of selection criteria, as described in Selsing et al. (2018). Out of the full statistical sample, afterglow observations were possible for 93 bursts. From this complete sample of 93 GRBs, 41 are at redshift $z > 1.7$, which allows measuring the neutral hydrogen column density (see their Table 4). From these 41 bursts, we further select the 33 bursts at redshift $z > 2.0$, where enough Lyman and Werner lines of molecular hydrogen are shifted into the X-shooter UVB arm, which is required to properly constrain the $H_2$ column density. This sample of 33 GRBs is presented in Table 1, where we list the individual redshifts as well as the instrumental resolution in each X-shooter arm, as taken from Selsing et al. (2018), and also the S/N ratio measured in the range of the expected $H_2$ absorption lines.

From these 33 bursts, however, only 22 have X-shooter spectra with a S/N ratio that is high enough to perform a rigorous analysis of the $H_2$ absorption lines (S/N $\geq$ 2). Therefore, 11 bursts were excluded from the final sample and are consequently listed at the bottom of the table, including a small note indicating the reason the S/N might be so low. This selection does not come without possible biases, which are important to address, and which we discuss below.

First of all, it is usually found that about 20% to 40% of the GRB population, the so-called *dark* bursts, are behind significant amounts of dust ($A_V > 0.5$ mag) along the GRB host galaxies' line of sight (Perley et al. 2011; Greiner et al. 2011; Covino et al. 2013), although we note that the fraction might get lower at redshift $z > 4$ (Bolmer et al. 2018; Zafar et al. 2018a). This is important, because the attenuation of the GRB afterglow by dust usually increases from red to blue wavelengths, which has a stronger effect on the observed spectrum with increasing redshift of the GRB, meaning that dusty and/or high redshift bursts are likely missed in spectral follow-up campaigns. For example, GRB 110128A, 130612A, and 161014A, that were excluded from the sample, were already very faint only several hours after the GRB trigger and might, therefore, be behind substantial dust columns. And, since all 22 GRBs in our sample are only behind small to moderate amounts of dust ($A_V < 0.5$ mag, see Table A.3), our sample might very well be biased toward sight lines with small foreground host galaxy dust columns. All the other burst that were excluded from the sample could only be observed later than $T_0 + 12$ hrs or were observed under poor seeing conditions, and their exclusion should therefore not add any biases to the sample selection.

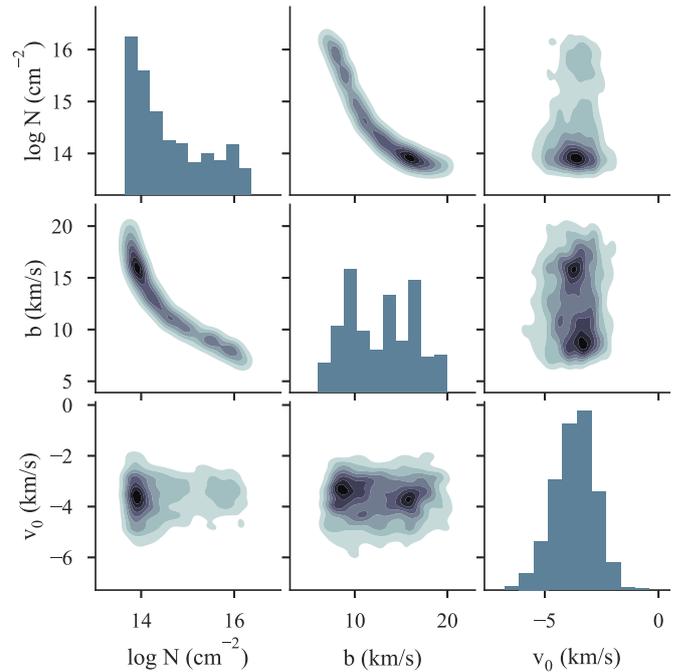

**Fig. 2.** Pairwise contour plots and posterior distributions for the column density, the broadening parameter, and the position of the saturated absorption line shown in Fig. 1.

Another bias, however, might be introduced by including GRBs at redshift $z > 4$, where the Lyman$-\alpha$ forest becomes an increasing hindrance in detecting the absorption lines from molecular hydrogen, which all fall blue-wards of 1215 Å. Also, at redshift $z > 4$ the Lyman blanketing significantly reduces the observed flux bluewards of Ly$\alpha$. This is the case for 6 of the 22 GRBs (27%), which will thus be discussed and labeled separately throughout the paper. Also, three of the excluded bursts are at redshift $z > 4$.

## 3. Methods

Kinematic or velocity profiles of absorption lines are a powerful tool to analyze the different gas phases in DLA absorbers. The predominant absorption features imprinted on the afterglow spectrum of the GRB are from singly-ionized metal species, residing in the cold, neutral gas of the GRB host galaxy. In most cases, also absorption lines from higher ionization species are seen in the spectra, which trace the warm, ionized gas and therefore usually have a different kinematic profile, offset to the singly-ionized lines (Fox et al. 2008; Wiseman et al. 2017a; Heintz et al. 2018). Since we are only interested in the neutral gas phase, these high ionization species will not be considered in this paper.

To analyze and fit the absorption lines in the afterglow spectra in our sample of 22 GRBs, we developed our own, state-of-the-art routines, which are entirely written in `python` and are based on the Markov chain Monte Carlo (MCMC) Bayesian inference library `PyMC 2.3.7`[2].

---

[2] `https://pymc-devs.github.io/pymc/`

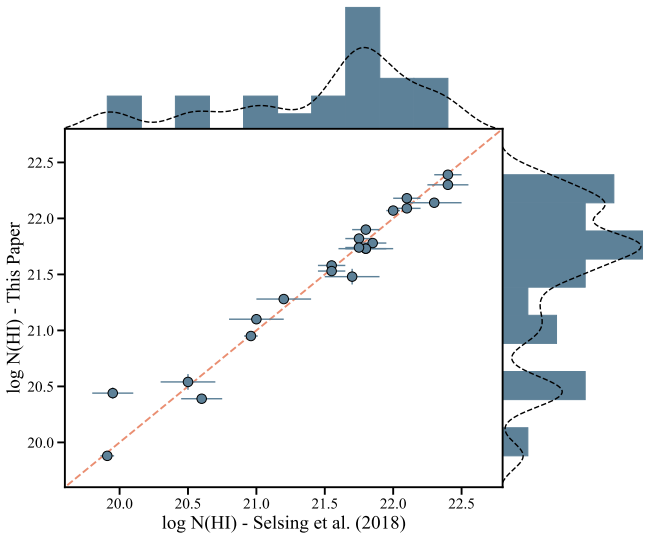

**Fig. 3.** The neutral hydrogen column densities determined in this paper versus those from Selsing et al. (2018). The dashed line is the line of equality. Additionally we show histograms and corresponding kernel density estimates.

Absorption lines are usually modeled with Voigt line profiles, which are fitted to the spectra simultaneously with the continuum flux of the GRB. The model for such an absorption profile in velocity space with a number of $n$ absorption components is described by

$$F_{\text{model}} = F_{\text{c}} \cdot \prod_{j=1}^{n} e^{-\tau_j}; \quad F_{\text{c}} = A + B(v - v_0) + C(v - v_0)^2 \quad (1)$$

where the optical depth $\tau$ is given by

$$\tau = \frac{\pi e^2}{m_e c} f_{ik} \lambda_{ik} N \cdot \phi_V \left(v - v_0, \frac{b}{\sqrt{2}}, \Gamma\right) \quad (2)$$

where $\phi_V$ is the Voigt profile in velocity space, and with $f_{ik}$, $\lambda_{ik}$ and $\Gamma$, being the oscillator strength, wavelength and damping constant for transition $i \rightarrow k$. The relative velocity of the absorption component $v_0$, the broadening parameter $b$, the column density $N$, and possibly also the number of components $n$, are the parameters of interest for our analysis. For the continuum flux of the GRB $F_{\text{c}}$, which is usually best described by a simple power law, we assume a second order polynomial, which introduces another three free parameters ($A$, $B$, and $C$). The Voigt profile $\phi_V$ is a convolution of a Gaussian (thermal and turbulent motion) and a Lorentzian (uncertainty of transition frequencies) profile, and is related to the real part of the Faddeeva function $w(z)$ that we imported from the Python package `SciPy` [3].

$$\phi_V = \frac{\text{Re}(w(z))}{b\sqrt{\pi}}; \quad z = \frac{v - v_0 + i\Gamma}{b} \quad (3)$$

Finally, the absorption line profile is convolved with a Gaussian with a FWHM of the instrumental resolution in the given arm, as determined by Selsing et al. (2018) (see Table 1).

Given this model, our parameters of interest are constrained in the following way. First, we select transitions that display a distinct absorption profile, i.e. unblended lines that fall in a segment of the spectrum with a good S/N, usually Si II 1808.0129, Fe II 1608.4509, or Mg II 2803.5311, and compare them in order to identify the number of required absorption components. This is done by fitting these lines with multiple MCMC runs for a realistic range of the number $n$ of possible absorption components with each of a total of 100 000 iterations and with a warm-up phase of 95 000 iterations. For DLAs with three or more distinct absorption components we doubled or tripled the number of iterations, and for the relative velocity of each component we start by putting a uniform prior on the whole input velocity range, which was usually $v_0 \pm 1000$ km/s. After each run, we analyzed the resulting fit and posterior distributions and compared the models for a different number of components by calculating $\chi^2$ until a convincing solution was found. Once the required number $n$ and the relative velocity $v_0$ of each absorption component was also comparing the results from different lines, we fixed $n$ and continued to fit the other lines by using smaller uniform priors on $v_0$, usually allowing for a small variation of $v_0 \pm 20$ km/s. For the broadening parameter, we generally also used uniform priors between the minimum resolvable value, usually 5 to 15 km/s, and a maximum value of 80 km/s. Finally, for the column density, we chose an exponential prior, which ensures that lower column densities are preferred when there is no information about absorption in the data, i.e. the flux around the position of the expected absorption line is consistent with the noise. As an example, in Fig. 1, we show the posterior distributions for the column density of a single absorption line in three different cases: a saturated line, a non-detection, and a line which can be used to measure a column density.

As expected for the resolution of X-shooter (R $\approx$ 4000 to 17 000 or equivalently 75 to 18 km/s), even mildly saturated lines show strong degeneracies between the column density and the broadening parameter (*hidden saturation;* Prochaska 2006; Wiseman et al. 2017b), which to identify requires sufficient sampling of the parameter space, which is growing exponentially with increasing number of absorption components. The effect of hidden saturation is demonstrated in Fig. 2, where we show pairwise contour plots and posteriors distributions of the three parameters from the saturated line shown in Fig. 1. Since the column density and the broadening parameter are highly correlated we use the adaptive Metropolis algorithm implemented in `PyMC` for these parameters [4]. For all other parameters, also the three parameters from the second order polynomial for the background flux (normal priors around $A = 1$, $B = 0$ and $C = 0$ with variance $\sigma = 0.5$), we use the Metropolis–Hastings step method. Finally, we want to note, that for the resolution of X-shooter, individual absorption components are often blended with each other, and therefore the decomposition of the profile is not necessarily unambiguous, such that the resultant $b$-values might not be physical.

# 4. Results

## 4.1. Neutral hydrogen and metal abundances

The most prominent absorption feature in GRB afterglow spectra is the Lyman-$\alpha$ line, which is used to measure the neutral hydrogen column density $N(\text{H I})$. For each of the 22 bursts in the selected sample, $N$(H I) was previously inferred by Selsing et al. (2018), by simply plotting synthetic absorption lines over the normalized spectra until a satisfying match was reached (which is a standard procedure, e.g. Fynbo et al. 2009). We are able to derive slightly more accurate values by fitting the Lyman-$\alpha$ line by also including all absorption lines in the red and blue damping wing. The results are listed in Table A.1 and shown in Fig. 4. Those parts of the spectrum that were ignored for the fit, mostly due to telluric and Ly$\alpha$ forest lines or too complex absorption structures, are marked by the gray shaded areas. To compare the results from both methods, in Fig. 3 we additionally plot the values derived in this paper versus those from Selsing et al. (2018). Besides one outlier, GRB120712A, which has the lowest S/N, all values are consistent within the errors, confirming the robustness of both methods.

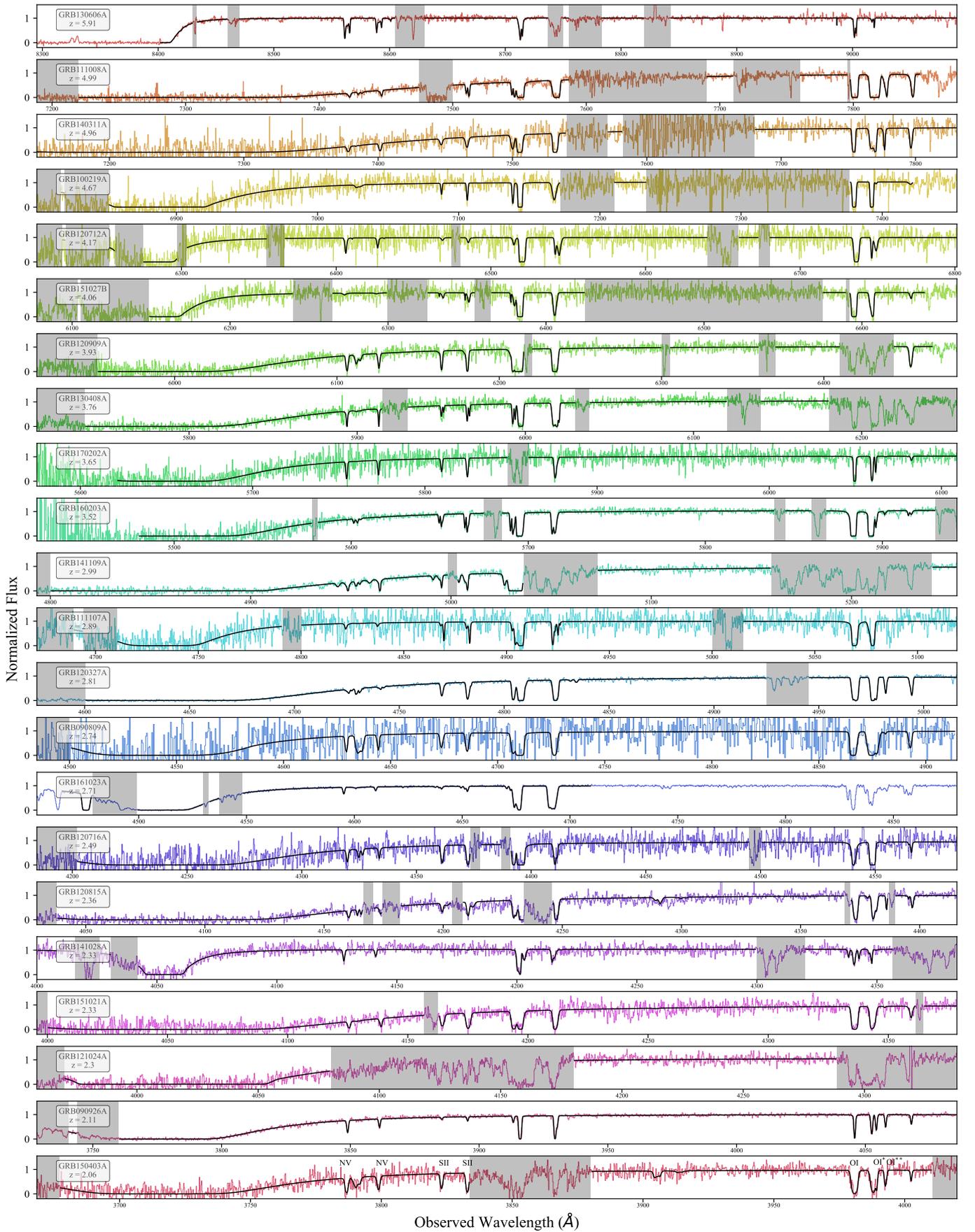

**Fig. 4.** Normalized X-shooter spectra around the Lyman-$\alpha$ line for the 22 GRBs analyzed in this paper with increasing redshift from the bottom to the top. The best fit model is indicated by the solid black line. The gray shaded areas show the regions that were ignored for the fit, mostly due to telluric lines or too complex absorption structures.

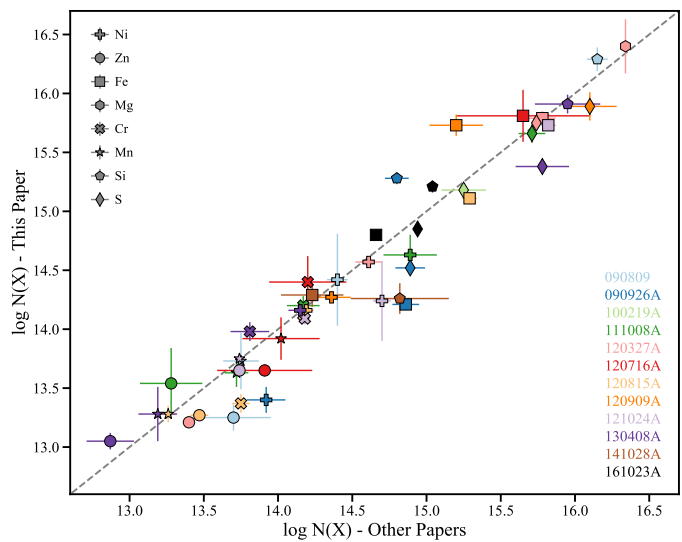

**Fig. 5.** Column densities as derived in this paper versus those published in Wiseman et al. (2017b), which are partially collected from various papers. Also plotted are the values for GRB 161023A from de Ugarte Postigo et al. (2018). The dashed line is the line of equality. The individual metal species X are indicated by different symbols (as labeled).

Furthermore, to measure the metallicity and study the dust depletion in each system, for all 22 GRB spectra we also fit lines and measure column densities of 10 different metal species, O I, Zn II, S II, Si II, Mg II, Mn II, Cr II, Ni II, Fe II, and Ti II, at least in those cases where it is possible and the spectrum is not affected by e.g. strong telluric lines. Besides Ti II, all of these absorption lines are commonly detected in GRB afterglow spectra (Christensen et al. 2011). In case of non-detections or saturated lines we determine 3-$\sigma$ upper or lower limits, respectively. All measured column densities, as well as upper or lower limits, are listed in Table A.1 and the corresponding plots showing the most constraining fits for each burst are shown in the Appendix in Fig. B.1 to B.22.

The values for the oscillator strength $f$ and damping constant $\Gamma$ are taken from the atomic data file (`atom.dat`) which is distributed in `VPfit` version 10.2 and originally based on Morton (2003). For Zn II and S II, however, we use the updated oscillator strength values from Kisielius et al. (2014, 2015) instead. While the difference in case of S II is almost negligible, the column densities of Zn II are about 0.1 dex lower when using the new values. We also note that in most cases, the three sulfur lines ($\lambda = 1250.58$, 1253.81, and 1259.52 Å) were fitted simultaneously with Ly$\alpha$, because they fall into the red part of the damping wing for column densities larger than log $N$(H I) ≳ 21.0, which is true for the majority of the bursts in the sample (see Fig. 4). Finally, since the two Zn II lines at $\lambda = 2026.14$, 2062.66 Å are blended with lines from Cr II and Mg I, we fit all of these lines simultaneously.

To compare the column densities measured in this paper with those previously published in the literature, in Fig. 5 we plot our values compared to those collected from Wiseman et al. (2017b). The latter are partially determined by Wiseman et al. (2017b) and otherwise taken from different papers, i.e.: GRB 090809 from Skúladóttir (2010), 090926A from D'Elia et al. (2010), 100219A from Thöne et al. (2013), 111008A from Sparre et al. (2014), 120327A from D'Elia et al. (2014), 120815A from Krühler et al. (2013), and 121024A from Friis et al. (2015). Additionally we also plot the values for GRB 161023A from de Ugarte Postigo et al. (2018). Since we are able to better identify any strong degeneracies in the $N$ versus $b$ parameters space, we are generally more conservative and put more upper or lower limits on the column densities compared to previous published values, which are not plotted in Fig. 5. Otherwise, our measurements are in reasonable agreement with the previous published values. The major differences can be explained by the updated oscillator strength values in case of Zn II, a better exploration of the parameter space in case of blended lines (Zn II and Cr II), or the in general advanced analysis method, which allows a better and more efficient exploration of the parameter space compared to the least squared approach of `VPfit`.

## 4.2. Dust depletion

Due to their apparent faintness, most of our knowledge on the elemental abundances in high redshift galaxies ($z > 2$), originates from absorption line studies of DLA systems in the spectra of QSOs and GRBs (Schady 2017)[5]. However, refractory elements, like Fe, Ni or Cr, can be heavily depleted by condensation onto interstellar dust grains, with different strength of the depletion depending on the element and the sight line (Savage & Sembach 1996; Jenkins 2009; De Cia et al. 2016). Thus, if some elements are partially locked onto dust grains, they are not completely accounted for when measuring the gas-phase abundances using absorption line analysis, and we, therefore, have to analyze the depletion in each sight line independently. The analysis of the dust depletion can thus be used to measure the dust content in the GRB host galaxies line of sight and to determine the dust-corrected metallicity [M/H] of the DLA.

As a first indicator of the strength of depletion, one can calculate the depletion factor [Y/Fe] $\equiv \log(N(Y)/N(Fe)) - \log(N(Y)/N(Fe))_{\odot}$, where Y is a non-refractory element that is only marginally depleted onto dust grains, like Zn or S, and is compared to iron, that is usually heavily depleted.

A more sophisticated way to study the dust depletion in each of the GRB-DLAs, is to compare the relative abundances of multiple refractory and non-refractory elements and use relations found in the Milky Way and QSO-DLAs to interpret the depletion pattern (Jenkins 2009; De Cia et al. 2013, 2016; Wiseman et al. 2017b). This method can be justified by the homogeneous properties regarding the metal and dust content found for different environments and redshift, that indicate that most of the dust content is built up from grain growth in the ISM, independent of the specific star formation history (De Cia 2018). To calculate

---

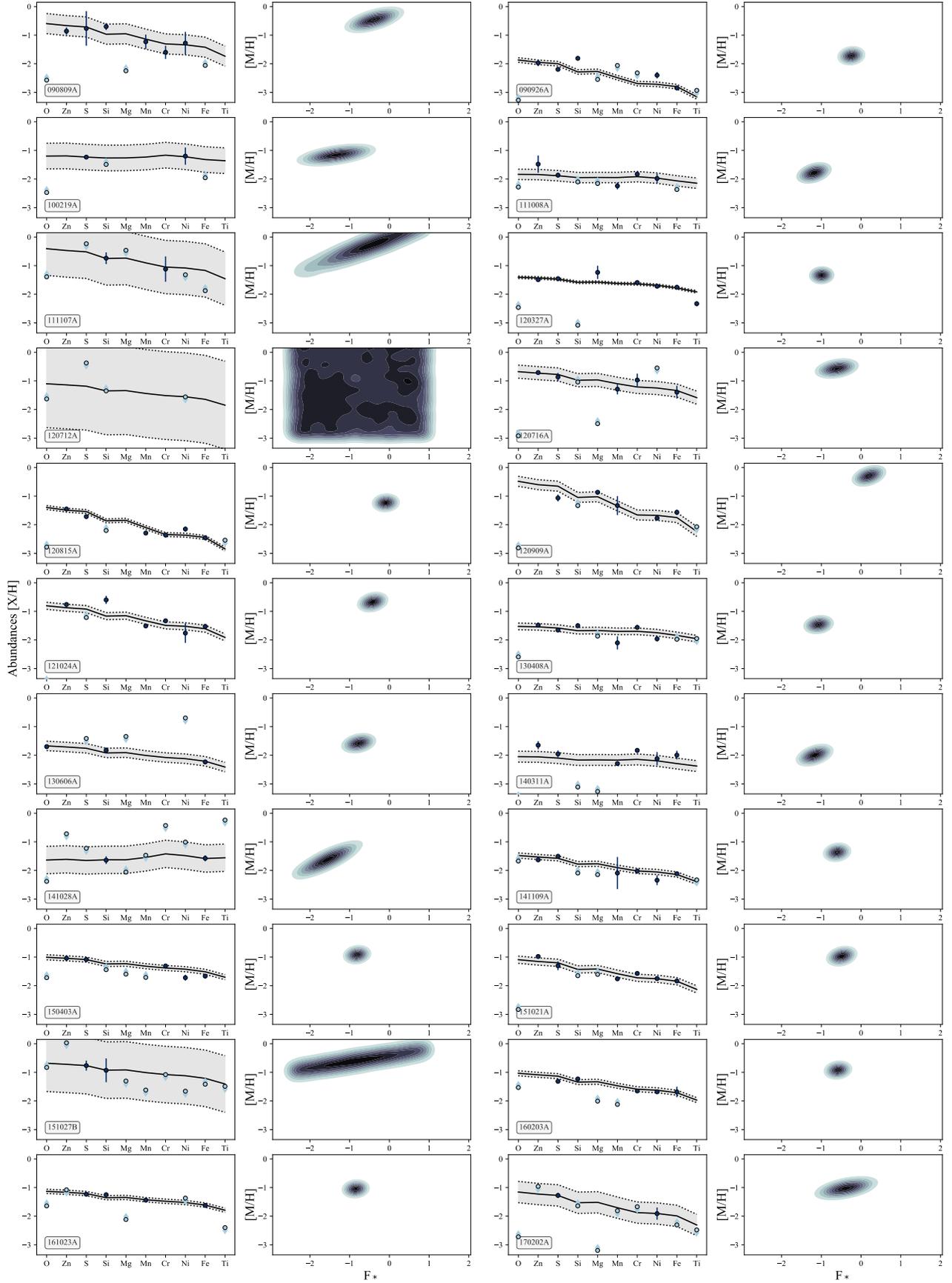

**Fig. 6.** Results from fitting the dust depletion sequences for the 22 GRBs in our H$_2$ sample. The relatives abundances of the elements used for the fit are plotted with dark blue dots. Upper and lower limits for the other elements are plotted in light blue. The black line indicates the best fit; and the dashed lines and gray shaded areas the corresponding 3-$\sigma$ confidence intervals. A contour plot of dust-corrected metallicity [M/H] versus $F_*$ is shown right next to every depletion sequence.

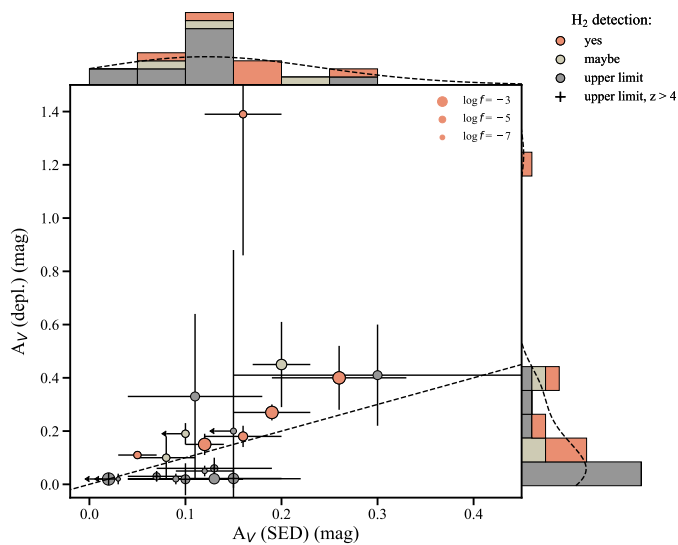

**Fig. 7.** Host intrinsic visual extinction as measured from the depletion analysis $A_V$ (depl.) versus the measurement from a fit to a spectral energy distribution from photometric data $A_V$ (SED). The dashed line is the line of equality.

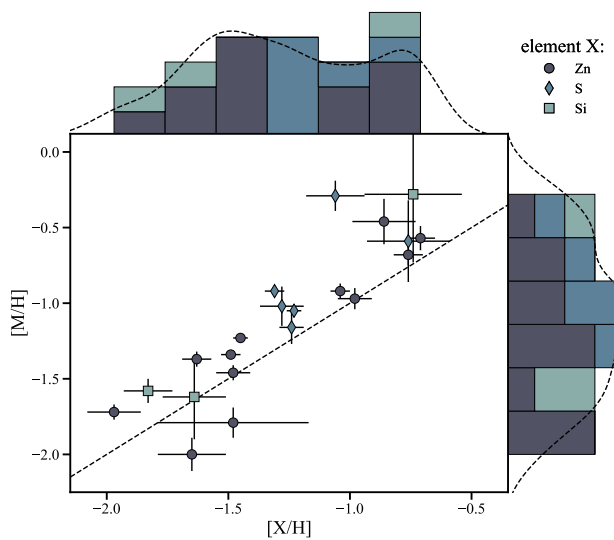

**Fig. 8.** Dust-corrected metallicity [M/H] as derived from the depletion analysis versus the metallicity derived using a single line of an element X with low depletion [X/H]. For each burst the used element is colored as labeled. The dashed line is the line of equality.

the relative abundances $[X/H] \equiv \log(N(X)/N(H)) - \log(X/H)_\odot$ we adopt the solar abundances from Asplund et al. (2009), following the recommendations of Lodders et al. (2009), as listed in Table 1 of De Cia et al. (2016). We also assume that our measurement of the column density of an element that is singly ionized is representative of the total gas phase abundance of that element. This is justified by the large pool of neutral gas $\log N(H\,I) \gtrsim 20.3$, found for all GRBs in our sample besides 130606A, indicating that only a small, negligible fraction of higher ionized elements should exist (Wolfe et al. 2005; Kanekar et al. 2011).

After calculating the relative abundances, our analysis of the dust depletion sequences was done following the work of Wiseman et al. (2017b), which is based on the results from De Cia et al. (2016). However, instead of using their minimum least squares approach, we again use a Bayesian framework and sample the posterior distributions using `PyMC`. For the depletion strength factor we use a uniform prior between $F_* = -2.5$ and 1.0 and for the dust-corrected metallicity of the system a uniform prior between [M/H] = $-3$ and 1.0. [6] We then run an MCMC with 100 000 iterations and sample the posterior distributions with the last 10 000 iterations after a warm-up phase of 90,000 iterations. The model for the expected relative abundance in element X is given by

$$[X/H]_{exp} = A + B \frac{F_* - 1.50}{1.48} + [M/H] \qquad (4)$$

where A and B are the updated linear depletion parameters taken from De Cia et al. (2016). The best fit $F_*$ and [M/H] can then be used to determine the dust-to-metals ratio $\mathcal{DTM}$ (normalized to the Milky Way), and further the expected visual extinction $A_V$ (depl.) in the host galaxy line of sight (see Sect. 4.2 and 6.2 in Wiseman et al. (2017b), and also Savaglio & Fall (2004) and De Cia et al. 2016). For all 22 GRBs, the results of the best fit depletion pattern, as well as the corresponding contour plots for the dust-corrected metallicity [M/H] versus the depletion strength factor $F_*$, are shown in Fig. 6 and the best fit results are also listed in Table A.3. Finally, we note that elements that are intrinsically under- or over-abundant, can badly influence the results of the depletion analysis in those cases where only a few elements are available. A detailed analysis of nucleosynthesis effects is beyond the scope of this paper and will be studied in a forthcoming paper (Thöne 2018, in prep.).

### 4.3. Visual extinction

Measurements of the host intrinsic visual extinction from fitting extinction curves to the afterglow SED, $A_V$ (SED), are available in the literature for all bursts from our $H_2$ sample. The individual values and corresponding references are listed in Table A.3. Most of them were taken from Bolmer et al. (2018), in case of the GRBs at $z > 4$, or from the measurements for the GROND 4h sample (Greiner et al. 2011; Greiner 2018, in prep.).

Following Wiseman et al. (2017b), in Fig. 7 we plot the host intrinsic visual extinction as derived from the depletion analysis $A_V$ (depl.) versus the measurement from the SED fit $A_V$ (SED). Similar to Wiseman et al. (2017b) and Savaglio & Fall (2004), we find that for some of the sight-lines the values are not consistent with being identical, and that data points can fall both below and above the line of equality. In contrast to Wiseman et al. (2017b), however, we find a better agreement between both methods, at least within the errors. Interestingly, most of the GRBs for which the extinction measured from the SED fit is higher, are from $z > 4$. This can thus be explained by the higher probability of intervening systems in the lines of sight, which can contain dust that only contributes to the extinction derived from the SED fit (Ellison et al. 2006; Ménard & Fukugita 2012). In addition, we note that the depletion-derived $A_V$ only probes non-carbonaceous dust, e.g. silicates or iron-rich grains, because it is based on the measurements of several metals, but not C. One possibility is therefore also that the $z > 4$ systems contain more C-rich dust, such as amorphous carbons or Polycyclic Aromatic Hydrocarbons (PAH).

---

[6] $F_*$ represents the strength of dust depletion in a system and was defined by Jenkins (2009) to range between 0 and 1 for the ISM in the Galaxy.

Data points that fall above the line of equality, i.e. sight-lines for which the depletion derived value is higher, could be the result of additional uncertainties coming from SED fitting; i.e. from the choice of extinction laws and whether the SED is fitted with a simple or broken power law (see also the discussion in Wiseman et al. 2017b). Due to these reasons, and also because the $A_V$ (SED) values are collected from different works and are thus not derived as consistently as the $A_V$ (depl.) values in this work, we believe that the latter is a better estimate for the GRB host galaxies' line of sight, and we will therefore in the following only discuss our results in respect to $A_V$ (depl.) where necessary.

### 4.4. Dust-corrected metallicities

In Fig. 8 we plot the dust-corrected metallicity [M/H] as derived from the depletion analysis versus the metallicity derived using a single line of an element X with low depletion [X/H]. As previously found by e.g. De Cia et al. (2018), we find that even Zn is slightly depleted onto dust grains, which explains that almost all of the data points fall above the line of equality. The two exception are GRB 111008A and 140311, which are both at redshift $z > 4$. Looking at the depletion pattern of these two bursts, it appears that Zn is intrinsically over-abundant compared to Fe, or more likely that Mn, which is an odd-Z element known to be intrinsically under-abundant in QSO-DLAs, is pulling the fit down. On average, the dust-corrected metallicities are with $\overline{[M/H]} = -1.09 \pm 0.50$ on about 0.2 dex higher than those derived from Zn or S ($\overline{[X/H]} = -1.26 \pm 0.37$), but also have larger errors. Nevertheless, since dust corrections were not applied to the QSO-DLA samples we want to compare our GRB-DLAs against, we will for the further analysis only use [X/H], but keep in mind that the actual metallicities might be slightly higher.

### 4.5. Molecular hydrogen

The Lyman and Werner absorption lines of molecular hydrogen, although numerous, are relatively hard to detect in the spectra of GRBs and QSOs, because they fall at restframe wavelengths $\lambda < 1150$ Å, in the Lyman-$\alpha$ forest, where the spectrum is speckled with random absorption lines with column densities of $\log N(\mathrm{H\,I}) \lesssim 17$ (e.g. Kim et al. 2002). The number of these forest lines usually increases with redshift, and in case of additional absorption from dust in the GRB host galaxy or the Milky Way, the flux is additionally suppressed, especially at these blue wavelengths.

We therefore searched for absorption from molecular hydrogen by creating synthetic spectra with $H_2$ absorption lines for different column densities $N(H_2)$ and around different redshifts known from the low-ionisation lines and compared them with the normalized GRB spectra in both wavelength and velocity space. In the case of the comparison in velocity space, we plotted the strongest absorption lines on top of each other in order to look for coincidences. Following this method, we confirm the presence of molecular hydrogen lines in the previously known spectra, GRB120327A, 120815A, and 121024A (Krühler et al. 2013; D'Elia et al. 2014; Friis et al. 2015). Additionally, we find conclusive evidence for absorption from $H_2$ in three more systems, which are GRB120909A, 141109A, and 150403A. For these 6 systems, we identified the most constraining lines in the rotational levels $J = 0$, 1, and 2, where most of the $H_2$ is expected, and fit them simultaneously. In cases where the observed velocity width of the $H_2$ absorption lines is close to the instrumental resolution, we decided to measure the column density for both, a fixed broadening parameter of $b = 2$ km/s and $b = 10$ km/s, in order to determine a realistic range for the total $H_2$ column density. This is the case for four GRBs, namely GRB120327A, 120909A, 121024A, and 141109A. For GRB121024A the total molecular hydrogen column density remains unchanged regardless of the broadening parameter assumed, but for the other three burst the total molecular hydrogen column density is about $\Delta \log N(H_2) \sim 2$ higher for $b = 2$ km/s. However, for the following analysis, we will adopt the column densities measured for $b = 2$ km/s, which is a more realistic value, given the temperatures of molecular gas. For the other two GRBs, 120815A and 150403A, because the lines are saturated, we are able to fit a large part of the spectrum with all rotational levels simultaneously and by including other blending lines from e.g. Fe II as well (see Fig. 9 and 10).

Most of the other spectra do not show convincing evidence for absorption from molecular hydrogen. Some, however, are still consistent with relatively high column densities, but can only be labeled as possible detections, mostly because of poor S/N or a strong Lyman-$\alpha$ forest. This is the case for three GRBs, namely GRB151021A, 160203A, and 170202A. In Fig. 11 we show part of the spectrum of GRB 151021A, demonstrating the difficulty in claiming a detection of $H_2$.

In these three cases and for the rest of the bursts for which we do not find any convincing evidence for absorption from molecular hydrogen in the spectra, we determine upper limits for the total molecular hydrogen column density with the following steps. First, we identify the three most constraining lines for the rotational levels $J = 0$, 1, and 2, fit them simultaneously, and determine the upper limits as we did for metals lines and as described in Sect. 3. We then use these upper limits to create a synthetic spectrum, which we plot over the whole range of covered $H_2$ lines in order to check if the determined upper limits are consistent with the rest of the spectrum. For some of the bursts we were not able to find regions of the spectrum where a proper fit is possible, mostly because the S/N is relatively poor, or the spectrum is covered with too many forest lines. In this case we create synthetic spectra with $H_2$ absorption lines at the position of the strongest component of the metal lines, and increase the column density until the synthetic spectrum is not consistent with the data anymore (e.g. the right panels in Fig. B.13). We here also assume a broadening parameter of $b = 2$ km/s as a conservative value, and find upper limits that are, depending on S/N, spectral coverage, and instrumental resolution, more or less constraining.

We note that we do not detect $H_2$ in the 6 GRB-DLAs at $z > 4$. However, the upper limits for most of them are not very constraining due to the strong Lyman-$\alpha$ forest. Especially for GRB 130606A, the forest is so strong that it is impossible to rule out that there is no absorption from molecular hydrogen. For this reason, the bursts at $z > 4$ are labeled separately in some plots. Additionally, Lyman blanketing strongly affects the reliability of the measurements for the GRBs at reshift $z > 4$. So far, $H_2$ has only been detected in one QSO-DLA at redshift $z > 4$ (Ledoux et al. 2006). GRB120909A is the $H_2$ bearing GRB-DLA with the highest redshift ($z = 3.929$) found to date.

The results of our search for $H_2$ are listed in Table A.2, which is divided into three parts: the six detections, the three tentative detections, and the rest of the bursts for which we do not find any convincing evidence for absorption from molecular hydrogen.

#### 4.5.1. Catalog of laboratory parameters

During our analysis, we found that some of the oscillator strength values for molecular hydrogen that we took from the compilation of Malec et al. (2010) are wrong, most likely due to a missing factor coming from the statistical weights. We thus calculated the correct values following Morton (2003) and created a new data file which is available under the following link: https://doi.org/10.5281/zenodo.1442558. While the values for the oscillator strength $f$ and the damping constant $\Gamma$, are still better than anything that can be currently measured, the wavelength or frequency positions are taken from the most recent lab experiments (Bailly et al. 2010, see their Table 11 & 12).

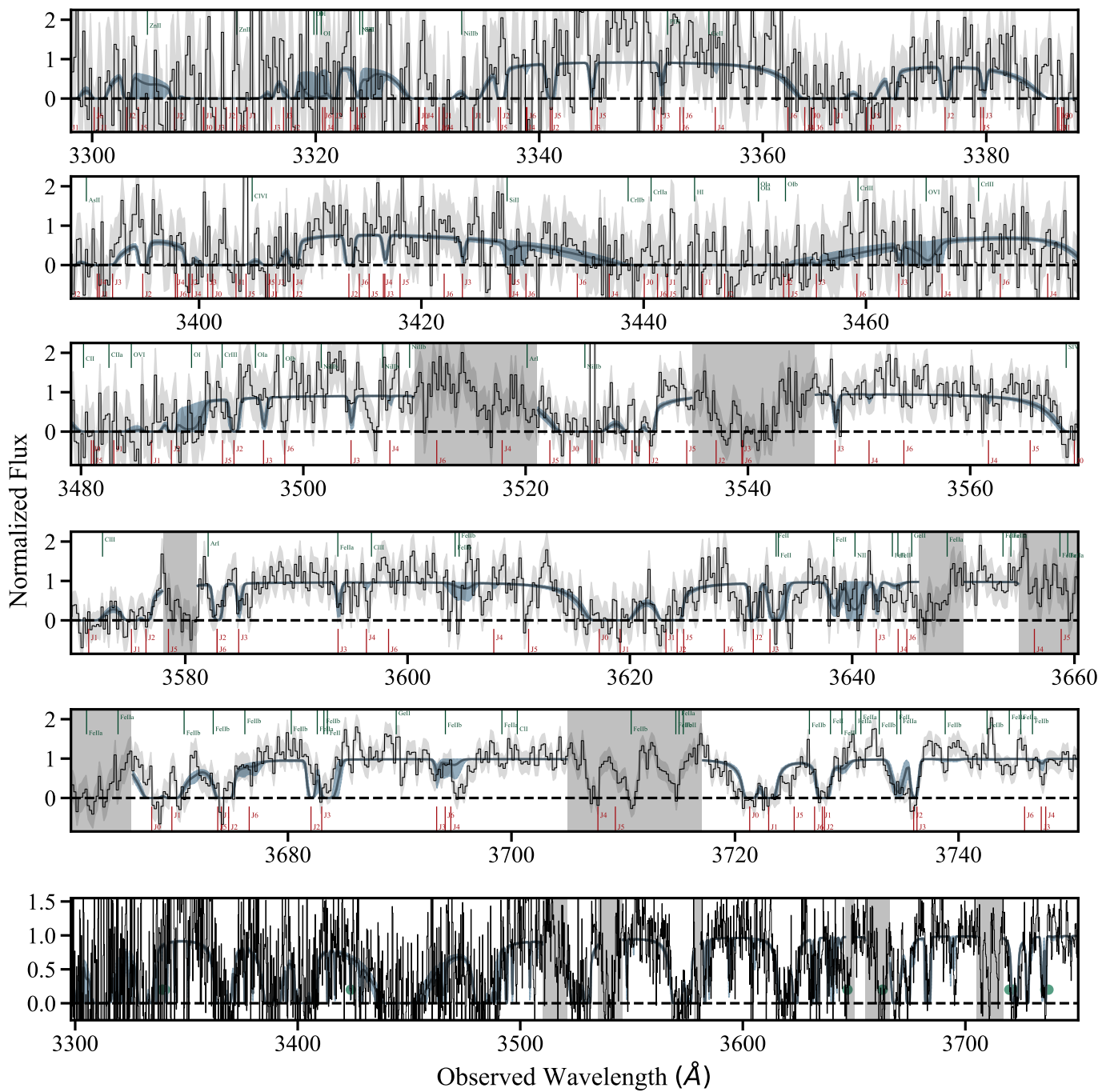

**Fig. 9.** X-shooter spectrum of GRB120815A covering the Lyman and Werner absorption lines of molecular hydrogen. The best fit model is indicated by the solid blue lines and the corresponding 3-$\sigma$ confidence intervals by the blue shaded regions. Gray shaded regions were ignored for the fit. The green circles in the bottom panel indicate the position of telluric lines.

### 4.6. Vibrationally-excited molecular hydrogen

The UV radiation from the GRB afterglow can theoretically pump any foreground $H_2$ into its vibrationally-excited levels that absorb at rest frame wavelengths $\lambda < 1650$Å. For the six $H_2$ bearing DLAs, as well as for the three tentative detections of $H_2$ and the bursts at $z > 4$, we thus additionally searched for vibrationally-excited molecular hydrogen $H_2^*$ by cross-correlating the observed spectra with the theoretical model from Draine (2000) and Draine & Hao (2002), as was done before for GRB120815A in Krühler et al. (2013) and for GRB121024A in Friis et al. (2015). We confirm the detection in GRB120815A and also find tentative evidence for vibrationally excited molecular hydrogen in the spectrum of GRB150403A (see Fig. 12). For GRB121024A, we confirm the non-detection and we also find no evidence for H$_2^*$ in 120327A, 120909A, 141109A, GRB151021A, 160203A, and 170202A. In the case of the six GRB at $z > 4$, we also find no convincing evidence for absorption from H$_2^*$.

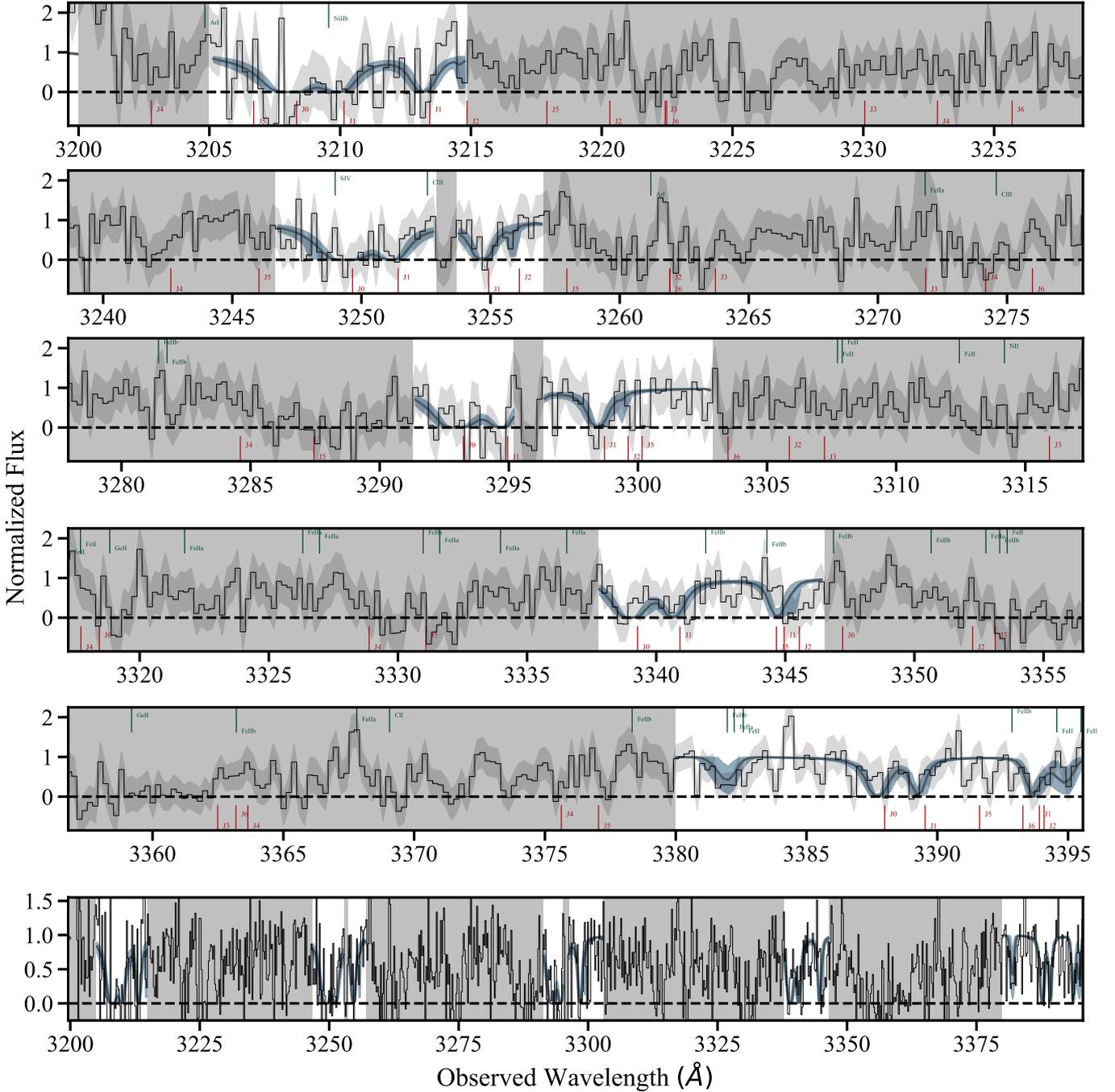

**Fig. 10.** X-shooter spectrum of GRB150403A covering the Lyman and Werner absorption lines of molecular hydrogen. The best fit model is indicated by the solid blue line and the corresponding 3-$\sigma$ confidence intervals by the blue shaded regions. Gray shaded regions were ignored for the fit.

### 4.7. Carbon monoxide

We do not find evidence for carbon monoxide absorption lines in any of the 22 GRB spectra. CO, which is primarily formed in the presence of H$_2$, was previously detected in the H$_2$ bearing DLA of GRB 080607 (Prochaska et al. 2009) and various H$_2$ bearing QSO-DLAs (e.g. Srianand et al. 2008; Noterdaeme et al. 2010, 2018). However, the presence of molecular hydrogen seems not to be a sufficient condition for the presence of CO (Noterdaeme et al. 2015b). To determine upper limits for the CO column density for the six H$_2$ bearing GRB-DLAs in our sample, we fitted the six strongest CO AX bandheads (CO

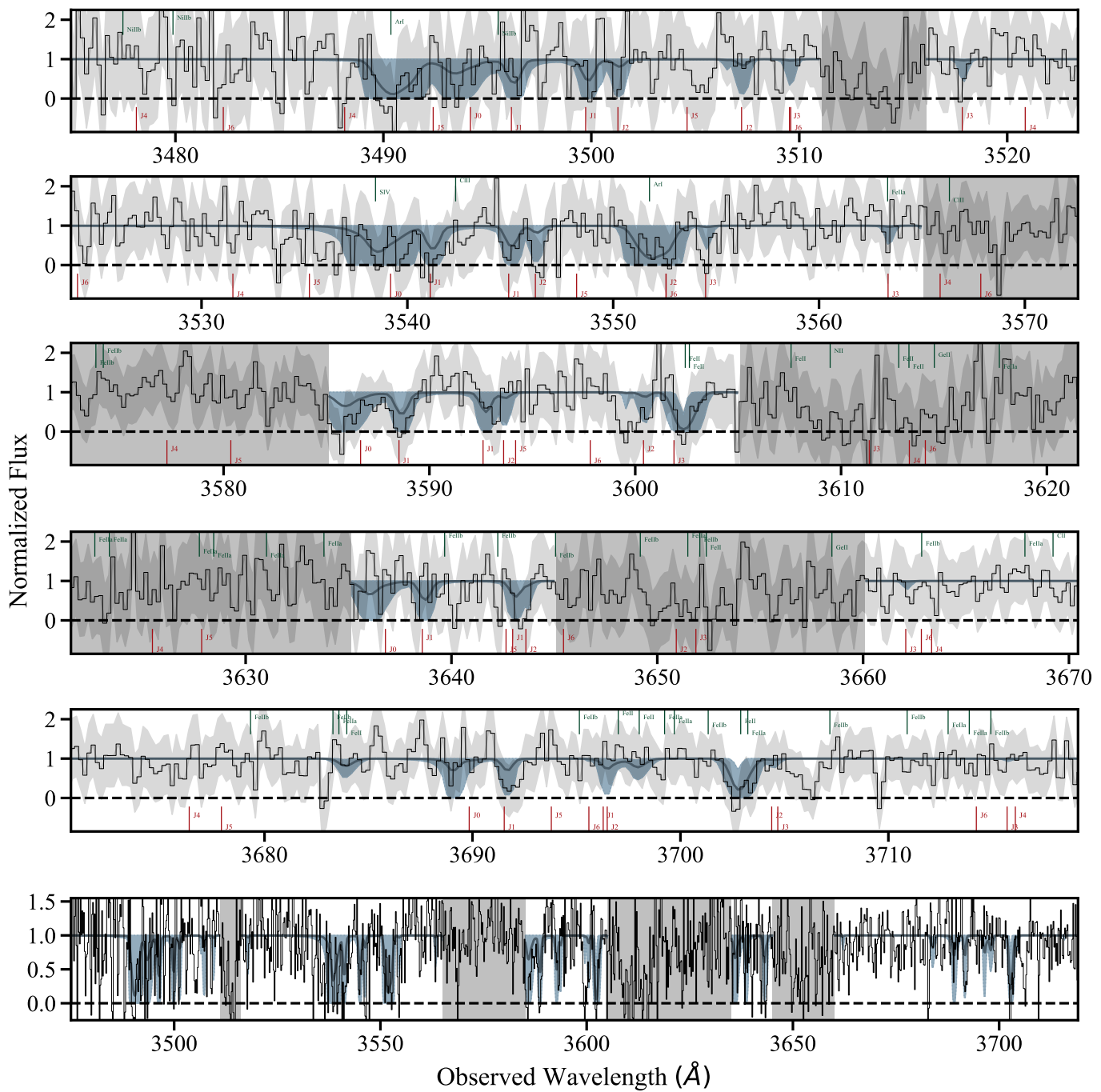

**Fig. 11.** X-shooter spectrum of GRB151021A covering the Lyman and Werner absorption lines of molecular hydrogen. The best fit model is indicated by the solid blue lines and the corresponding 3-$\sigma$ confidence intervals by the blue shaded regions. Gray shaded regions were ignored for the fit.

AX(0-0) to CO AX(5-0)) simultaneously, under the conservative assumption that $b = 2.0$ km/s. The resulting upper limits and also the corresponding upper limits for the $N$(CO) to $N$(H$_2$) ratio are listed in Table 2. The upper limits on the $N$(CO)/$N$(H$_2$) ratio range from $< 10^{-2.1}$ to $< 10^{-5.5}$ and are consistent with values found for diffuse molecular clouds, or translucent clouds, which are in the transition region between diffuse and dark (Burgh et al. 2010). This is further supported by the relatively low extinction on the line of sight to the GRBs in our sample ($A_V$ (depl.) < 1.5 mag).

### 4.8. Overall results

To summarize our results, we analyzed X-shooter spectra of 22 GRB afterglows in order to derive the column densities of 10 different metal species as well as those of neutral atomic and molecular hydrogen in each GRB host galaxies line of sight (see Table A.1 and A.2). By using these measurements we derive the elements relative abundances [X/H] and further analyze their depletion pattern in order to infer dust-corrected metallicities [M/H], the dust-to-metals ratio $\mathcal{DTM}$, and the expected host intrinsic visual extinction $A_V$ (depl.). Additionally, we collected the visual extinction as derived from SED fits $A_V$ (SED), compared

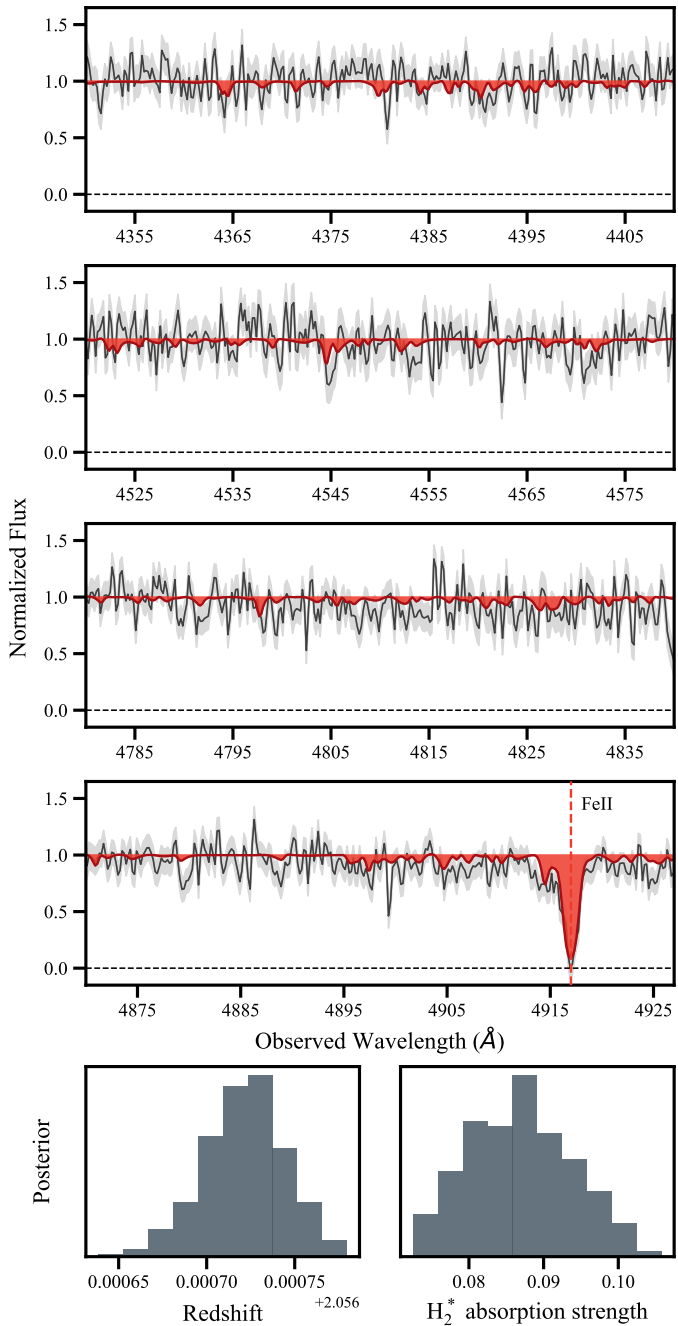

**Fig. 12.** Normalized spectrum of GRB150403A showing tentative evidence for absorption from vibrationally-excited molecular hydrogen $H_2^*$. The model is indicated by the solid red line. In the insets at the bottom we show the posterior distribution of the redshift $z$ and the $H_2^*$ absorption strength from the Draine & Hao (2002) model that we fitted to the data. An intervening absorption line from Fe II is indicated by the dashed orange line.

it with $A_V$ (depl.), and argued that the latter provides the better estimate. We also calculate the depletion factor [Y/Fe] as another tracer of dust. Finally, given the column densities or their derived upper limits of molecular hydrogen, we calculate the molecular gas fraction $f(H_2) = 2N(H_2)/(2N(H_2)+N(H I))$. The complete data set is presented in Table A.3 and in Fig. 13 we show pairwise scatter plots and histograms of the most important properties. By color and symbol, we differentiate between four

**Table 2.** Upper limits on the CO column density in the six $H_2$ bearing GRB DLAs as well as the $N(CO)/N(H_2)$ ratios. Also listed are the values for GRB 080607 as taken from Prochaska et al. (2009).

| GRB<br>yymmdd# | log $N$(CO)<br>AX(0-0) to (5-0) | $N$(CO)/$N$($H_2$) |
|---|---|---|
| 120327A | < 15.3 | < $10^{-2.1}$ |
| 120815A | < 15.0 | < $10^{-5.4}$ |
| 120909A | < 14.2 | < $10^{-3.2}$ |
| 121024A | < 14.4 | < $10^{-5.5}$ |
| 141109A | < 15.9 | < $10^{-2.1}$ |
| 150403A | < 14.9 | < $10^{-5.0}$ |
| 080607 | 16.5 ± 0.3 | ∼ $10^{-4.7}$ |

**Notes.** The CO upper limits (3-$\sigma$) were determined assuming $b = 2.0$ km/s as a conservative value. In case of GRB120327A, 120909A and 141109A, the $N$(CO)/$N$($H_2$) ratio was determined taking the $H_2$ column densities measured for $b = 2.0$ km/s as given in Tab. A.2.

different cases: (1) Detections and (2) possible detections of absorption from molecular hydrogen as well as non-detections for (3) GRBs at redshift $z < 4$ and the non-detections for (4) GRBs at redshift $z > 4$.

The mean neutral hydrogen column density for the GRB-DLAs in our sample is log $\overline{N(H I)} = 21.50\pm0.67$, and the individual values range from log $N(H I) = 19.88 \pm 0.01$ (GRB130606, the only sub-DLA) to log $N(H I) = 22.39 \pm 0.01$ for GRB 111008A, which is among the highest values observed to date (Prochaska et al. 2009; Sparre et al. 2014; Noterdaeme et al. 2015b). DLAs with such high column densities are expected to be very rare, because at log $N(H I) > 21.5$ the conversion from H I to $H_2$ effectively steepens the column density distribution (Schaye 2001; Altay et al. 2011).

On average, the 22 GRB-DLAs have a metallicity of $\overline{[X/H]} = -1.27 \pm 0.37$, which is comparable to metallicities found for QSO-ESDLAs (Noterdaeme et al. 2015a). The average metallicity of the GRB-DLAs at $z > 4$ is $\overline{[X/H]} = -1.39 \pm 0.42$, which is slightly lower than the mean metallicity for the bursts at redshift $2 < z < 4$ ($\overline{[X/H]} = -1.23 \pm 0.37$). This is in accordance with the results from Rafelski et al. (2012) and De Cia et al. (2018), that DLA metallicities decrease by a factor of 50-100 from redshift $z = 0$ to $z = 5$. This might also explain why we do not find evidence for absorption from molecular hydrogen in the high redshift bursts, especially in case of GRB111008A with its very high H I column density.

Interestingly we find that, while $H_2$ is detected over almost the whole range of covered metallicities [X/H], all detections and possible detections of molecular hydrogen are for DLAs with neutral hydrogen column densities larger than log $N(H I) > 21.7$. However, not all GRB-DLAs with log $N(H I) > 21.7$ show absorption from $H_2$. Also, as indicated by the dashed blue lines, which are the result of a Bayesian linear regression as described in Sect. 5, we find positive correlations for [Y/Fe] and log $A_V$ (depl.) versus metallicity, as well as for $\mathcal{DTM}$ versus depletion [Y/Fe] and log $A_V$ (depl.). A weak correlation is also found for $\mathcal{DTM}$ versus metallicity, which is, however, not as tight as the one determined by Wiseman et al. (2017b). Also, all $H_2$ bearing systems have dust-to-metals ratios larger than $\mathcal{DTM} > 0.4$ and significant line-of-sight dust columns ($A_V$ (depl.) > 0.1 mag). See also Fig. 7, which shows that for all $H_2$ bearing GRB-DLAs as well as the tentative detections $A_V$ (depl.) is higher than $A_V$ (SED), indicating that the phase of the (high redshift) ISM that is amenable to $H_2$ formation might have a different dust composition or grain sizes. Furthermore, we do not find any evidence for absorption of CO in any of the spectra. And finally, we confirm the detection of vibrationally excited $H_2^*$ in GRB120815A and additionally find tentative evidence for $H_2^*$ in GRB150403A.

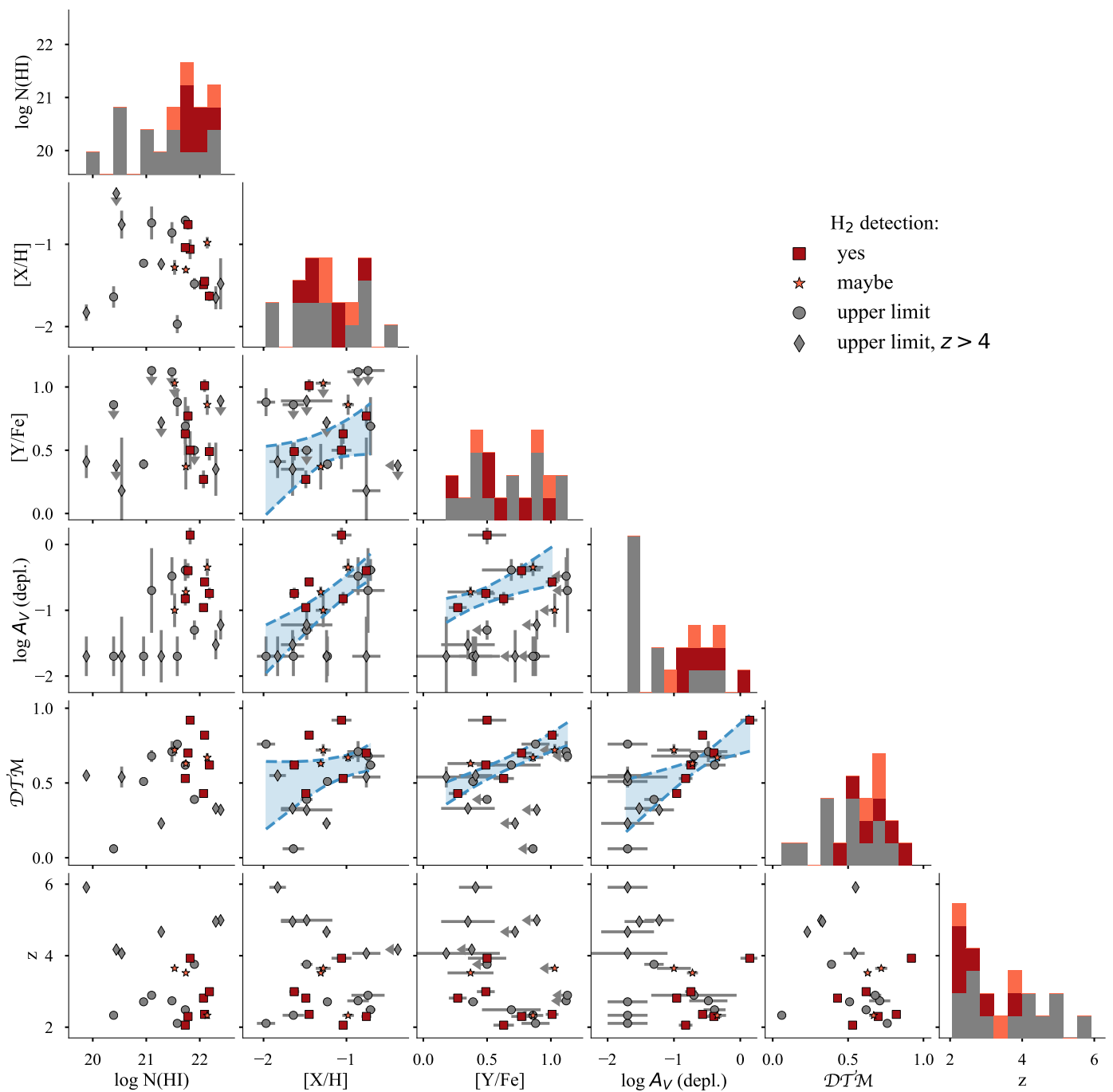

**Fig. 13.** Pairwise scatter plots and histograms of the most important properties of our sample of 22 GRB-DLAs at $z > 2$. In case absorption from molecular hydrogen is detected, the data points are squares colored in red. Possible detections are indicated by orange stars, and upper limits are gray diamonds in case of DLAs at $z > 4$ and circles otherwise. In blue, we show the results of a Baysian linear regression analysis as described in Sect. 5.

## 5. Discussion

The two main questions we aimed to address in this paper are: (1) if (and why) there might be a lack of $H_2$ bearing GRB-DLAs, and (2) how GRB-DLAs compare to QSO-DLAs. To answer the first question: there is no lack of $H_2$ bearing GRB-DLAs. While the fraction of $H_2$ bearing systems in the general population of QSO-DLAs is only about 10% or less (Ledoux et al. 2003; Noterdaeme et al. 2008; Balashev et al. 2014; Jorgenson et al. 2014), we here find evidence for absorption from $H_2$ in 6/22 (27%) or even 9/22 (41%) of our GRB spectra, when including the tentative detections. Furthermore, when only comparing those QSO and GRB-DLAs that are at redshift $2 < z < 4$ and behind large neutral hydrogen column densities $\log N(\text{H\,\textsc{i}}) > 21.7$, and therefore likely probe sight-lines that are associated with the central region of a galaxy, the fractions are very similar: 7/9 (78%) $H_2$ bearing systems for QSO-DLAs (Noterdaeme et al. 2015b,a) and 6/10 (60%) for GRB-DLAs, or 8/10 (80%), when including the possible detections. Additionally, we want to stress that, although this is the first time a systematic search for $H_2$ could be performed in such a large sample of 22 GRB-DLAs, we are still biased against dusty sight-lines ($A_V > 0.5$ mag). Since the production of molecular hydrogen is enabled by its formation on dust grains and because dust and $HI$ provide shielding against $H_2$ dissociating radiation, the true fraction of $H_2$ bearing GRB-DLAs might be even higher; also because our upper limits on the $H_2$ column density are not in all cases very constraining, meaning that the relatively low spectral resolution of X-shooter and sometimes a poor S/N and/or a strong Lyman-$\alpha$ forest make it difficult to detect the possibly narrow absorption lines of molecular hydrogen. We also note that the lack of $H_2$ bearing systems found by Ledoux et al. (2009) in a sample of 7 GRB-DLAs is in fact consistent with our results. Only 5 of their 7 DLAs have hydrogen column densities larger than log $N(HI) > 21.0$, and only three have log $N(HI) > 21.7$. The latter three DLAs have lower metallicities ($\overline{[X/H]} < -1.5$) than the average metallicity of our 6 detections ($\overline{[X/H]} = -1.24 \pm 0.33$) and also relatively low depletion factors $[Y/Fe] < 0.2$, compared to the average of our $H_2$ bearing DLAs ($\overline{[Y/Fe]} = 0.62 \pm 0.26$).

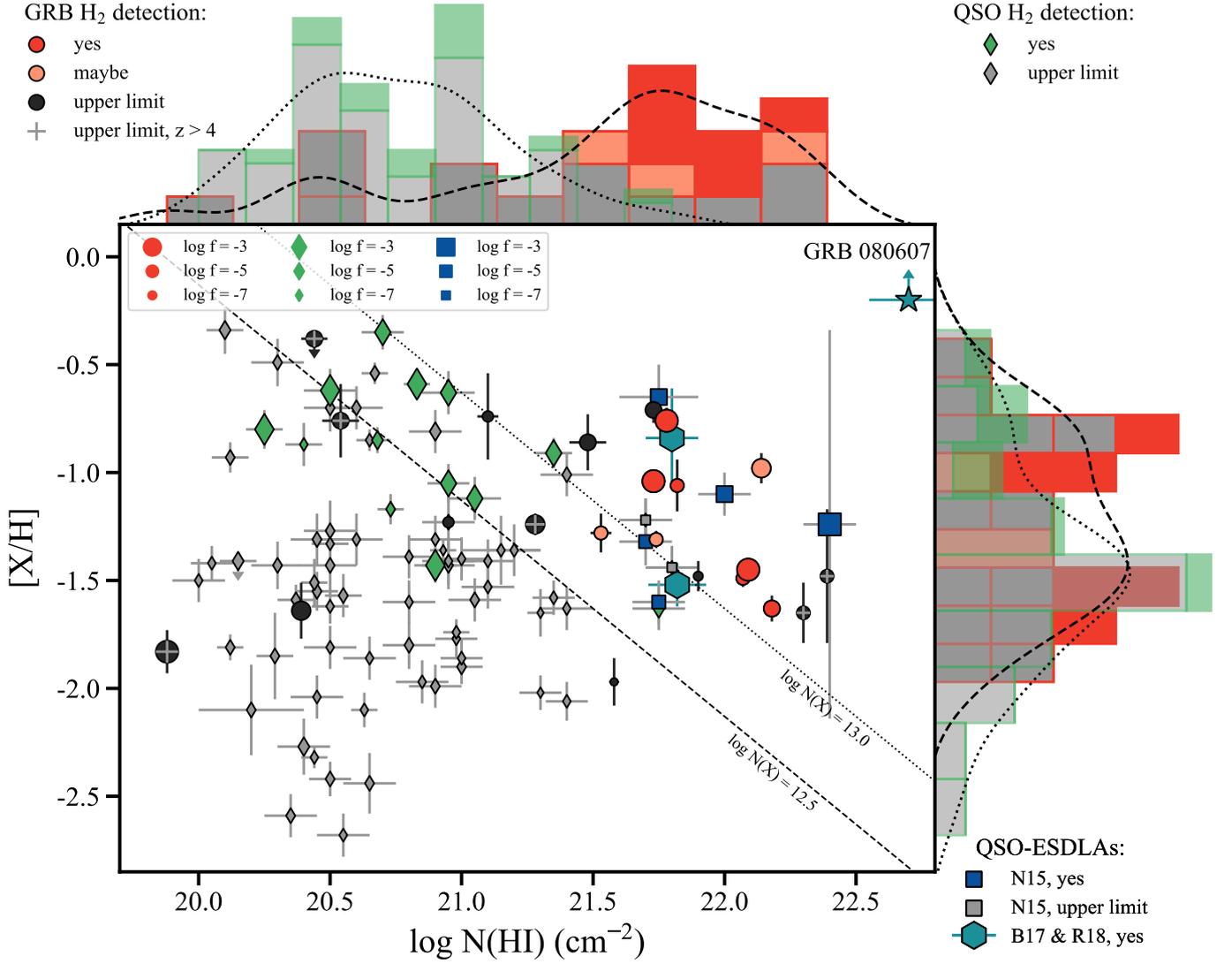

**Fig. 14.** Metallicity versus neutral hydrogen column density for the GRB-DLAs in our sample and the QSO-DLAs from Noterdaeme et al. (2008) and Noterdaeme et al. (2015a) (N15). Also shown are the two extreme QSO-DLAs from Balashev et al. (2017) (B17) and Ranjan et al. (2018) (B18). In case molecular hydrogen is detected, the data points are colored (GRB-DLAs in red, QSO-DLAs in green and blue, as labeled). The symbol size represents the overall molecular fraction log $f$, and in case of a non-detection the corresponding upper limit. The GRB-DLAs at $z > 4$ are additionally marked with a cross. The dotted and dashed lines represent a constant metal column of log $N(X) = 12.5$ and log $N(X) = 13.0$, respectively. Above and to the right of the scatter plot we also show the histograms and corresponding kernel density estimates for our GRB and the N08 QSO-DLA samples. Again, systems with $H_2$ detections are colored as labeled.

A natural explanation for the higher fraction of $H_2$ bearing systems in our 22 GRB-DLAs compared to the general population of QSO-DLAs, and the similar fraction at high $N(HI)$ column densities, is the impact parameter (Arabsalmani et al. 2015). While higher $HI$ column densities are on average related to absorbing cold neutral gas within a galaxy, low column densities likely trace cool, circumgalactic environments (Pontzen et al. 2008; Prochaska et al. 2011; Noterdaeme et al. 2015b). As background sources, QSOs naturally probe a wide range of impact parameters and therefore $N(HI)$ column densities, but GRBs are usually found to originate from the central parts of their host galaxies (Fruchter et al. 2006; Lyman et al. 2017), and are therefore usually behind relatively large $N(\text{H\,\textsc{i}})$ column densities (Fynbo et al. 2009; Selsing et al. 2018).

Theoretical models predict that the $N(\text{H\,\textsc{i}})$ threshold for the conversion from atomic to molecular hydrogen increases with decreasing metallicity (Savage et al. 1977; McKee & Krumholz 2010; Sternberg et al. 2014). This is because systems with higher metallicity usually contain more dust (see Fig. 15), which provides shielding and enables the production of molecular hydrogen. This explains why the $H_2$ bearing QSO-DLAs from the sample of Noterdaeme et al. (2008) (from now on N08) at $\log N(\text{H\,\textsc{i}}) < 21.5$ have relatively high metallicities and fall in a region that is barely probed by our GRB-DLAs (consistent with previous findings that GRBs are generally found in low metallicity environments Krühler et al. 2015; Japelj et al. 2016; Perley et al. 2016). Only two GRBs fall in that regime, 120712A and 151027B, which are both at redshift $z > 4$ where $H_2$ is hard to detect anyway. Therefore, and also due to the low number statistics in this regime, it is not surprising that we do not detect any $H_2$ bearing GRB-DLAs with $\log N(\text{H\,\textsc{i}}) < 21.5$.

Since molecular gas has a small covering fraction, also neutral carbon lines can be used as a tracer of cold gas and dust (Ledoux et al. 2015). In fact, Heintz (2018, in subm.) systematically searched for absorption from C\,\textsc{i} in a similar sample of GRB-DLAs. They find conclusive or tentative evidence for C\,\textsc{i} in the spectra of GRB 120815A, 121024A, and 150403A, which are three of our six $H_2$ bearing systems [7]. We note that C\,\textsc{i} is not detected in the $H_2$ bearing system in the line of sight toward GRB 141109A and in neither of our three possible detections (151021A, 160203A, 170202A). GRB 120909A didn't enter their sample because the C\,\textsc{i} lines are blended with tellurics. C\,\textsc{i} absorbers are usually found in systems with higher visual extinction and frequently present the 2175Å absorption feature, and given the first ionization potential (11.2 eV) C\,\textsc{i} is an indicator of molecular hydrogen (Srianand et al. 2005). Heintz (2018, in subm.) find a fraction of 25% C\,\textsc{i} bearing GRB-DLAs, or 46% for DLAs with $\log N(\text{H\,\textsc{i}}) > 21.7$. This is lower than our fraction of $H_2$ bearing systems with $\log N(\text{H\,\textsc{i}}) > 21.7$, which is likely a result of the low metallicities. For example, the DLA in the line of sight towards GRB 141109A is with $[\text{X/H}] \sim -1.6$ one of the most metal-poor absorber in the sample.

That we don't find a lack of $H_2$ bearing GRB-DLAs is also in agreement with the recent results from Arabsalmani et al. (2018), who report the detection of a molecular gas rich GRB host galaxy at $z = 2.086$ based on the observation of CO emission lines. They showed that the GRB hosts with measured molecular gas mass are quite normal in their gas contents when compared to the general population, and suggested that there is no sign of molecular gas deficiency, in contradiction with the previously proposed results from other, low redshift GRB hosts (Hatsukade et al. 2014; Stanway et al. 2015; Michałowski et al. 2018).

To answer the second question, about how GRB-DLAs and QSO-DLAs compare, in Fig. 14 we plot metallicity $[\text{X/H}]$ versus neutral hydrogen column density for the GRB-DLAs in our sample and also for the QSO-DLAs from N08 and Noterdaeme et al. (2015a) (N15). Additionally, we plot two other QSO-ESDLAs from Balashev et al. (2017) (B17) and Ranjan et al. (2018) (R18).

The histograms over metallicity for our GRB sample and the QSO sample from N08 show that both, GRB and QSO-DLAs, cover more or less the same range of metallicities, with the only difference being the tail over lower metallicities covered by QSO-DLAs. This is likely a result of the different range of impact parameters, or because GRBs trace starforming galaxies, while QSO-DLAs do not necessarily form stars and produce metals. The metallicity distribution of DLAs is predicted to center at $[\text{X/H}] = -1.5$ to $-1$ (Cen 2012), which is true for our sample of 22 GRB-DLAs and also for the QSO-DLA sample from N08. Also, the overall metallicity should floor at $[\text{X/H}] \sim -3$ at $z = 1.6 - 4$, which is exactly what we see as the tail over lower metallicities covered by the N08 QSO-DLAs. These systems are likely associated with the cold gas outside the galactic disk or halo, where the metallicity is low and not a lot of star formation takes place. In contrast, GRBs generally tend to probe the inner regions of their host galaxies, where the metallicity is expected to be higher (e.g. Belfiore et al. 2017; Sánchez-Menguiano et al. 2018).

Furthermore, all $H_2$ bearing systems, for both GRBs and QSOs, are found at metallicities $[\text{X/H}] \geq -1.7$, but cover the whole range of neutral hydrogen column densities. However, only QSO-DLAs that contain $H_2$ are found below $\log N(\text{H\,\textsc{i}}) < 21.7$. In general, the majority of $H_2$-bearing DLAs is found at metal columns $\log N(X) > 12.5$, and $\log N(X) > 13.0$ for $\log N(\text{H\,\textsc{i}}) > 21.5$, as indicated by the dashed and dotted line in Fig. 14. For the QSO-DLA sample from N08 the mean neutral hydrogen column density is $\log \overline{N(\text{H\,\textsc{i}})} = 20.75 \pm 0.41$, which is much lower than the one of our GRB-DLAs $\log \overline{N(\text{H\,\textsc{i}})} = 21.50 \pm 0.67$. The range of metallicities is more comparable; $\overline{[\text{X/H}]} = -1.45 \pm 0.53$ for N08 and $\overline{[\text{X/H}]} = -1.22 \pm 0.53$ for the $H_2$ GRB sample.

Looking at only the detections, i.e. the colored data points in Fig. 14 one can see a slight negative trend, that for lower neutral hydrogen column densities the $H_2$ bearing systems have on average higher metallicities. Higher metallicities are usually found for dustier systems (see Fig. 13), so that one expects more efficient $H_2$ production and shielding. Also, at higher H\,\textsc{i} column densities a lower metallicity is required to have the same metal column (as indicated by the dashed and dotted line in Fig. 14), and thus dust, which needs a cold and dense environment but also a high cross section of metals to form.

This becomes also evident from Fig. 15, which is similar to Fig. 14, but this time we plot the dust depletion factor, $[\text{Y/Fe}]$, versus the systems' metal column densitiy $\log N(\text{H\,\textsc{i}}) + [\text{X/H}]$. One can see that $H_2$ bearing systems are exclusively found at $[\text{Y/Fe}] \gtrsim 0.2$ and $\log N(\text{H\,\textsc{i}}) + [\text{X/H}] \gtrsim 19.5$. Also, the molecular fraction $f$ and the fraction of $H_2$ bearing systems increase with the dust depletion factor and metal column density. At metal column densities larger than $\log N(\text{H\,\textsc{i}}) + [\text{X/H}] > 20.0$ and redshifts $2 < z < 4$, we detect $H_2$ in 6 of 13 (46%) systems, or 9 of 13 (69%) when including the tentative detections. For the QSO-DLAs from N08 this fraction is 4/9 (44%) and for the QSO-ESDLAs from N15 it is 5/7 (71%). So, also when selecting GRB and QSO-DLAs based on their metal column densities, the fraction of $H_2$ bearing systems are similar. Additionally, looking at the data points from both QSO and GRB-DLAs, there seems to be a slight, positive trend of increasing dust depletion with increasing metal column density. The red line indicates the best linear fit and the dashed lines the corresponding 3-$\sigma$ confidence intervals. The fit was performed with a Bayesian linear regression and the posteriors on slope and intercept were computed using PyMC. To account for outliers we use a Cauchy distribution for the errors on [Y/Fe]. For slope and intercept we use normal priors centered on the best estimates from a simple first fit using `scipy.stats.linregress`. Upper limits were not included in the fit. It is interesting to note that almost all the DLAs that fall as outliers above this trend are strong H$_2$ bearing systems. This is likely a result of the dust production that is enabled and the shielding provided by the dust in those systems, which protects the H$_2$ from being photo-dissociated by Lyman-Werner photons. Therefore, most DLAs that follow the trend supposedly trace diffuse gas in the intercloud medium, whereas the outliers above trace more likely diffuse molecular clouds.

---

[7] Heintz (2018, in subm.) also find C\,\textsc{i} in the spectra of GRB 120119A and 180325A. GRB 120119A, however, is not included in our sample because of its redshift of $z \sim 1.73$. GRB 180325A at $z = 2.248$ is not part of our sample either, because it is not in the complete sample presented in Selsing et al. (2018), and its high visual extinction prevents the search for $H_2$.

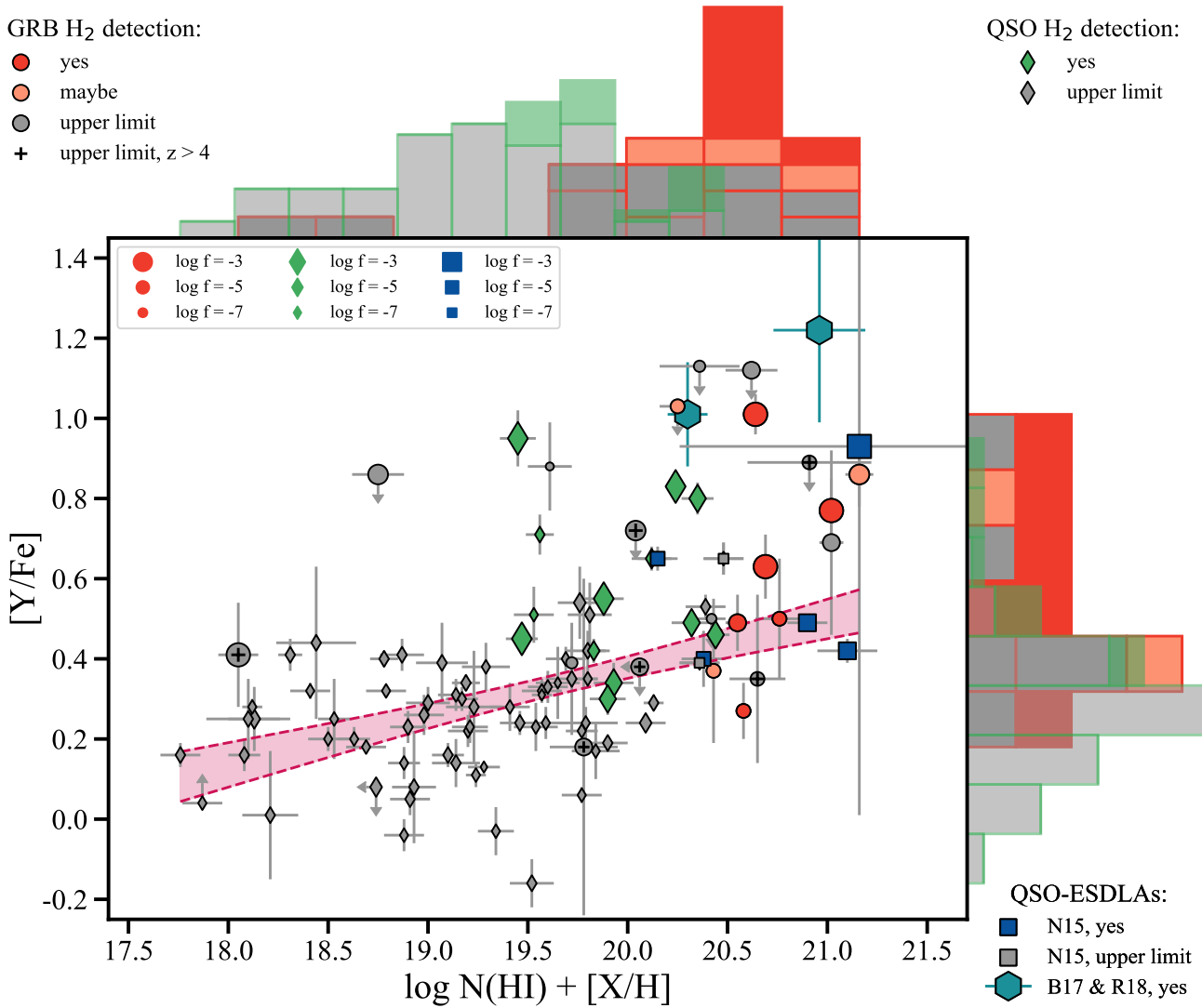

**Fig. 15.** Depletion factor, [Y/Fe], versus metal column density, log $N$(H I) + [X/H], for the GRB-DLAs in our sample and the QSO-DLAs from Noterdaeme et al. (2008) and Noterdaeme et al. (2015a) (N15). Also shown are the two QSO-ESDLAs from Balashev et al. (2017) (B17) and Ranjan et al. (2018) (B18). In case molecular hydrogen is detected, the data points are colored (GRB-DLAs in red, QSO-DLAs in green and blue, as labeled). The symbol size represents the overall molecular fraction log $f$, and in case of a non-detection the corresponding upper limit. The GRB-DLAs at $z > 4$ are additionally marked with a cross. The solid and dashed lines indicate the best fit linear model and corresponding 3-$\sigma$ errors. Above and to the right of the scatter plot we also show the histograms for our GRB and the N08 QSO-DLA samples. Again, systems with H$_2$ detections are colored as labeled. Upper limits are not included for the histograms.

### 5.1. Nucleosynthesis signatures in the ISM

The gas-phase metal abundances [X/H] that we observe in GRB-DLAs do not directly represent the metallicity of the absorbing systems, because of dust depletion. Indeed dust depletion can dramatically lower the observed abundances, such as that it must be properly taken into account to be able to recover the real (dust-corrected) metallicity, [M/H], and any nucleosynthetic signatures in the ISM. In Sect. 4.2 we fit the observed abundances of several metals with the depletion patterns found in the Galaxy and QSO-DLAs by De Cia et al. (2016), and following Wiseman et al. (2017b). Any deviations from this fit could in principle indicate peculiar abundances, for example those produced by specific nucleosynthetic processes, such as $\alpha$-element enhancement. As notable in Fig. 6, the fit to the depletion pattern is overall very good for most GRBs, with little deviations from QSO-DLAs. This is remarkable, given the different star-formation histories of galaxy counterparts of GRB- and QSO-DLAs. In particular, because of the higher star-formation rates of GRB host galaxies (e.g. Krühler et al. 2015), $\alpha$-element enhancement could be expected. We only observe a tentative case of $\alpha$-element enhancement for GRB 121024A, with a Si overabundance of $\sim 0.5$ dex, but no confirmation from other $\alpha$-elements such as O, Mg, Ti, nor S. On the other hand, there is a tentative $\sim 0.2$ over-abundance of the Fe-group elements Fe and Cr, and the somewhat related Zn, but with no apparent overabundance of Ni, albeit the large error bar. Titanium is unfortunately constrained in only a few cases, because of the weakness of the Ti II absorption lines. Nevertheless, for two cases, namely GRB120327A and GRB 161023A the observed Ti columns seem significantly lower than what can be expected from the depletion patterns. In both cases, no deviations of the other $\alpha$-elements are observed. One possibility is that this is indeed a peculiar underabundance of Ti. Alternatively, and perhaps more likely, this may signal that the values for Ti in the depletion patterns may be imprecise. Indeed, the depletion sequences were not characterized for DLAs for Ti and Ni, but instead their slope was extrapolated to low-metallicity systems by De Cia et al. (2016) from the Galactic values of Jenkins (2009). This might have caused a somewhat shallower slope of the depletion sequence of Ti, and thus a significant effect in the Ti depletion pattern, because Ti depletes very heavily. Quantifying this possible effect requires analyzing a large sample of QSO-DLAs and is beyond the scope of this paper.

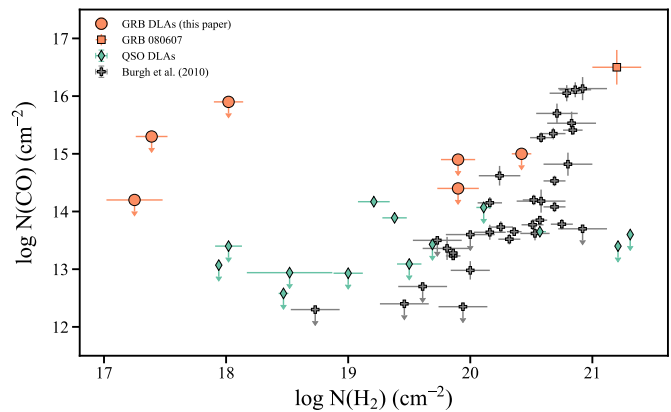

**Fig. 16.** CO versus H$_2$ column densities for the H$_2$ bearing GRB-DLAs presented in this paper compared to H$_2$ bearing QSO-DLAs from Balashev et al. (2017); Noterdaeme et al. (2018); Ranjan et al. (2018). Additionally we plot the results from Burgh et al. (2010) for sight lines through diffuse and translucent clouds in the Milky Way and the result from Prochaska et al. (2009) for the DLA toward GRB 080607.

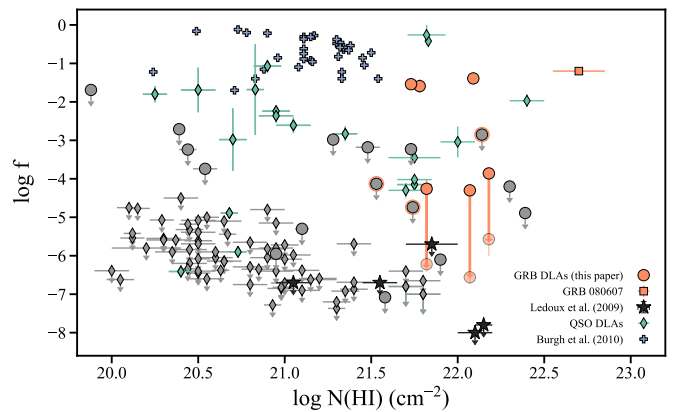

**Fig. 17.** Molecular fraction versus neutral hydrogen column density for the GRB-DLAs from this paper and Ledoux et al. (2009); Prochaska et al. (2009) in comparison to QSO-DLAs (Noterdaeme et al. 2008, 2015a; Balashev et al. 2017; Ranjan et al. 2018) and Galactic sight lines probing diffuse and translucent clouds (Burgh et al. 2010). H$_2$ bearing DLAs are plotted with colored symbols and the non-detections in gray. For GRB 120327A, 120909A, and 141109A we plot the molecular fraction for both $b = 2$ and $b = 10$ km/s.

### 5.2. The lack of carbon monoxide

The lack of carbon monoxide could be related to the low metallicity and/or the relatively low dust content ($A_V$(depl.) < 1.5 mag) of the systems, meaning that not enough oxygen and carbon are present to form CO, or not enough shielding against the photodissociating radiation is provided (Bolatto et al. 2013; Glover & Clark 2016; Balashev et al. 2017). Alternatively, the lack of CO absorption could be the result of a geometrical effect, because CO should only reside in the inner core of the molecular cloud, where more shielding is provided. In this scenario, the GRB line-of-sight would cross through the outer edge of the cloud where only H$_2$ is present. In Fig. 16 we plot the CO versus the H$_2$ column densities for GRB and QSO-DLAs as well as for Galactic sight lines. All our upper limits on the CO column density are consistent with the relation found by Burgh et al. (2010) for diffuse and translucent clouds in the Milky Way.

### 5.3. Comparison with Galactic sight lines

In general, a comparison with Galactic sight lines is complicated, due to the fact that QSO and GRB-DLAs do not have metallicities close to solar, but rather relatively low metallicities of the order of one tenth of solar. Also the dust-to-metal ratios are usually lower than in the Galaxy (De Cia et al. 2013, 2016; Wiseman et al. 2017b), so that even for the same amount of metals, one would expect less dust and thus less molecules. Nevertheless, in Fig. 17 we plot the molecular fraction $f$ versus the molecular hydrogen column density for GRB and QSO-DLAs in comparison to Galactic sight lines. One can see that only about half of the H$_2$ bearing DLAs with log $N$(HI) > 21.7 have molecular fractions consistent with the galactic diffuse and translucent clouds from (Burgh et al. 2010). For the rest of the DLAs, the molecular fraction is much lower, which is also the case for the upper limits we could put on the non-detections. This could be a result of the generally low metallicities and low dust columns.

### 5.4. Distances and nature of the absorbing clouds

To determine the distance of the absorbing cloud to the GRB explosion site, it is necessary to observe absorption-line variability to trace the electronic populations in different energy levels. Such an analysis has only been possible for a handful of cases and the distances were all found to be 50 to more than several hundred parsecs (Vreeswijk et al. 2007; Ledoux et al. 2009; D'Elia et al. 2009; Vreeswijk et al. 2011; Hartoog et al. 2013). Alternatively, we could use the H$_2^*$ detections and non-detections of our six H$_2$ bearing GRB-DLAs, to determine the required UV flux to populate the vibrationally-excited levels. This flux could be compared with the actual UV flux measured for the afterglow light curve in order to put a constraint or a lower limit to the distance of the absorbing cloud. Nevertheless, this exercise is beyond the scope of this paper, and we just note that the detection of molecular hydrogen alone, should imply a distance of at least $\gtrsim 0.5$ to 1 kpc between the GRB and the absorbing gas to avoid photo-ionization. For example, Ledoux et al. (2009) find for GRB 050730 that H$_2$ photo-dissociation can only be effective in clouds with distances smaller than $d < 500$ pc (see their Fig. 7).

Additionally, we detect the fine-structure levels of O$\,$I, O$\,$I$^*$ and O$\,$I$^{**}$, in almost all of the GRB spectra (see Fig. 4). These lines are commonly detected in GRB-DLAs and indicative of high densities and temperatures, and therefore add to the picture of the absorbing gas having mixed phases and being under the influence of a local radiation field, i.e. diffuse molecular clouds.

Our results are thus pointing to a picture where GRB and QSO-DLAs with very high neutral column densities ($\log N(H\,I) > 21.7$) or metal column densities ($\log N(H\,I)$+ [X/H] $> 20$) are associated with sight lines passing close to the galactic center, where the gas pressure is higher (Blitz & Rosolowsky 2006; Balashev et al. 2017), the conversion of H$\,$I to H$_2$ steepens the column density distribution (Altay et al. 2011; Noterdaeme et al. 2014), and thus a higher fraction of H$_2$ bearing systems is expected. And, since the absorbing gas is likely at distances of several hundred parsec from the GRB, the H$_2$ bearing GRB-DLAs are not associated to the star-forming region where the GRB progenitor was born, but to diffuse molecular clouds in the host galaxies line of sight.

# 6. Summary

For the first time, we have been able to use a large, less-biased sample of 22 GRBs, to perform a systematic search for molecular hydrogen in DLAs associated with GRB host galaxies, and to study the effects of metallicity and dust depletion in the diffuse interstellar medium of these galaxies at high redshift. The main results derived from our analysis can be summarized as follows:

(i) There is no lack of H$_2$ in GRB-DLAs. We find evidence for absorption from molecular hydrogen in 6 out of 22 GRB-DLAs and claim three additional tentative detections. This makes a fraction of 27% (41%), which is three to four times larger than for the general population of QSO-DLAs ($\leq 10\%$), but comparable to the fraction of H$_2$-bearing QSO-DLAs that cover the same range of neutral hydrogen column densities and redshifts, the so-called extremely strong DLAs (ES-DLAs).

(ii) Both, QSO- and GRB-DLAs, with column densities $\log N(H\,I) > 21.7$ at redshifts $2 < z < 4$ appear to probe similar systems, i.e. sight-lines with small impact parameters that are passing close to the galactic centers, where the gas pressure is higher and the transition of H$\,$I to H$_2$ is easier. In this range, the fraction of H$_2$-bearing systems is much higher and comparable in both QSO- and GRB-DLAs (60 to 80%). These systems are likely diffuse molecular clouds, which in the case of GRBs, are unrelated to the star-forming region where the explosion occurred.

(iii) At lower column densities ($\log N(H\,I) < 21.7$), QSO-DLAs need on average a higher metallicity to contain significant amounts of molecular hydrogen. This can naturally be explained by the fact that systems with a higher metallicity contain on average more dust, which provides shielding and enables the production of H$_2$. Our sample of GRB-DLAs does not significantly probe low hydrogen column densities (i.e. only 22% of the sample has $\log N(H\,I) < 21.0$), because GRBs are found to originate from the central regions of their host galaxies, and additionally because GRBs tend to probe systems with relatively low metallicities. As random background sources, QSOs naturally sample a broad range of impact parameters.

(iv) The fraction of H$_2$-bearing GRB-DLAs might be even higher than the one we have found because our sample is probably still biased against dusty sight lines (no sightlines with $A_V$(SED) > 0.5 mag). More dust provides more shielding, adds more catalyst to aid H$_2$ production, and indicates an advanced grain chemistry and thus a higher chance to form molecules. Also, our upper limits on the H$_2$ column density are not in all cases very constraining, meaning that the relatively low spectral resolution of X-shooter and sometimes a poor S/N and/or a strong Lyman-$\alpha$ forest make it difficult to detect the narrow absorption lines of molecular hydrogen.

(v) In the future, it is important to expand the current surveys of GRBs with prompt optical spectroscopy in order to increase sample sizes. With a new class of 30m telescopes like the ELT we should also be able to detect GRB afterglows behind translucent ($1 < A_V < 5$ mag) and possibly also Giant Molecular Clouds ($A_V > 5$ mag), in order to study the physical conditions in dense molecular gas at high redshift.

*Acknowledgements.* The lead author acknowledges support from a studentship at the European Southern Observatory in Chile and thanks the many astronomers who dedicated their time observing the numerous GRBs with VLT/X-shooter. P.S. acknowledges support through the Sofja Kovalevskaja Award from the Alexander von Humboldt Foundation of Germany, and J.B. acknowledges support through this award. K.E.H. and P.J. acknowledge support by a Project Grant (162948—051) from The Icelandic Research Fund. J.J. acknowledges support from NOVA and NWO-FAPESP grant for advanced instrumentation in astronomy. Finally, we are indebted with Thomas Krühler for providing code to create synthetic GRB afterglow spectra.

# Appendix A: Additional tables

# Appendix B: Additional plots

**Table A.1.** Hydrogen and metal abundances for all 22 GRBs in the sample.

| GRB | z | log$N$(H I) | log$N$(O) | log$N$(Zn) | log$N$(S) | log$N$(Si) | log$N$(Mg) | log$N$(Mn) | log$N$(Cr) | log$N$(Ni) | log$N$(Fe) | log$N$(Ti) |
|---|---|---|---|---|---|---|---|---|---|---|---|---|
| 090809A | 2.7373 | 21.48 ± 0.07 | > 15.6 | 13.25 ± 0.11 | 15.85 ± 0.60 | 16.29 ± 0.10 | > 14.8 | 13.73 ± 0.24 | 13.52 ± 0.22 | 14.42 ± 0.39 | > 14.9 | - |
| 090926A | 2.1069 | 21.58 ± 0.01 | > 15.0 | 12.24 ± 0.11 | 14.52 ± 0.04 | 15.28 ± 0.05 | > 14.6 | < 13.0 | < 12.9 | 13.40 ± 0.11 | 14.21 ± 0.02 | < 11.6 |
| 100219A | 4.6676 | 21.28 ± 0.02 | > 15.5 | - | 15.18 ± 0.05 | > 15.3 | - | - | - | 14.30 ± 0.30 | > 14.8 | - |
| 111008A | 4.9910 | 22.39 ± 0.01 | > 16.8 | 13.54 ± 0.30 | 15.66 ± 0.08 | > 15.8 | > 15.8 | 13.63 ± 0.12 | 14.20 ± 0.08 | 14.63 ± 0.17 | > 15.5 | - |
| 111107A | 2.8930 | 21.10 ± 0.04 | > 16.4 | - | < 16.0 | 15.87 ± 0.20 | < 16.2 | - | 13.62 ± 0.44 | < 14.0 | > 14.7 | - |
| 120327A | 2.8143 | 22.07 ± 0.01 | > 16.3 | 13.21 ± 0.04 | 15.75 ± 0.02 | > 14.5 | 16.40 ± 0.23 | - | 14.12 ± 0.02 | 14.57 ± 0.04 | 15.79 ± 0.06 | 12.69 ± 0.08 |
| 120712A | 4.1719 | 20.44 ± 0.05 | > 15.5 | - | < 15.2 | > 14.6 | - | - | - | < 13.1 | - | - |
| 120716A | 2.4874 | 21.73 ± 0.03 | > 15.5 | 13.65 ± 0.05 | 16.00 ± 0.14 | > 16.2 | > 14.8 | 13.92 ± 0.18 | 14.40 ± 0.22 | < 15.4 | 15.81 ± 0.22 | - |
| 120815A | 2.3582 | 22.09 ± 0.01 | > 16.0 | 13.27 ± 0.02 | 15.51 ± 0.05 | > 15.4 | - | 13.28 ± 0.07 | 13.37 ± 0.08 | 14.16 ± 0.08 | 15.11 ± 0.05 | < 12.5 |
| 120909A | 3.9290 | 21.82 ± 0.02 | > 15.7 | - | 15.89 ± 0.12 | > 16.0 | 16.52 ± 0.06 | 13.97 ± 0.33 | - | 14.27 ± 0.05 | 15.73 ± 0.09 | < 12.7 |
| 121024A | 2.3005 | 21.78 ± 0.02 | > 15.0 | 13.65 ± 0.06 | > 15.7 | 16.69 ± 0.14 | > 17.8 | 13.75 ± 0.03 | 14.09 ± 0.05 | 14.24 ± 0.34 | 15.73 ± 0.05 | - |
| 130408A | 3.7579 | 21.90 ± 0.01 | > 16.0 | 13.05 ± 0.07 | 15.38 ± 0.05 | 15.91 ± 0.08 | > 15.6 | 13.28 ± 0.23 | 13.98 ± 0.08 | 14.16 ± 0.09 | > 15.4 | < 12.9 |
| 130606A | 5.9127 | 19.88 ± 0.01 | 14.87 ± 0.07 | - | < 13.6 | 13.56 ± 0.10 | < 14.1 | < 14.4 | - | < 13.4 | 13.12 ± 0.08 | - |
| 140311A | 4.9550 | 22.30 ± 0.02 | > 15.5 | 13.28 ± 0.14 | 15.48 ± 0.12 | > 14.7 | > 14.6 | 13.49 ± 0.03 | 14.11 ± 0.08 | 14.40 ± 0.23 | 15.78 ± 0.15 | - |
| 141028A | 2.3333 | 20.39 ± 0.03 | > 14.7 | < 12.3 | < 14.3 | 14.26 ± 0.13 | > 13.9 | < 12.4 | < 13.6 | < 13.6 | 14.29 ± 0.10 | < 13.1 |
| 141109A | 2.9940 | 22.18 ± 0.02 | > 17.2 | 13.18 ± 0.06 | 15.81 ± 0.05 | > 15.6 | > 15.6 | 13.57 ± 0.56 | 13.80 ± 0.03 | 14.06 ± 0.17 | 15.54 ± 0.04 | < 12.8 |
| 150403A | 2.0571 | 21.73 ± 0.02 | > 16.7 | 13.32 ± 0.04 | 15.78 ± 0.08 | > 15.8 | > 15.7 | > 13.5 | 14.06 ± 0.03 | 14.23 ± 0.10 | 15.54 ± 0.07 | - |
| 151021A | 2.3297 | 22.14 ± 0.03 | > 16.0 | 13.79 ± 0.06 | 15.97 ± 0.15 | > 16.0 | > 16.1 | 13.86 ± 0.03 | 14.21 ± 0.04 | 14.61 ± 0.07 | 15.78 ± 0.06 | - |
| 151027B | 4.0650 | 20.54 ± 0.07 | > 16.4 | < 13.2 | 14.91 ± 0.16 | 15.12 ± 0.41 | < 14.8 | < 12.4 | < 13.1 | < 13.1 | 15.07 ± 0.39 | < 12.0 |
| 160203A | 3.5187 | 21.74 ± 0.02 | > 16.9 | - | 15.56 ± 0.04 | 16.02 ± 0.02 | > 15.3 | > 13.1 | 13.73 ± 0.04 | 14.28 ± 0.05 | 15.53 ± 0.18 | - |
| 161023A | 2.7100 | 20.95 ± 0.01 | > 16.0 | < 12.5 | 14.85 ± 0.03 | 15.21 ± 0.05 | > 14.4 | 12.99 ± 0.06 | - | < 13.8 | 14.80 ± 0.03 | < 11.5 |
| 170202A | 3.6456 | 21.53 ± 0.04 | > 15.5 | < 13.2 | 15.39 ± 0.08 | > 15.4 | > 13.9 | < 13.2 | < 13.5 | 13.84 ± 0.21 | > 14.7 | < 12.0 |

**Notes.** Upper and lower limits are given at the 3$\sigma$ confidence level.

**Table A.2.** Molecular hydrogen column densities and upper limits

| GRB | log$N$(H$_2$) total | J0 | J1 | J2 | $b$ (km/s) | fitted and/or most constraining transitions J0 | J1 | J2 |
|---|---|---|---|---|---|---|---|---|
| 120327A | $17.39 \pm 0.13$ | $< 13.8$ | $17.29 \pm 0.08$ | $< 17$ | 2 | L4R(0); L10R(0); L2R(0) | W0Q(1); L4R(1); L4P(1) | W2Q(2); L3R(2); L5P(5) |
| | $15.21 \pm 0.17$ | $< 12.9$ | $15.07 \pm 0.20$ | $14.63 \pm 0.05$ | 10 | L4R(0); L10R(0); L2R(0) | W0Q(1); L4R(1); L4P(1) | W2Q(2); L3R(2); L5P(5) |
| 120815A | $20.42 \pm 0.08$ | $20.09 \pm 0.08$ | $20.14 \pm 0.07$ | $17.82 \pm 0.49$ | $14 \pm 3$ | all lines between 982Å to 1117Å | | |
| 120909A | $17.25 \pm 0.23$ | $< 14.4$ | $17.23 \pm 0.23$ | $< 16.1$ | 2 | L8R(0); L4R(0); L0R(0) | W2Q(1); W1R(1); W0Q(1) | W2R(2); W0R(2); L10R(2) |
| | $15.23 \pm 0.25$ | $< 14.0$ | $15.14 \pm 0.24$ | $< 14.7$ | 10 | L8R(0); L4R(0); L0R(0) | W2Q(1); W1R(1); W0Q(1) | W2R(2); W0R(2); L10R(2) |
| 121024A | $19.90 \pm 0.17$ | $19.78 \pm 0.15$ | $19.29 \pm 0.22$ | $< 18.3$ | 2 | L1R(0) | L2P(1) | |
| | $19.87 \pm 0.19$ | $19.77 \pm 0.16$ | $19.19 \pm 0.29$ | $< 18.3$ | 10 | L1R(0) | L2P(1) | |
| 141109A | $18.02 \pm 0.12$ | $17.21 \pm 0.28$ | $17.94 \pm 0.09$ | $< 16.4$ | 2 | W1R(0); L4R(0); L10R(0) | W1R(1); W0Q(1); L10R(1) | W1R(2); L10R(2); L5P(2) |
| | $16.31 \pm 0.42$ | $15.37 \pm 0.35$ | $16.25 \pm 0.43$ | $< 14.4$ | 10 | W1R(0); L4R(0); L10R(0) | W1R(1); W0Q(1); L10R(1) | W1R(2); L10R(2); L5P(2) |
| 150403A | $19.90 \pm 0.14$ | $19.69 \pm 0.13$ | $19.47 \pm 0.15$ | $< 17.8$ | $5 \pm 3$ | all lines between 1046Å to 1110Å | | |
| 151021A | $< 18.99 \pm 1.28$ | $< 19.1$ | $18.35 \pm 1.89$ | $< 18.4$ | 2 | all lines between 1043Å to 1117Å | | |
| 160203A | $< 16.7$ | $< 14.0$ | $< 16.7$ | $< 14.0$ | 2 | L7R(0) | W1Q(1); W3Q(1); L9R(1) | W3R(2); L8P(2) |
| 170202A | $< 17.1$ | $< 16.4$ | $< 17.0$ | $< 14.9$ | 2 | W1R(0); L4R(0); L2R(0) | W1R(1); L7R(1); L2P(1) | W0Q(2); L9R(2) |
| 090809A | $< 18.0$ | $< 18.0$ | $< 16.4$ | $< 16.9$ | 2 | W1R(0); L8R(0); L0R(0) | W1Q(1); W1R(1); L0R(1) | W2R(2); W3R(2); L7R(2) |
| 090926A | $< 14.2$ | $< 13.6$ | $< 14.0$ | $< 13.4$ | 2 | L0R(0); L2R(0) | L4R(1); L3R(1); L2R(1) | L4R(2); L3R(2) |
| 100219A | $< 18.0$ | $< 17.6$ | $< 17.6$ | $< 17.4$ | 2 | W0R(0); L8R(0); L4R(0) | W0R(1); L4R(1); L5P(1) | L7R(2); L8P(2); L4P(2) |
| 111008A | $< 17.2$ | $< 16.5$ | $< 17.0$ | $< 16.5$ | 2 | L9R(0); L4R(0); L12R(0) | W0Q(1); L9R(1); L8P(1) | L4R(2); L4P(2); W1P(2) |
| 111107A | $< 15.5$ | $< 14.8$ | $< 15.0$ | $< 15.2$ | 2 | W0R(0); L9R(0); L4R(0) | W1Q(1); W0Q(1); L7R(1) | W0Q(2); L7R(2); L9R(2) |
| 120712A | $< 16.9$ | $< 16.2$ | $< 16.6$ | $< 16.2$ | 2 | W1R(0); W3R(0); L7R(0) | W0Q(1); L7R(1); L10R(1) | W1R(2); W0Q(2); L8R(2) |
| 120716A | $< 18.2$ | $< 17.6$ | $< 18.0$ | $< 17.4$ | 2 | L2R(0); L1R(0); L0R(0) | L3P(1); L2P(1); L0P(1) | L2R(2); L2P(2); L1R(2) |
| 130408A | $< 15.5$ | $< 14.1$ | $< 15.2$ | $< 15.1$ | 2 | W1R(0); L2R(0); L12R(0) | L9R(1); L0R(1); L0P(1) | W1R(2); L7R(2); L4P(2) |
| 130606A | $< 17.9$ | $< 17.5$ | $< 17.5$ | $< 17.4$ | 2 | L5R(0); L10R(0); L0R(0) | L5R(1); L5P(1); L12R(1) | L5R(2); L6P(2); L7P(2) |
| 140311A | $< 17.8$ | $< 17.2$ | $< 17.5$ | $< 17.0$ | 2 | W1R(0); L2R(0); L1R(0) | W1Q(1); L5R(1); L5P(1) | W1R(2); L7R(2); L4R(2) |
| 141028A | $< 17.4$ | $< 16.4$ | $< 17.1$ | $< 17.0$ | 2 | L7R(0), L4R(0) | L2R(1) | L4R(2) |
| 151027B | $< 16.5$ | $< 16.0$ | $< 16.2$ | $< 15.8$ | 2 | W1R(0); W3R(0); L7R(0) | W1R(1); L5R(1); L10R(1) | L7R(2); L10R(2); L5P(2) |
| 161023A | $< 14.7$ | $< 14.6$ | $< 14.0$ | $< 13.5$ | 2 | W1R(0); W2R(0); L8R(0) | W0R(1); L7R(1); L3P(1) | L4R(2); L3P(2) |

**Notes.** Upper limits are given at $3\sigma$ confidence and are determined assuming a broadening parameter of $b = 2$ km/s. In case of 120327A, 120909A, 121024A and 141109A the observed velocity width is close to the instrumental resolution, and we therefore measured the column density for $b = 2$ km/s and $b = 10$ km/s in order to determine a realistic range for the total H$_2$ column density.

**Table A.3.** Overall properties of the $H_2$ GRB sample.

| GRB yymmdd# | Redshift (z) | $A_V$ (SED) (mag) | $\log N(\text{H\,\textsc{i}})$ (cm$^{-2}$) | $\log N(\text{H}_2)$ (cm$^{-2}$) | $\log f(\text{H}_2)$ | [X/H] | Ion X | [Y/Fe] | Ion Y | $F_*$ | [M/H] | $\mathcal{DTM}$ | $A_V$ (depl.) (mag) |
|---|---|---|---|---|---|---|---|---|---|---|---|---|---|
| 090809A | 2.7373 | $0.11^{+0.04}_{-0.04}$ (2) | $21.48 \pm 0.07$ | < 18.0 | < −3.18 | $-0.86 \pm 0.13$ | Zn | < 1.12 | Zn | $-0.39 \pm 0.31$ | $-0.46 \pm 0.15$ | $0.71 \pm 0.11$ | $0.33 \pm 0.31$ |
| 090926A | 2.1069 | < 0.03 (3) | $21.58 \pm 0.01$ | < 14.2 | < −7.08 | $-1.97 \pm 0.11$ | Zn | $0.88 \pm 0.11$ | Zn | $-0.25 \pm 0.07$ | $-1.72 \pm 0.05$ | $0.76 \pm 0.02$ | $0.02 \pm 0.02$ |
| 100219A | 4.6676 | $0.15^{+0.04}_{-0.05}$ (1) | $21.28 \pm 0.02$ | < 18.0 | < −2.98 | $-1.24 \pm 0.05$ | S | < 0.72 | S | $-1.32 \pm 0.44$ | $-1.16 \pm 0.11$ | $0.23 \pm 0.26$ | $0.01 \pm 0.03$ |
| 111008A | 4.9910 | $0.13^{+0.03}_{-0.07}$ (1) | $22.39 \pm 0.01$ | < 17.2 | < −4.89 | $-1.48 \pm 0.31$ | Zn | < 0.89 | Zn | $-1.18 \pm 0.15$ | $-1.79 \pm 0.10$ | $0.32 \pm 0.09$ | $0.06 \pm 0.04$ |
| 111107A | 2.8930 | < 0.15 (2) | $21.10 \pm 0.04$ | < 15.5 | < −5.30 | $-0.74 \pm 0.20$ | Si | < 1.13 | Si | $-0.47 \pm 0.81$ | $-0.28 \pm 0.45$ | $0.68 \pm 0.28$ | $0.20 \pm 0.68$ |
| 120327A | 2.8143 | $0.05^{+0.02}_{-0.02}$ (2) | $22.07 \pm 0.01$ | $17.39 \pm 0.13$ | $-4.38 \pm 0.14$ | $-1.49 \pm 0.04$ | Zn | $0.27 \pm 0.07$ | Zn | $-0.99 \pm 0.04$ | $-1.34 \pm 0.02$ | $0.43 \pm 0.02$ | $0.11 \pm 0.01$ |
| 120712A | 4.1719 | $0.08^{+0.03}_{-0.08}$ (1) | $20.44 \pm 0.05$ | < 16.9 | < −3.24 | $-0.38$ | S | < 0.38 | S | - | - | - | - |
| 120716A | 2.4874 | $0.30^{+0.15}_{-0.15}$ (2) | $21.73 \pm 0.03$ | < 18.2 | < −3.23 | $-0.71 \pm 0.06$ | Zn | $0.69 \pm 0.23$ | Zn | $-0.61 \pm 0.22$ | $-0.57 \pm 0.08$ | $0.62 \pm 0.09$ | $0.41 \pm 0.19$ |
| 120815A | 2.3582 | $0.19^{+0.04}_{-0.04}$ (4) | $22.09 \pm 0.01$ | $20.42 \pm 0.08$ | $-1.39 \pm 0.09$ | $-1.45 \pm 0.03$ | Zn | $1.01 \pm 0.05$ | Zn | $-0.09 \pm 0.06$ | $-1.23 \pm 0.03$ | $0.82 \pm 0.02$ | $0.27 \pm 0.03$ |
| 120909A | 3.9290 | $0.16^{+0.04}_{-0.04}$ (2) | $21.82 \pm 0.02$ | $17.25 \pm 0.23$ | $-4.27 \pm 0.25$ | $-1.06 \pm 0.12$ | S | $0.50 \pm 0.15$ | S | $+0.20 \pm 0.14$ | $-0.29 \pm 0.10$ | $0.92 \pm 0.04$ | $1.39 \pm 0.53$ |
| 121024A | 2.3005 | $0.26^{+0.07}_{-0.07}$ (2) | $21.78 \pm 0.02$ | $19.90 \pm 0.17$ | $-1.59 \pm 0.18$ | $-0.76 \pm 0.06$ | Zn | $0.77 \pm 0.08$ | Zn | $-0.42 \pm 0.10$ | $-0.68 \pm 0.07$ | $0.70 \pm 0.04$ | $0.40 \pm 0.12$ |
| 130408A | 3.7579 | $0.12^{+0.03}_{-0.03}$ (2) | $21.90 \pm 0.01$ | < 15.5 | < −6.10 | $-1.48 \pm 0.07$ | Zn | < 0.50 | Zn | $-1.07 \pm 0.10$ | $-1.46 \pm 0.05$ | $0.39 \pm 0.06$ | $0.05 \pm 0.02$ |
| 130606A | 5.9127 | < 0.02 (1) | $19.88 \pm 0.01$ | < 17.9 | < −1.69 | $-1.83 \pm 0.10$ | Si | $0.41 \pm 0.13$ | Si | $-0.77 \pm 0.15$ | $-1.58 \pm 0.08$ | $0.55 \pm 0.07$ | $0.02 \pm 0.02$ |
| 140311A | 4.9550 | $0.07^{+0.03}_{-0.03}$ (1) | $22.30 \pm 0.02$ | < 17.8 | < −4.20 | $-1.65 \pm 0.14$ | Zn | $0.34 \pm 0.2$ | Zn | $-1.16 \pm 0.16$ | $-2.00 \pm 0.11$ | $0.33 \pm 0.09$ | $0.03 \pm 0.02$ |
| 141028A | 2.3333 | $0.13^{+0.09}_{-0.09}$ (2) | $20.39 \pm 0.03$ | < 17.4 | < −2.71 | $-1.64 \pm 0.13$ | Si | < 0.86 | Zn | $-1.56 \pm 0.39$ | $-1.62 \pm 0.28$ | $0.06 \pm 0.27$ | $0.02 \pm 0.02$ |
| 141109A | 2.9940 | $0.16^{+0.04}_{-0.04}$ (2) | $22.18 \pm 0.02$ | $18.02 \pm 0.12$ | $-3.86 \pm 0.14$ | $-1.63 \pm 0.06$ | Zn | $0.49 \pm 0.07$ | Zn | $-0.60 \pm 0.07$ | $-1.37 \pm 0.05$ | $0.62 \pm 0.03$ | $0.18 \pm 0.05$ |
| 150403A | 2.0571 | $0.12^{+0.02}_{-0.02}$ (6) | $21.73 \pm 0.02$ | $19.90 \pm 0.14$ | $-1.54 \pm 0.15$ | $-1.04 \pm 0.04$ | Zn | $0.63 \pm 0.08$ | Zn | $-0.81 \pm 0.07$ | $-0.92 \pm 0.05$ | $0.53 \pm 0.03$ | $0.15 \pm 0.04$ |
| 151021A | 2.3297 | $0.20^{+0.03}_{-0.03}$ (6) | $22.14 \pm 0.03$ | $18.99 \pm 1.28$ | < −2.85 | $-0.98 \pm 0.07$ | Zn | $0.86 \pm 0.08$ | Zn | $-0.48 \pm 0.10$ | $-0.97 \pm 0.07$ | $0.67 \pm 0.04$ | $0.45 \pm 0.16$ |
| 151021B | 4.0650 | $0.10^{+0.05}_{-0.06}$ (1) | $20.54 \pm 0.07$ | < 16.5 | < −3.74 | $-0.76 \pm 0.17$ | S | $0.18 \pm 0.42$ | S | $-0.78 \pm 0.95$ | $-0.59 \pm 0.27$ | $0.54 \pm 0.36$ | $0.22 \pm 0.06$ |
| 160203A | 3.5187 | < 0.10 (2) | $21.74 \pm 0.02$ | < 16.7 | < −4.74 | $-1.31 \pm 0.04$ | S | $0.37 \pm 0.18$ | S | $-0.58 \pm 0.07$ | $-0.92 \pm 0.04$ | $0.63 \pm 0.03$ | $0.19 \pm 0.04$ |
| 161023A | 2.7100 | $0.09^{+0.03}_{-0.03}$ (5) | $20.95 \pm 0.01$ | < 14.7 | < −5.95 | $-1.23 \pm 0.03$ | S | $0.39 \pm 0.04$ | S | $-0.85 \pm 0.06$ | $-1.05 \pm 0.04$ | $0.51 \pm 0.03$ | $0.02 \pm 0.02$ |
| 170202A | 3.6456 | $0.08^{+0.03}_{-0.03}$ (6) | $21.53 \pm 0.04$ | < 17.1 | < −4.13 | $-1.28 \pm 0.09$ | S | < 1.03 | S | $-0.37 \pm 0.34$ | $-1.02 \pm 0.13$ | $0.72 \pm 0.12$ | $0.10 \pm 0.08$ |

**Notes.** The last four columns are the results from fitting the depletion sequences following Wiseman et al. (2017b). The Molecular fraction is $f(\text{H}_2) = 2N(\text{H}_2)/(2N(\text{H}_2)+N(\text{H}))$

**References.** (1) Bolmer et al. (2018); (2) Greiner (2018, in prep.); (3) D'Elia et al. (2010); (4) Zafar et al. (2018b); (5) de Ugarte Postigo et al. (2018); (6) Heintz (2018, in subm.)

1 European Southern Observatory, Alonso de Córdova 3107, Vitacura, Casilla 19001, Santiago 19, Chile
2 Max-Planck-Institut für extraterrestrische Physik, Giessenbachstraße, 85748 Garching, Germany
3 School of Physics and Astronomy, University of Southampton, Southampton, SO17 1BJ, UK
4 European Southern Observatory, Karl-Schwarzschild Str. 2, 85748 Garching bei München, Germany
5 The Cosmic Dawn Center, Niels Bohr Institute, Copenhagen University, Juliane Maries Vej 30, DK-2100 Copenhagen Ø, Denmark
6 Department of Physics, University of Bath, Claverton Down, Bath, BA2 7AY, UK
7 Physics Department, University of Calabria, I-87036 Arcavacata di Rende, Italy University of Calabria, Rende, Italy
8 INAF-Osservatorio Astronomico di Roma, via Frascati 33, 00040, Monteporzio Catone, Italy
9 Astroparticule et Cosmologie, Université Paris Diderot, CNRS/IN2P3, CEA/Irfu, Observatoire de Paris, Sorbonne Paris Cité, 10, Rue Alice Domon et Léonie Duquet, 75205, Paris Cedex 13, France
10 Department of Physics and Astronomy, Clemson University, Clemson, SC29634-0978, USA
11 Centre for Astrophysics and Cosmology, Science Institute, University of Iceland, Dunhagi 5, 107 Reykjavík, Icelandmpe
12 Anton Pannekoek Institute for Astronomy, University of Amsterdam, Science Park 904, 1098 XH Amsterdam, The Netherlands
13 Department of Physics and Astronomy and Leicester Institute of Space and Earth Observation, University of Leicester, University Road, Leicester LE1 7RH, UK
14 Benoziyo Center for Astrophysics and the Helen Kimmel Center for Planetary Science, Weizmann Institute of Science, 76100 Rehovot, Israel
15 Australian Astronomical Observatory, PO Box 915, North Ryde, NSW 1670, Australia

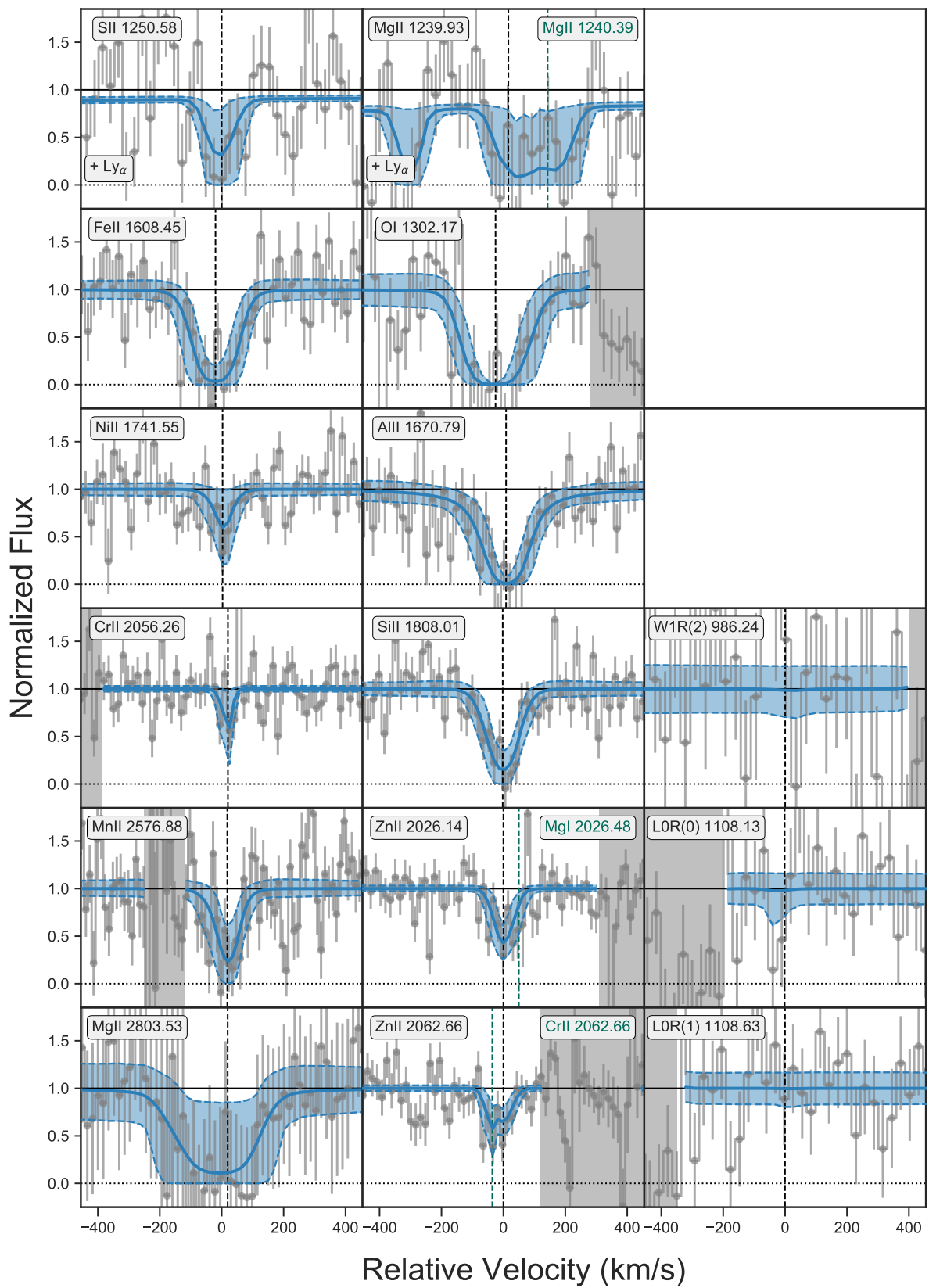

**Fig. B.1.** Results from fitting the absorption lines in the X-shooter spectrum of GRB090809A. As shown to the right, we do not find any evidence for absorption from molecular hydrogen. This X-shooter spectrum was previously also analysed by Skúladóttir (2010).

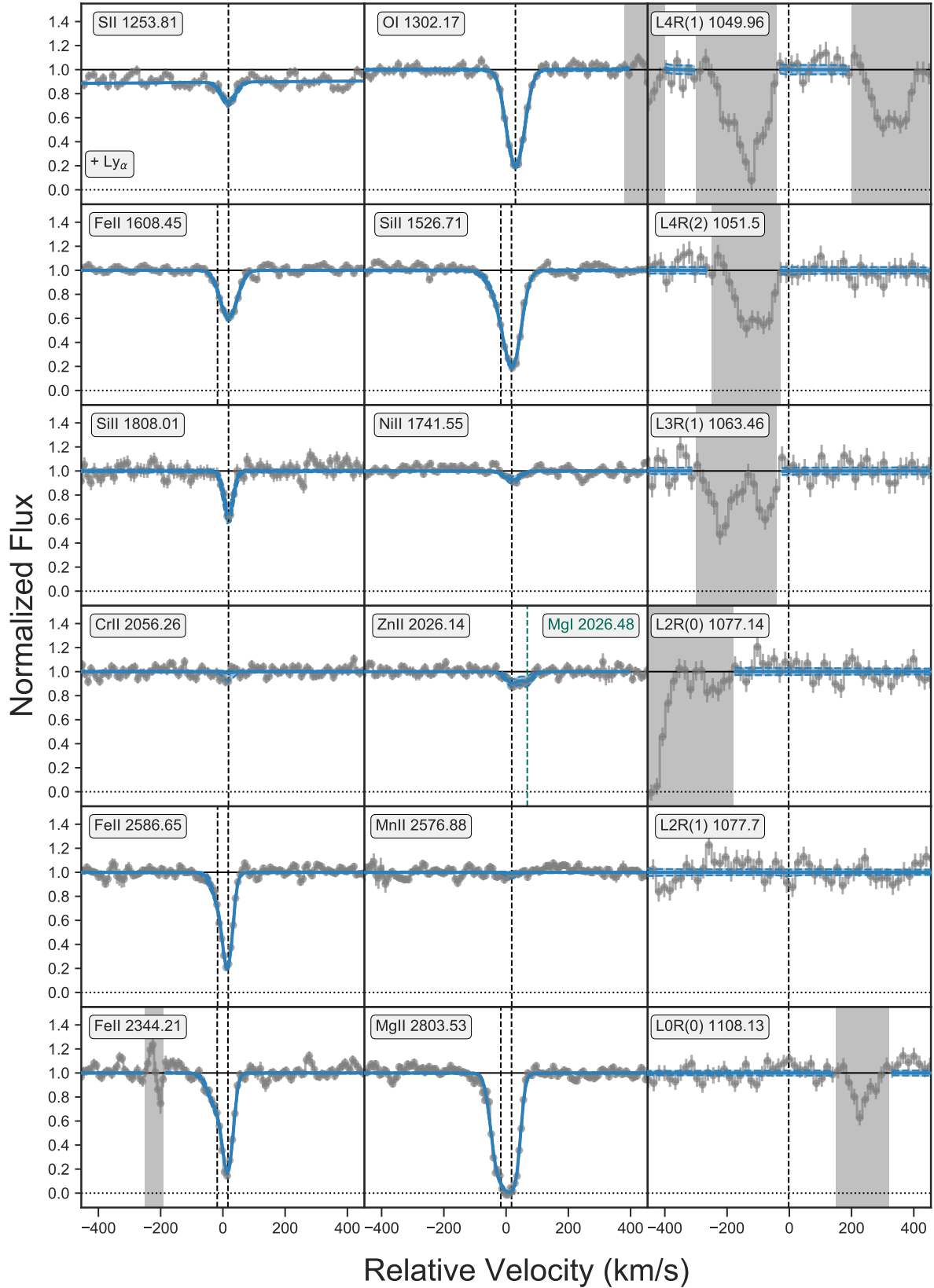

**Fig. B.2.** Results from fitting the absorption lines in the X-shooter spectrum of GRB090926A. As shown to the right, we do not find any evidence for absorption from molecular hydrogen. This X-shooter spectrum was previously also analysed by D'Elia et al. (2010).

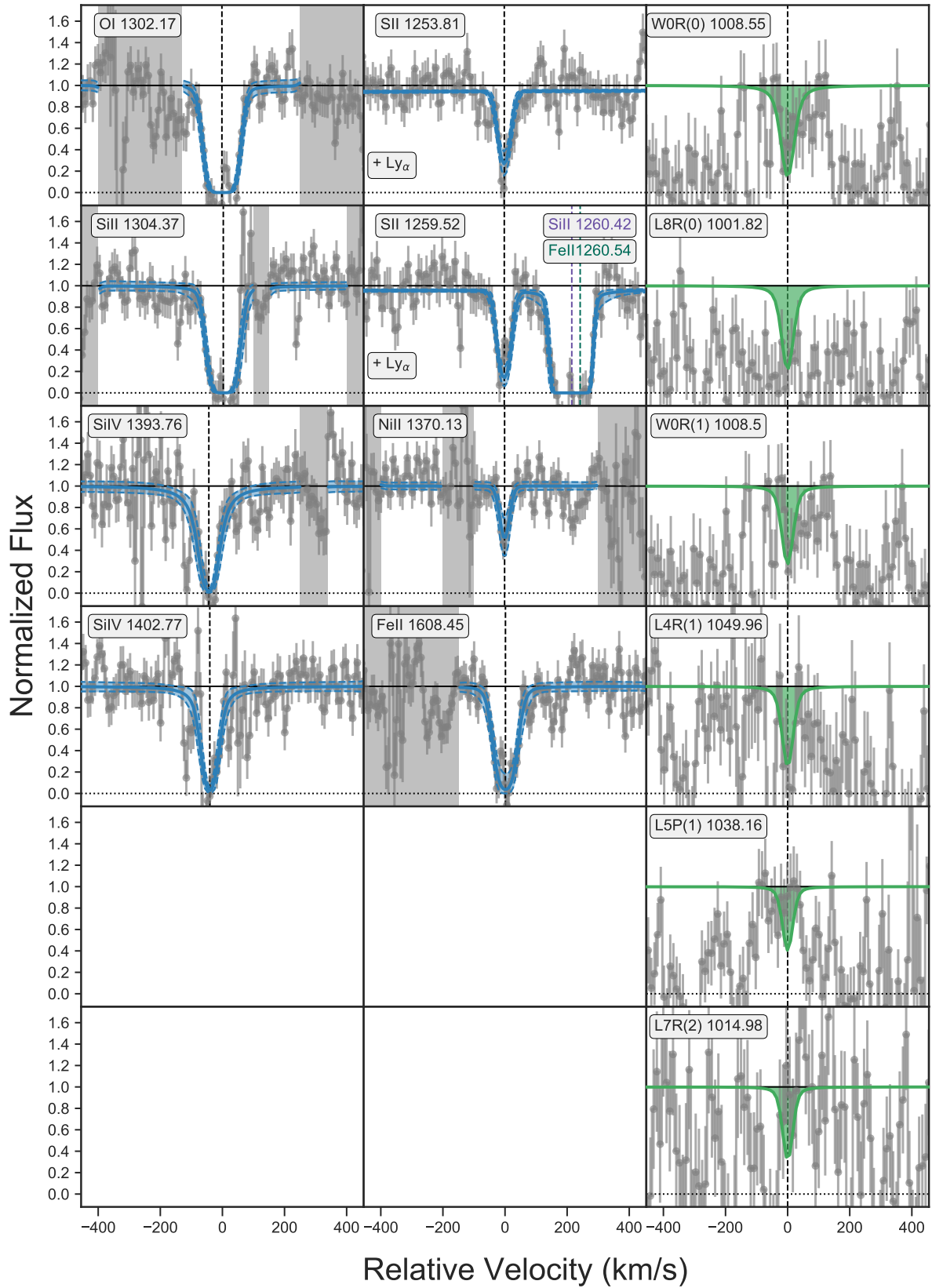

**Fig. B.3.** Results from fitting the absorption lines in the X-shooter spectrum of GRB100219A. As shown to the right, we do not find compelling evidence for absorption from molecular hydrogen, however, due to the strong Lyman-α forest only down to relatively loose upper limits (see Table A.2). These upper limits are over-plotted in green for the 6 most constraining lines. This X-shooter spectrum was previously also analysed by Thöne et al. (2013).

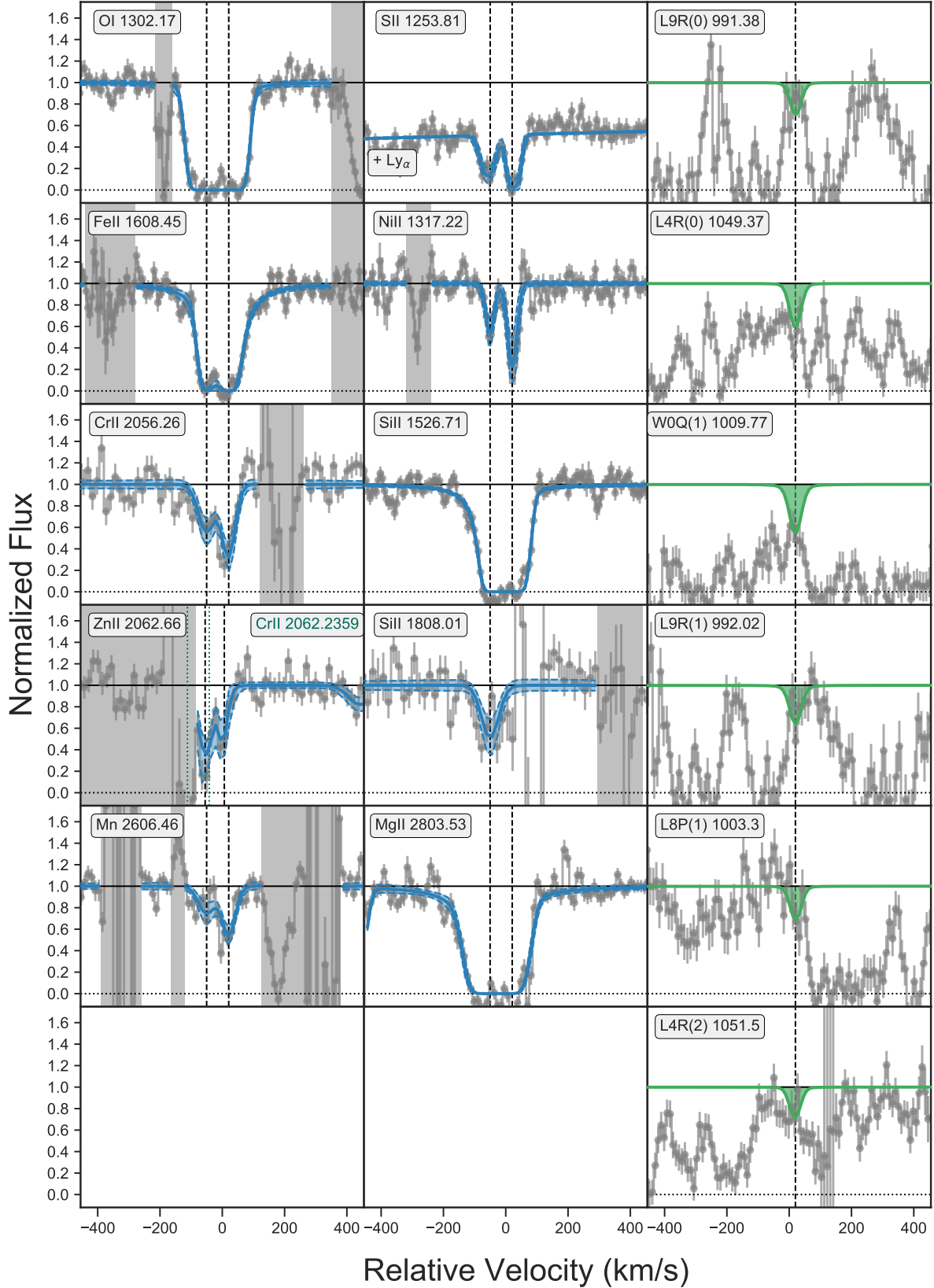

**Fig. B.4.** Results from fitting the absorption lines in the X-shooter spectrum of GRB111008A. As shown to the right, we do not find compelling evidence for absorption from molecular hydrogen, however, due to the strong Lyman-α forest only down to relatively loose upper limits (see Table A.2). These upper limits are over-plotted in green for the 6 most constraining lines. This X-shooter spectrum was previously also analysed by Sparre et al. (2014).

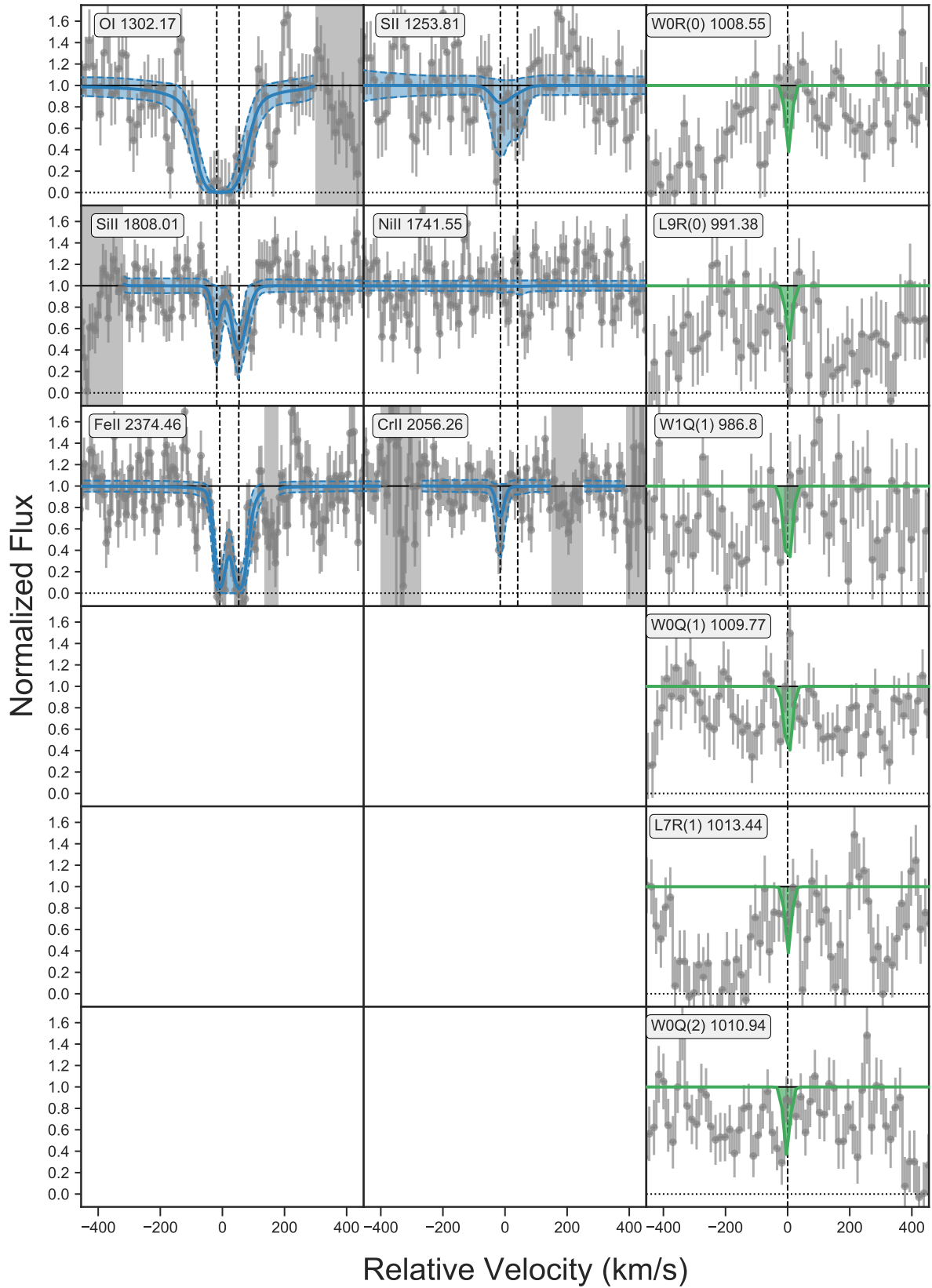

**Fig. B.5.** Results from fitting the absorption lines in the X-shooter spectrum of GRB111107A. As shown to the right, we do not find evidence for absorption from molecular hydrogen down to the upper limits given in Table A.2. These upper limits are over-plotted in green for the 6 most constraining lines.

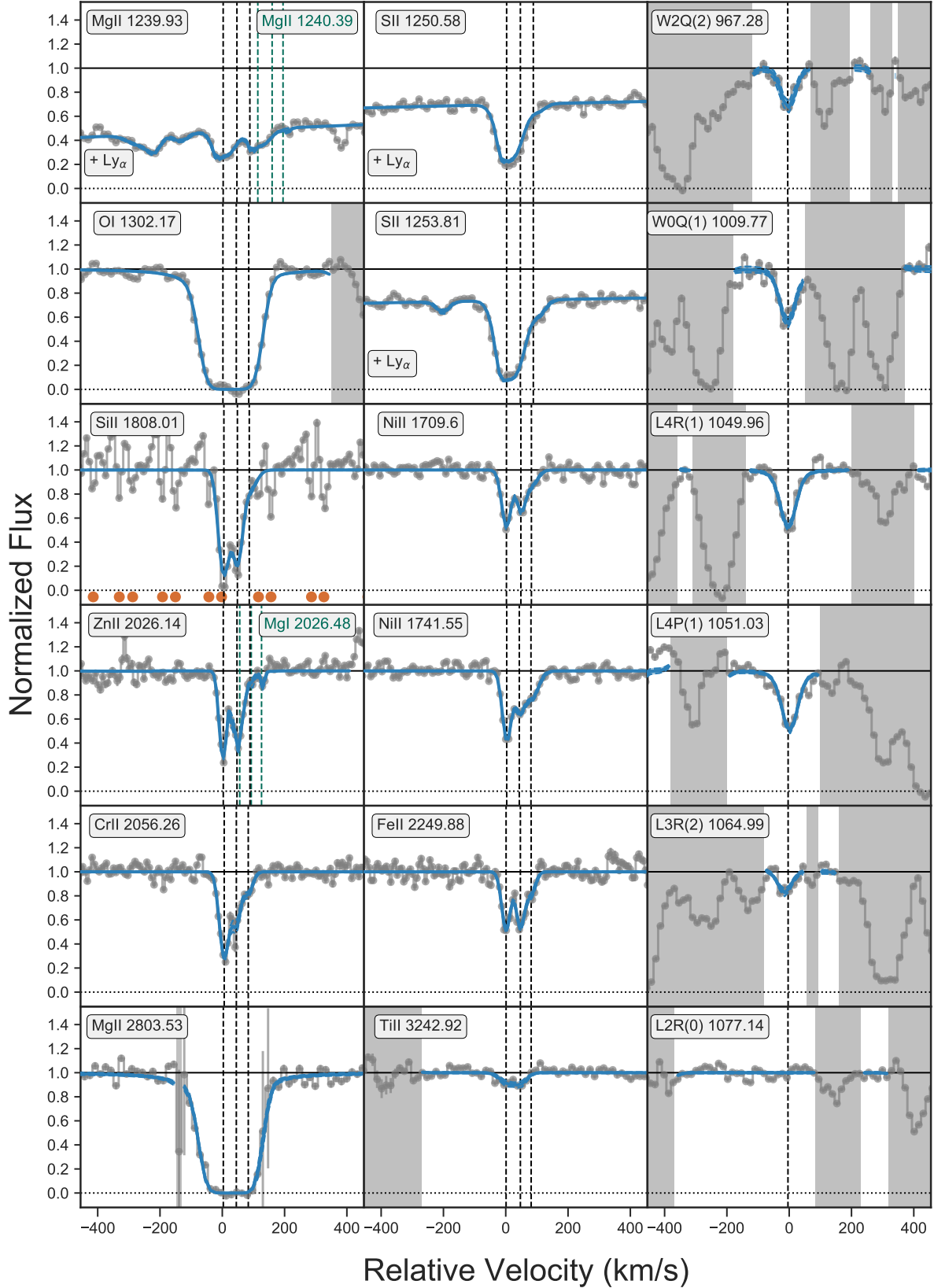

**Fig. B.6.** Results from fitting the absorption lines in the X-shooter spectrum of GRB120327A. The position of telluric lines, which were corrected, are marked by red dots. As shown to the right, we find absorption lines from molecular hydrogen consistent with strongest component. This X-shooter spectrum was previously also analysed by D'Elia et al. (2014).

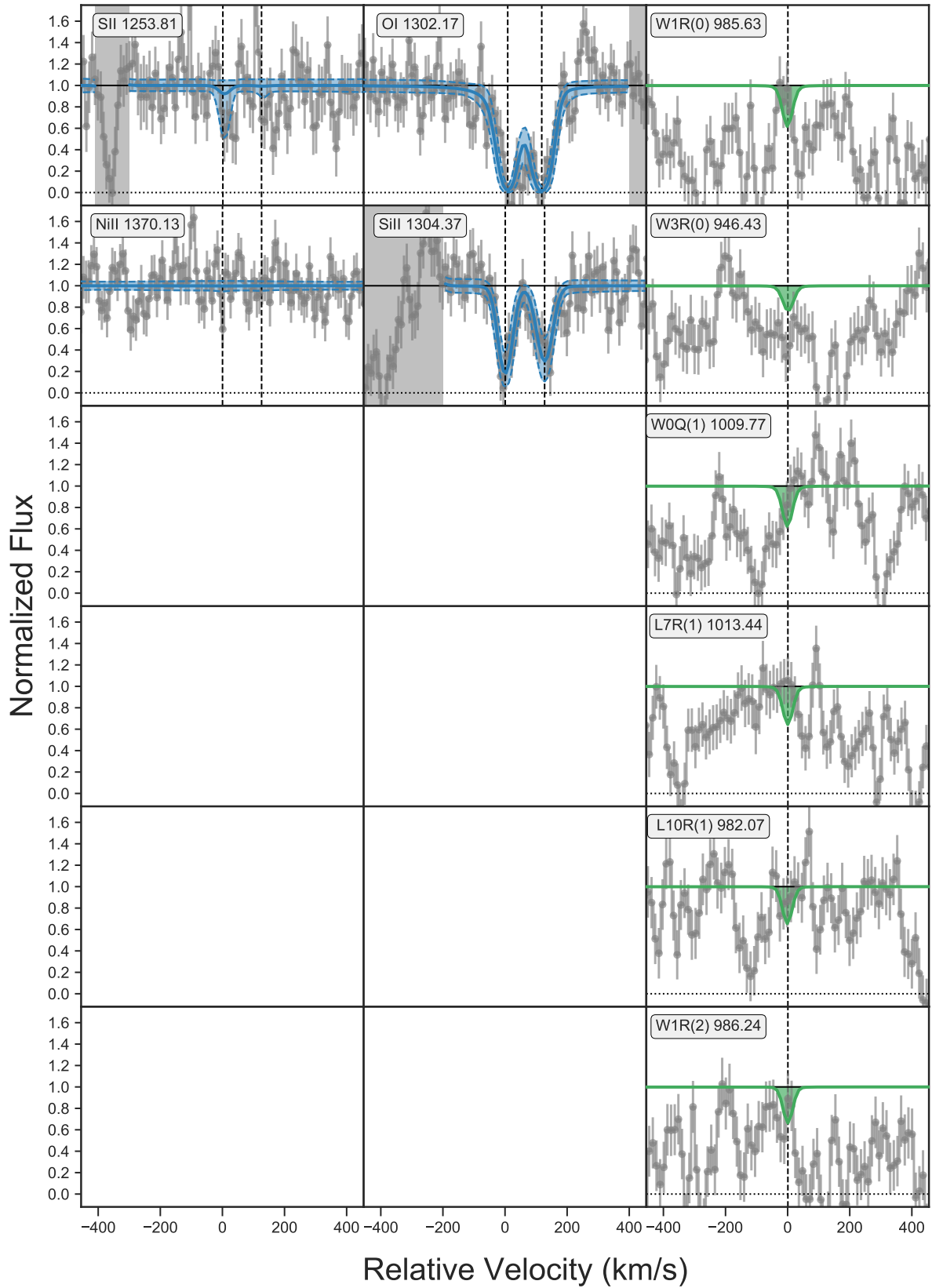

**Fig. B.7.** Results from fitting the absorption lines in the X-shooter spectrum of GRB120712A. As shown to the right, we do not find convincing evidence for absorption from molecular hydrogen down to the upper limits given in Table A.2. These upper limits are over-plotted in green for the 6 most constraining lines.

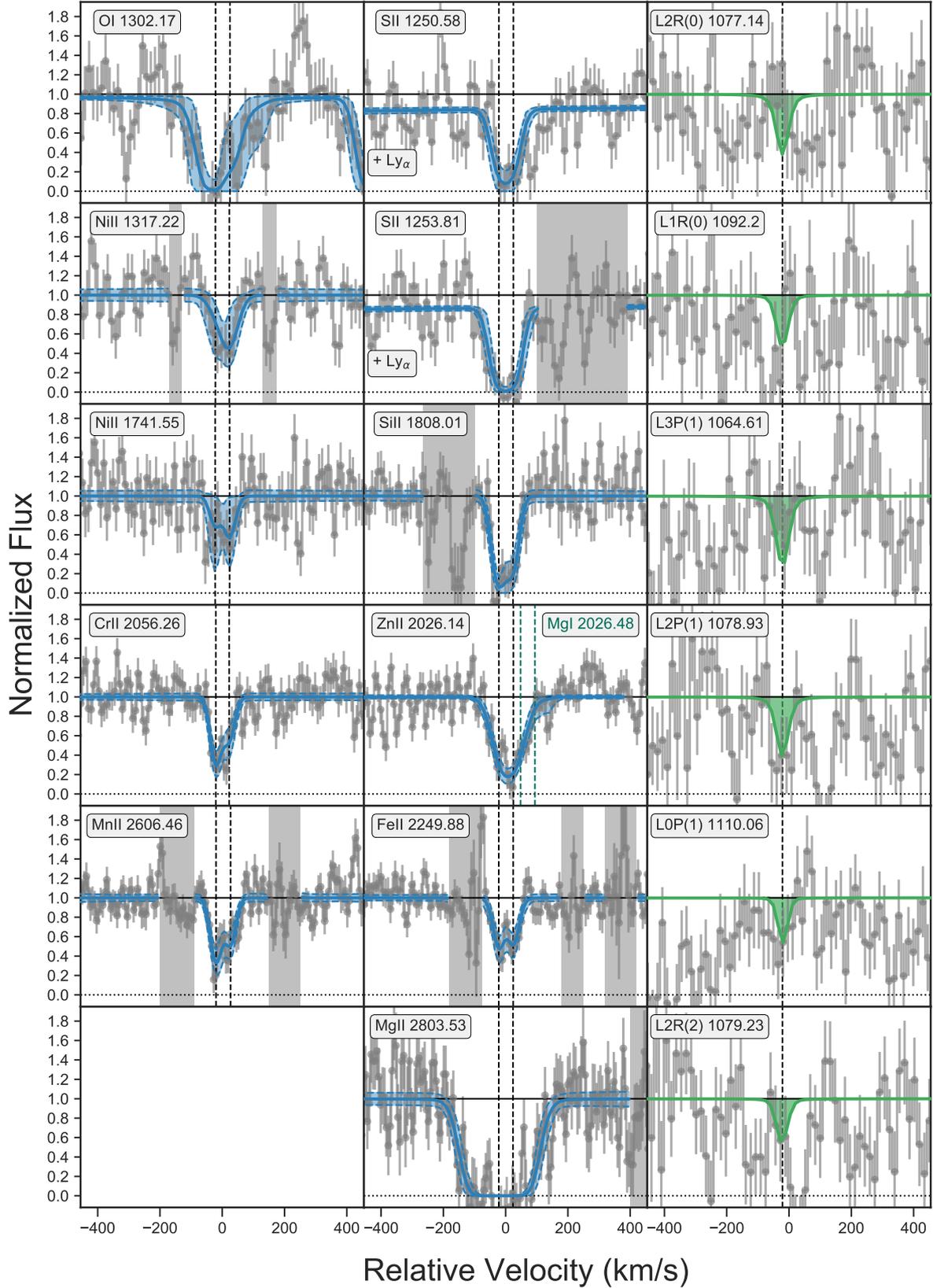

**Fig. B.8.** Results from fitting the absorption lines in the X-shooter spectrum of GRB120716A. As shown to the right, we do not find compelling evidence for absorption from molecular hydrogen, however, due to the poor S/N only down to relatively loose upper limits (see Table A.2). These upper limits are over-plotted in green for the 6 most constraining lines. This X-shooter spectrum was previously also analysed by Wiseman et al. (2017b).

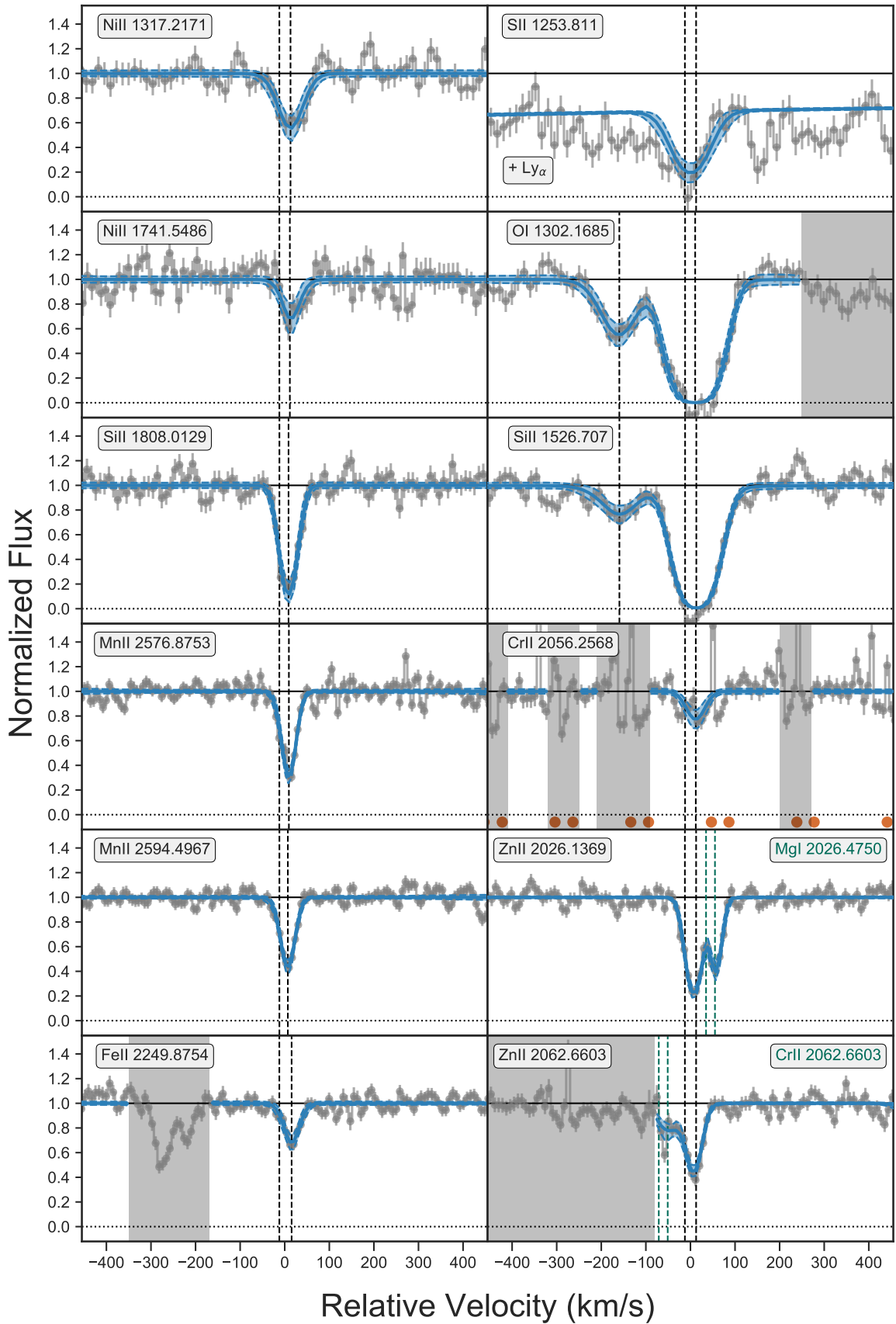

**Fig. B.9.** Results from fitting the absorption lines in the X-shooter spectrum of GRB120815A. The result of fitting the H$_2$ lines is shown in Fig. 9. This X-shooter spectrum was previously also analysed by Krühler et al. (2013).

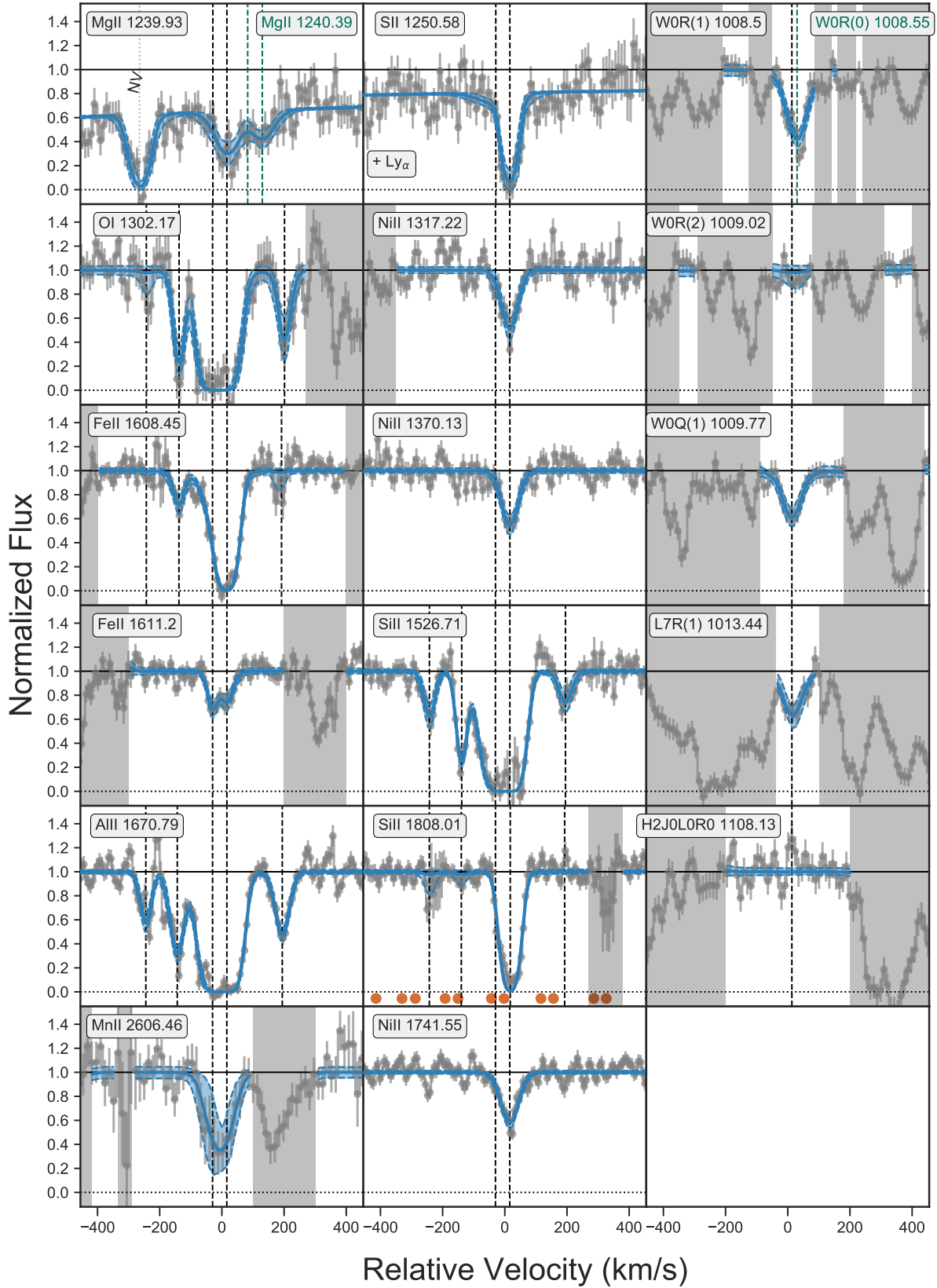

**Fig. B.10.** Results from fitting the absorption lines in the X-shooter spectrum of GRB120909A. As shown to the right, we find evidence for absorption from molecular hydrogen consistent with the strongest component. This X-shooter spectrum was previously also analysed by Wiseman et al. (2017b).

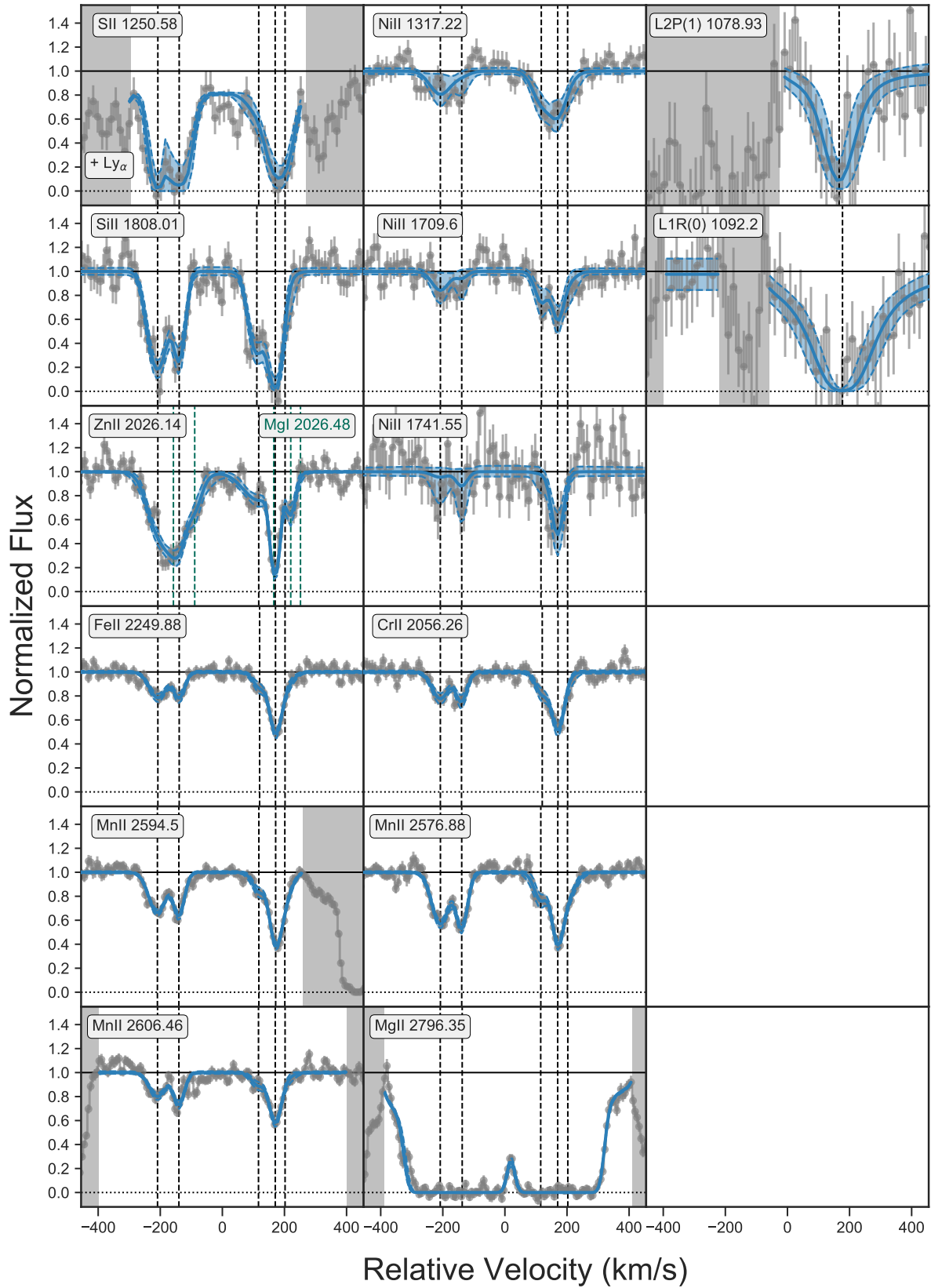

**Fig. B.11.** Results from fitting the absorption lines in the X-shooter spectrum of GRB121024A. As shown to the right, we find evidence for absorption from molecular hydrogen consistent with the three red components. This X-shooter spectrum was previously also analysed by Friis et al. (2015).

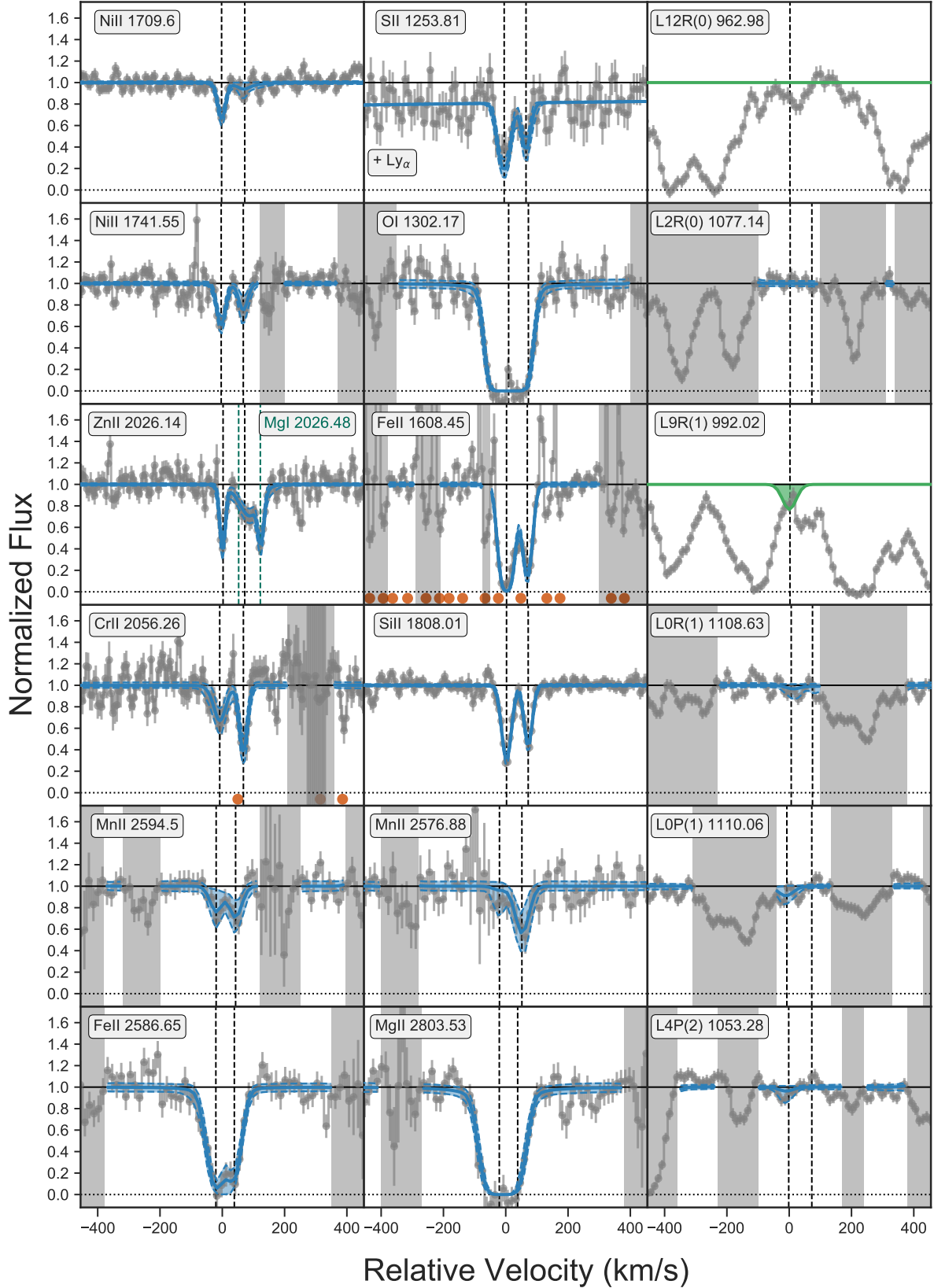

**Fig. B.12.** Results from fitting the absorption lines in the X-shooter spectrum of GRB130408A. As shown to the right, we do not find evidence for absorption from molecular hydrogen down to the upper limits given in Table A.2. These upper limits were derived from the fits shown in blue and are additionally over-plotted for two additional lines in green. This X-shooter spectrum was previously also analysed by Wiseman et al. (2017b).

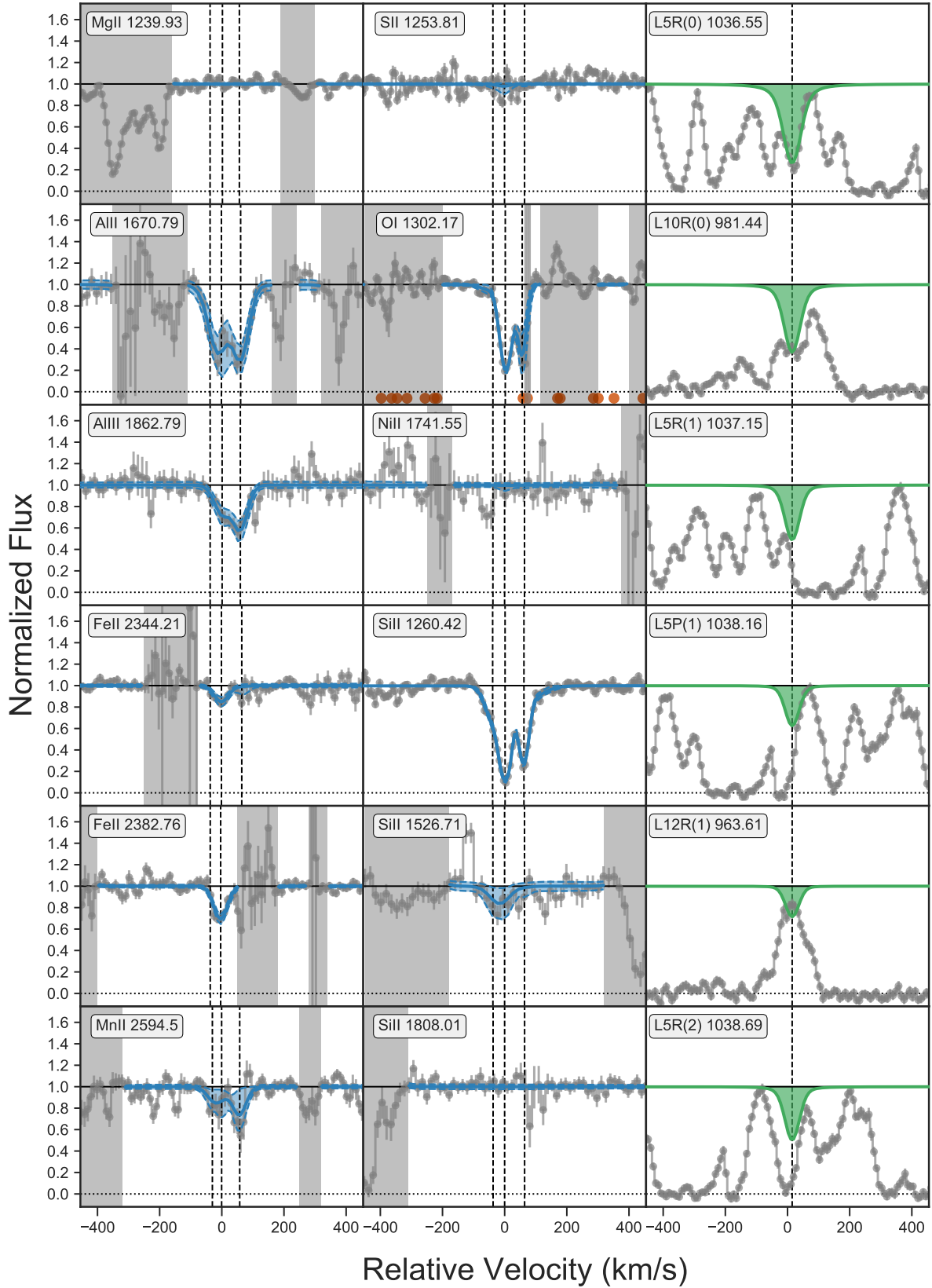

**Fig. B.13.** Results from fitting the absorption lines in the X-shooter spectrum of GRB130606A. As shown to the right, we do not find compelling evidence for absorption from molecular hydrogen, however, due to the strong Lyman-$\alpha$ forest only down to relatively loose upper limits (see Table A.2). These upper limits are over-plotted in green for the 6 most constraining lines. This X-shooter spectrum was previously also analysed by Hartoog et al. (2015).

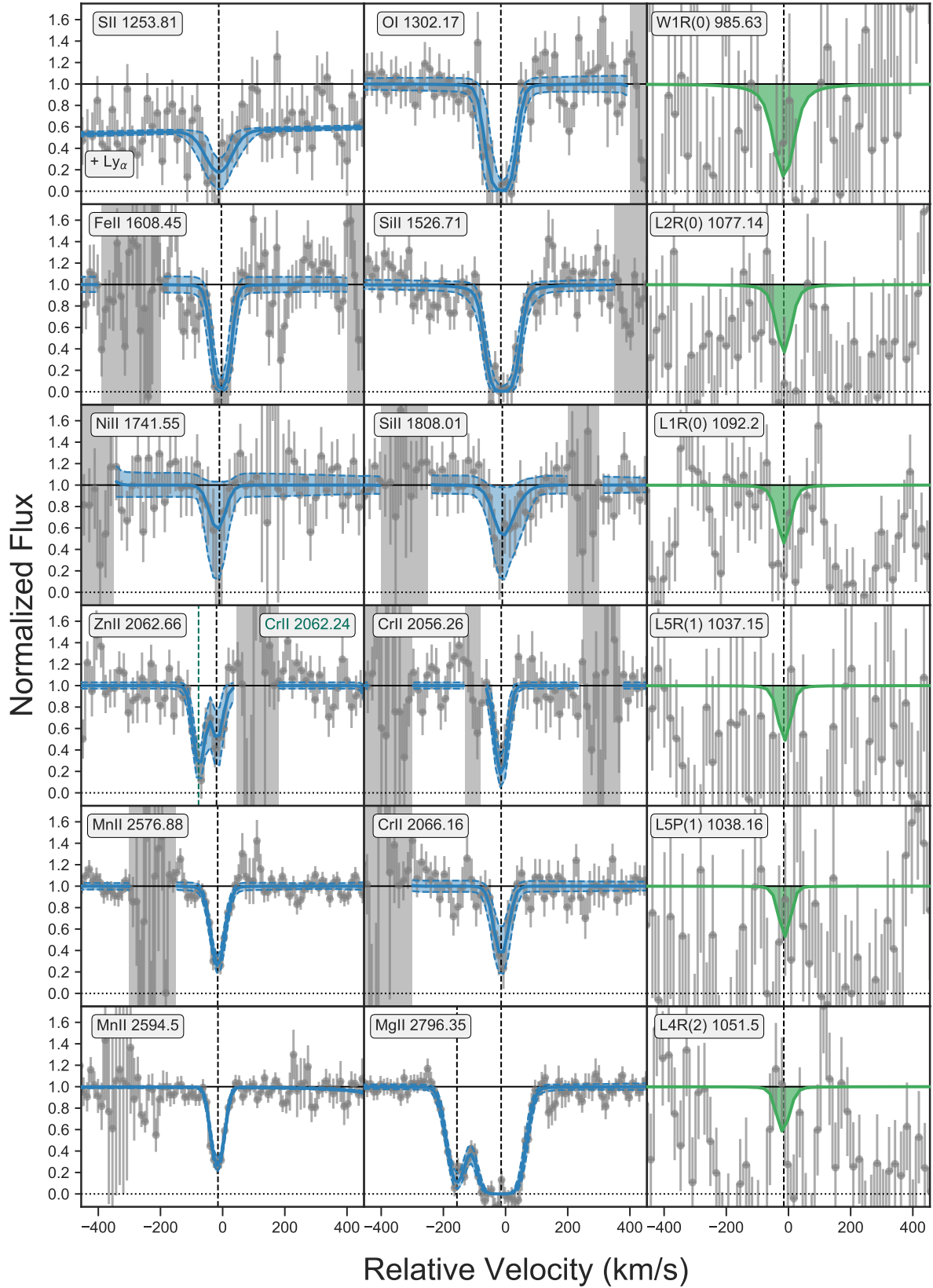

**Fig. B.14.** Results from fitting the absorption lines in the X-shooter spectrum of GRB140311A. As shown to the right, we do not find compelling evidence for absorption from molecular hydrogen, however, due to the poor S/N only down to relatively loose upper limits (see Table A.2). These upper limits are over-plotted in green for the 6 most constraining lines.

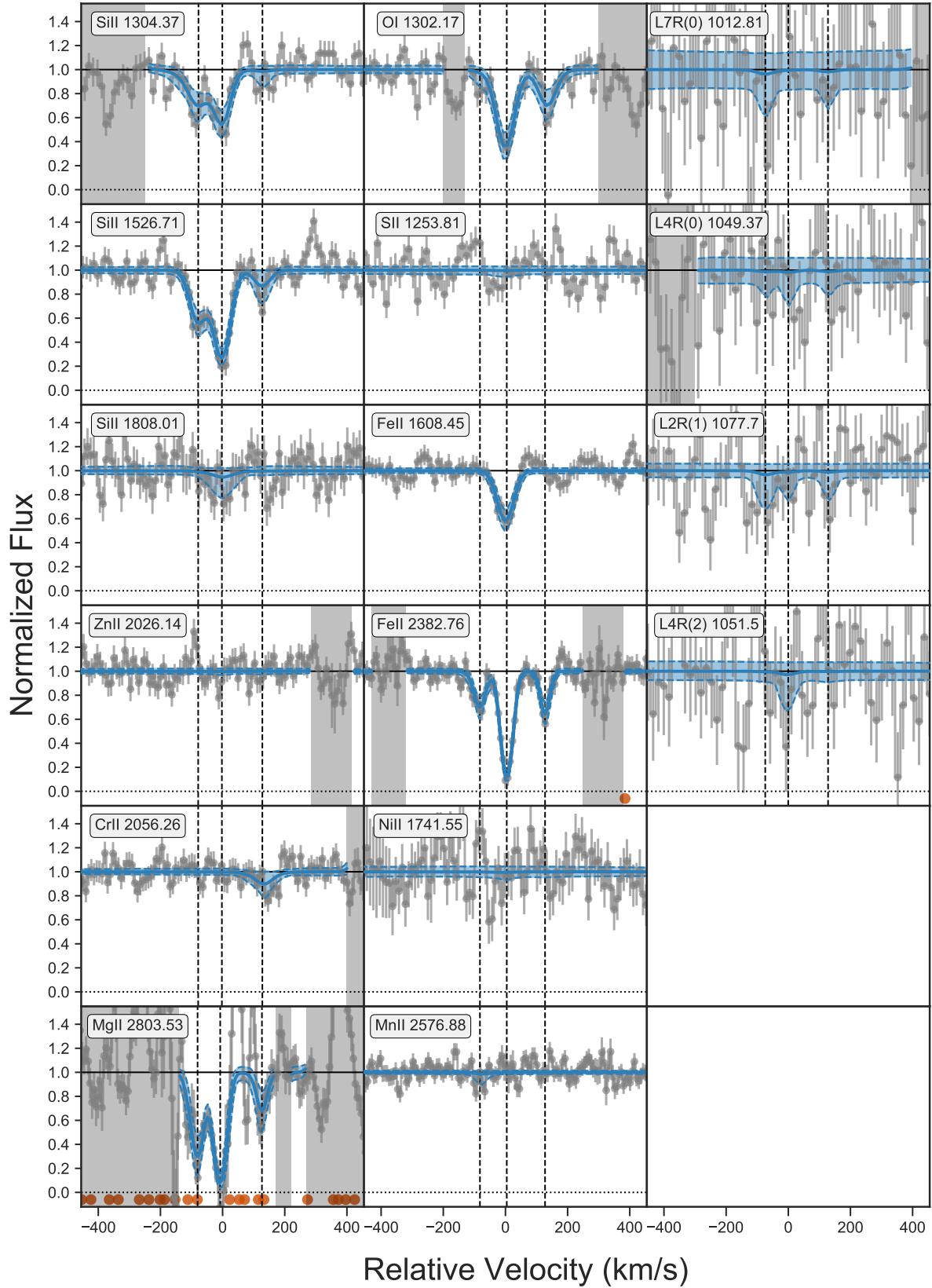

**Fig. B.15.** Results from fitting the absorption lines in the X-shooter spectrum of GRB141028A. The position of telluric lines, which were corrected, are marked by red dots. As shown to the right, we do not find evidence for absorption from molecular hydrogen, however, due to the poor S/N only down to relatively loose upper limits (see Table A.2). This X-shooter spectrum was previously also analysed by Wiseman et al. (2017b).

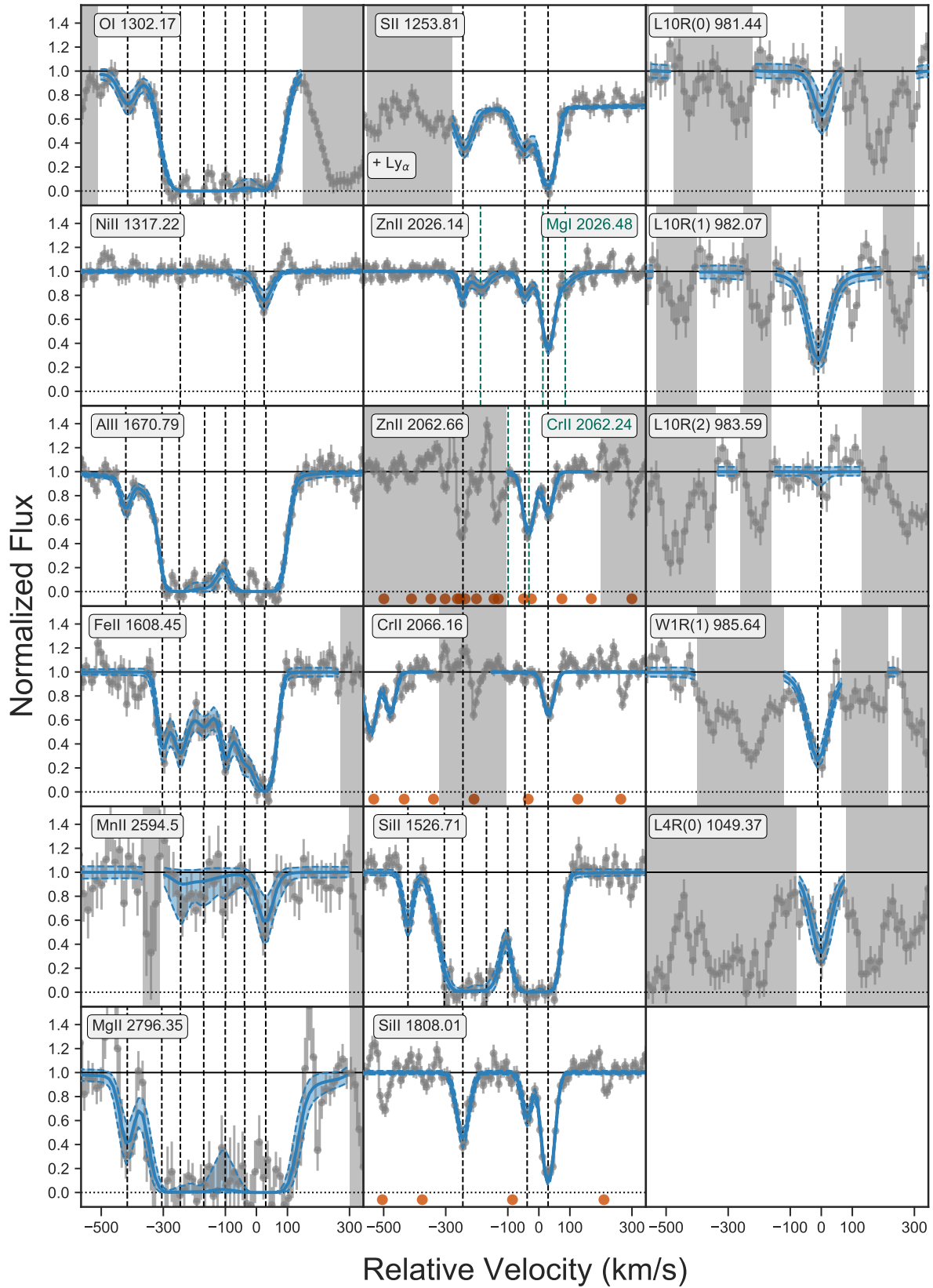

**Fig. B.16.** Results from fitting the absorption lines in the X-shooter spectrum of GRB141109A. The position of telluric lines, which were corrected, are marked by red dots. As shown to the right, we find absorption lines from molecular hydrogen consistent with the two red components.

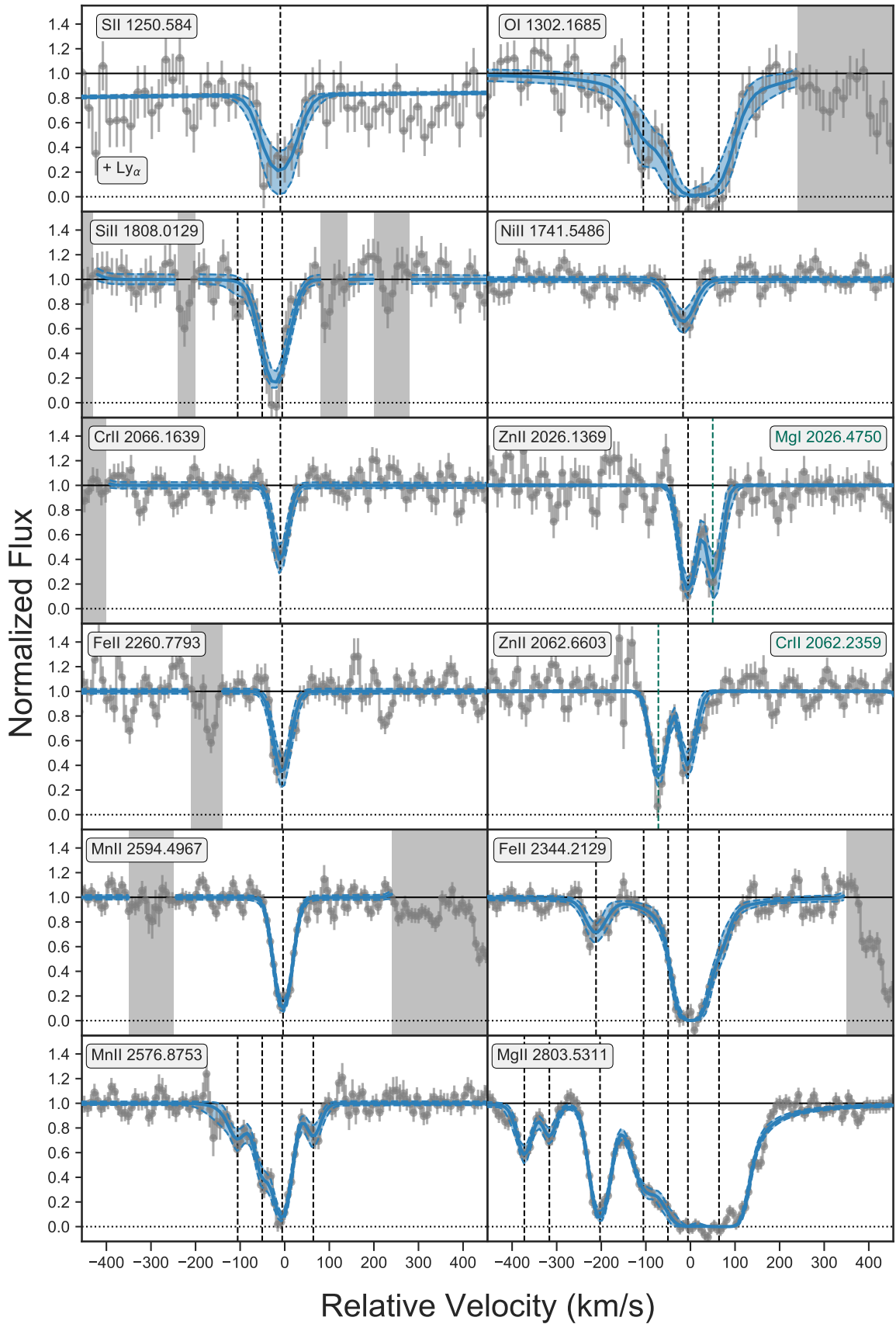

**Fig. B.17.** Results from fitting the absorption lines in the X-shooter spectrum of GRB150403A. The result of fitting the H$_2$ lines is shown in Fig. 10.

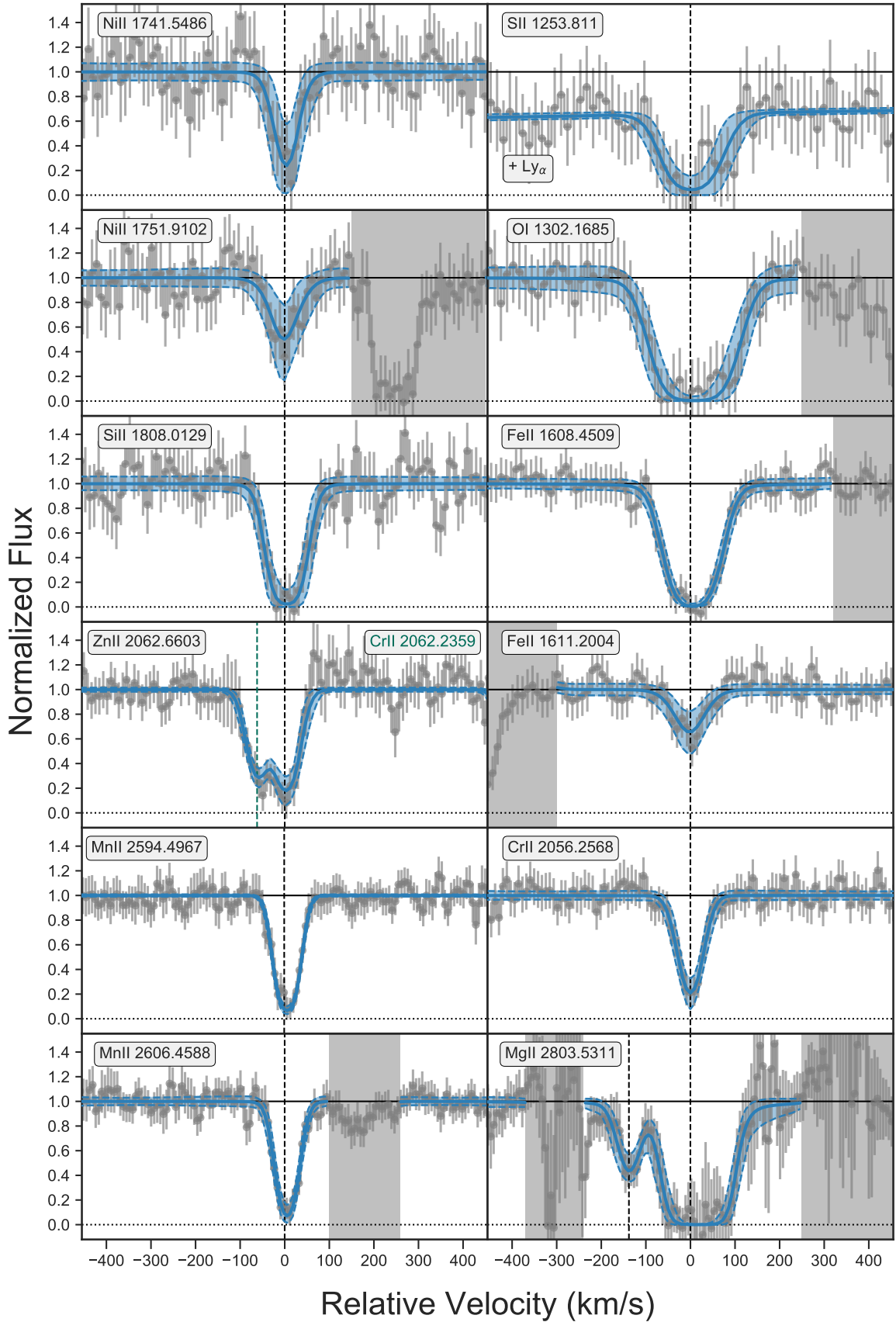

**Fig. B.18.** Results from fitting the absorption lines in the X-shooter spectrum of GRB151021A. The result of fitting the $H_2$ lines is shown in Fig. 11.

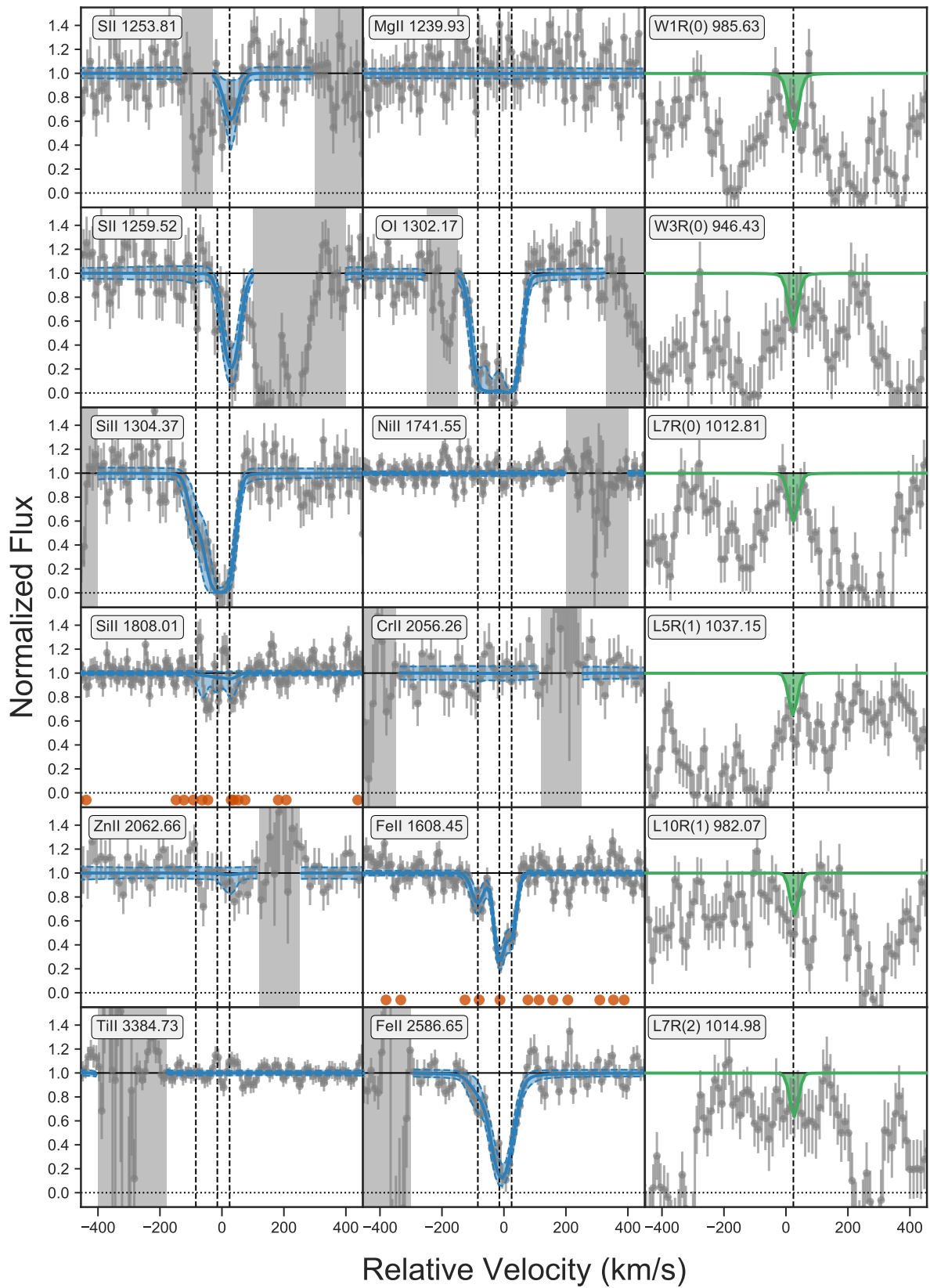

**Fig. B.19.** Results from fitting the absorption lines in the X-shooter spectrum of GRB151027B. The position of telluric lines, which were corrected, are marked by red dots. We do not find compelling evidence for absorption from molecular hydrogen down to the limits given in Table A.2. To the right, in green, we plot synthetic spectra for these limits

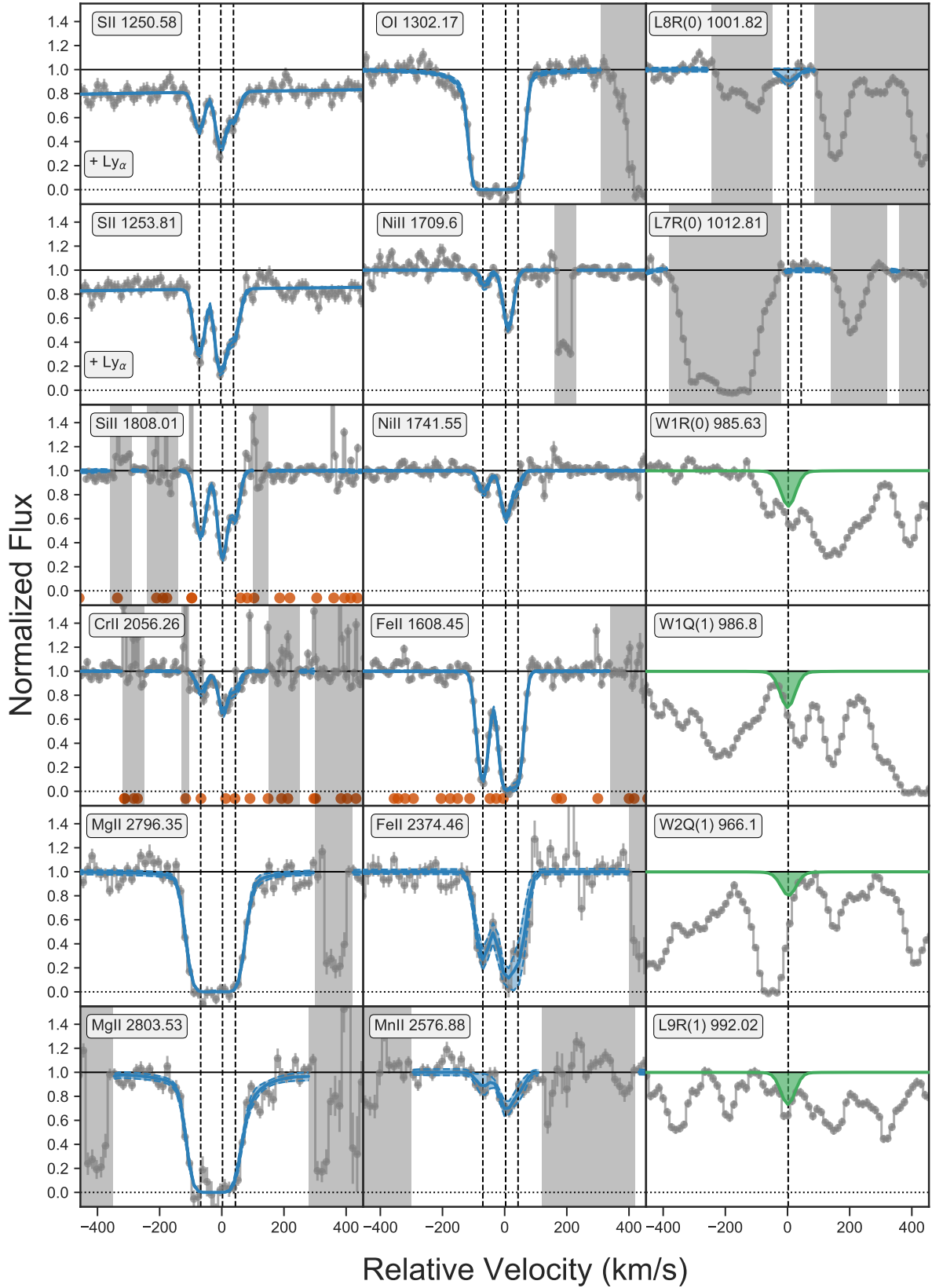

**Fig. B.20.** Results from fitting the absorption lines in the X-shooter spectrum of GRB160203A. The position of telluric lines, which were corrected, are marked by red dots. As shown in the 6 subplots to the right, the spectrum is consistent with absorption from molecular hydrogen, but only to the agree of claiming a *possible* detection. In green we plot synthetic spectra for the upper limits given in Table A.2.

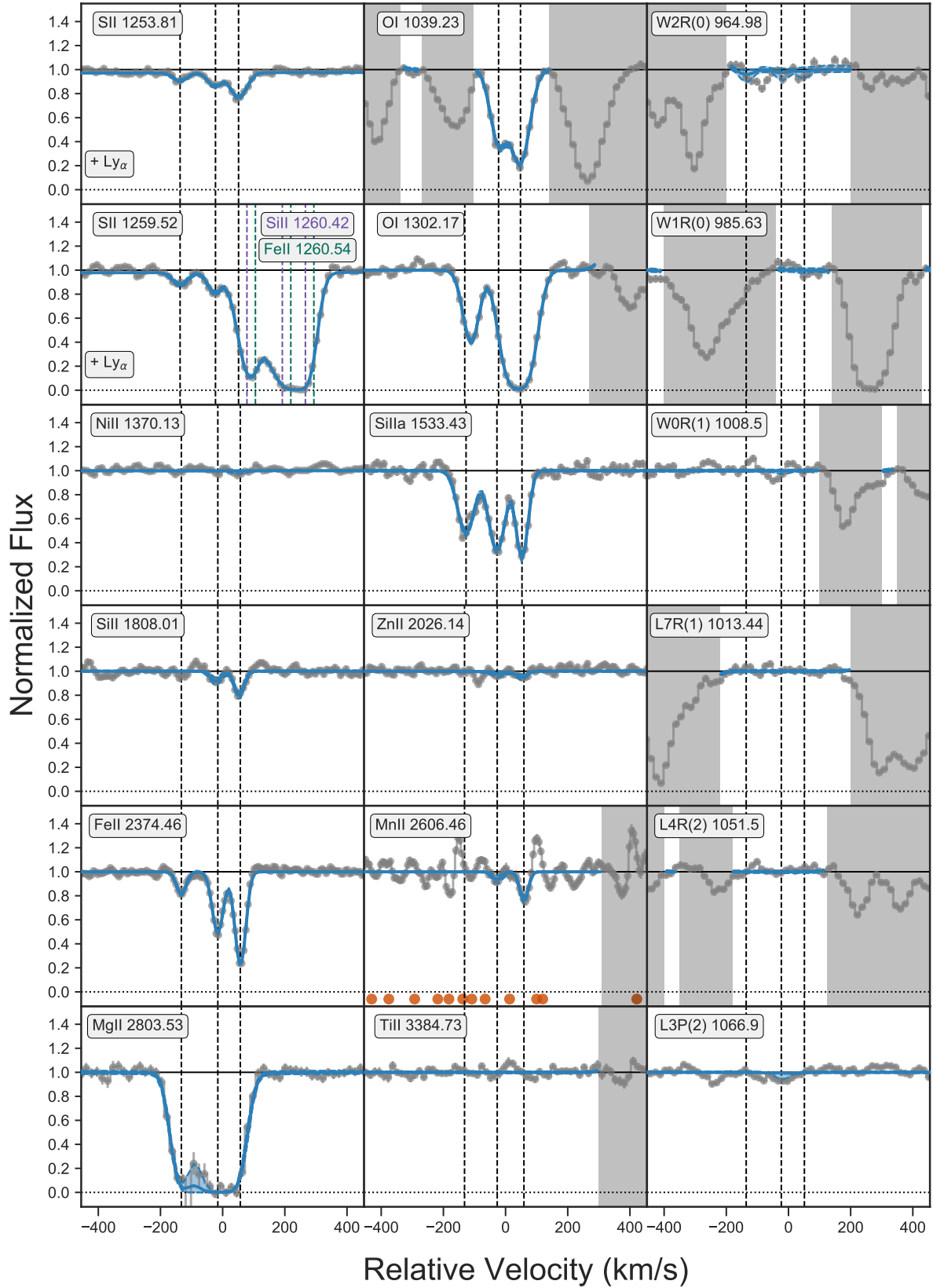

**Fig. B.21.** Results from fitting the absorption lines in the X-shooter spectrum of GRB161023A. The position of telluric lines, which were corrected, are marked by red dots. As shown to the right, we do not find any evidence for absorption from molecular hydrogen. This X-shooter spectrum was previously also analysed by de Ugarte Postigo et al. (2018).

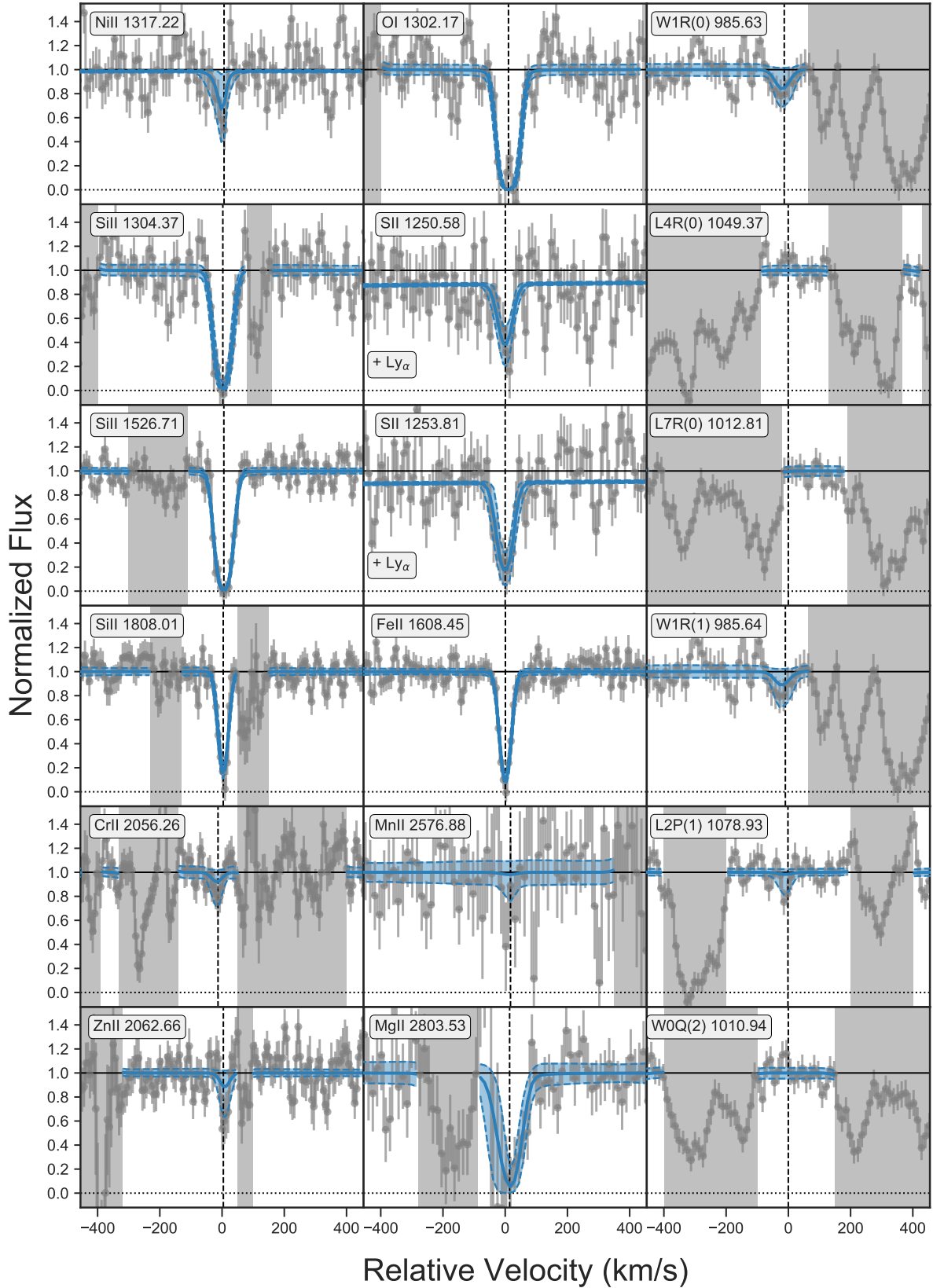

**Fig. B.22.** Results from fitting the absorption lines in the X-shooter spectrum of GRB170202A. As shown in the 6 subplots to the right, the spectrum is consistent with absorption from molecular hydrogen in the strongest J = 0 and J = 1 at $\lambda \sim 985.6$ Å, which led us to claim a *possible* detection.